\documentclass[aps,pre,preprint,numerical,a4paper,superscriptaddress]{revtex4-1}

\usepackage{chngcntr}
\usepackage{graphicx}
\usepackage[refpage,cfg,prefix]{nomencl}
\usepackage{parskip}
\usepackage{soul}
\usepackage{xcolor}
\usepackage{rotating}
\usepackage{wrapfig}
\usepackage{graphicx}
\usepackage{amsmath}
\usepackage{bbm}
\usepackage{amsfonts}
\usepackage{amssymb}
\usepackage{amsbsy}
\usepackage[T1]{fontenc}
\usepackage{theorem}
\usepackage{stmaryrd}
\usepackage{tipa}
\usepackage{textcomp}
\usepackage{siunitx}
\usepackage{subfigure}
\usepackage{epsfig}
\usepackage{lmodern}
\usepackage{framed}
\usepackage[subfigure]{tocloft}
\usepackage{subeqnarray}
\usepackage{alphalph}
\usepackage[refpage,cfg,prefix]{nomencl}
\usepackage{multirow}
\usepackage{xifthen}
\usepackage{enumerate}
\usepackage{enumitem}
\usepackage{color}
\usepackage{wasysym}
\usepackage{bibentry}
\usepackage{lipsum}
\usepackage{relsize}
\usepackage{verbatim}
\usepackage{hyperref}
\hypersetup{
colorlinks=true,
urlcolor=blue,
linkcolor=black,
citecolor=black,
breaklinks=true
}
\usepackage{breakurl}
\usepackage{url}
\usepackage{setspace}

\usepackage{tabularx}
\usepackage{booktabs}
\usepackage{subfiles}
\makenomenclature

\newcommand{\D}[2]{\frac{\partial #1}{\partial #2}}

\newcommand{\listofalgorithms}{\textbf{\Huge{List of Algorithms}}}
\newlistof{Algorithm}{exp}{\listofalgorithms}
\newcounter{instructioncounter}

\allowdisplaybreaks
\usepackage{mathtools}
\usepackage[margin={1in, 1in}]{geometry}
\usepackage[UKenglish]{babel}
\usepackage[UKenglish]{isodate}
\usepackage{appendix}
\usepackage{lastpage}
\usepackage{fancyhdr}

\newcommand*{\blankpage}{%
\vspace*{\fill}
{\fontfamily{phv}\selectfont{\centering This page intentionally left blank.\par}}
\vspace{\fill}}
\makeatletter
\renewcommand*{\cleardoublepage}{\clearpage\if@twoside \ifodd\c@page\else
\blankpage
\thispagestyle{empty}
\newpage
\if@twocolumn\hbox{}\newpage\fi\fi\fi}
\makeatother

\begin{document}

\title{Swapping in lattice-based cell migration models}

\author{Shahzeb Raja Noureen}
\email[Corresponding author: ]{srn32@bath.ac.uk}
\affiliation{Centre for Mathematical Biology, University of Bath, Claverton Down, Bath, BA2 7AY, UK.}

\author{Jennifer P. Owen}
\affiliation{Centre for Mathematical Biology, University of Bath, Claverton Down, Bath, BA2 7AY, UK.}

\author{Richard L. Mort}
\affiliation{Division of Biomedical and Life Sciences, Faculty of Health and Medicine, Furness Building, Lancaster University, Bailrigg, Lancaster, LA1 4YG, UK.}

\author{Christian A. Yates}
\affiliation{Centre for Mathematical Biology, University of Bath, Claverton Down, Bath, BA2 7AY, UK.}

\date{\today}

\begin{abstract}
Cell migration is frequently modelled using on-lattice agent-based models (ABMs) that employ the excluded volume interaction. However, cells are also capable of exhibiting more complex cell-cell interactions, such as adhesion, repulsion, pulling, pushing and swapping. Although the first four of these have already been incorporated into mathematical models for cell migration, swapping has not been well studied in this context. In this paper, we develop an ABM for cell movement in which an active agent can `swap' its position with another agent in its neighbourhood with a given swapping probability. We consider a two-species system for which we derive the corresponding macroscopic model and compare it with the average behaviour of the ABM. We see good agreement between the ABM and the macroscopic density. We also analyse the movement of agents at an individual level in the single-species as well as two-species scenarios to quantify the effects of swapping on an agent's motility.
\\

\end{abstract}

\maketitle

\section{Introduction}

Cell migration is an essential biological process required for the correct development of tissues and organs during embryonic development and their proper maintenance, through wound healing and tissue homeostasis, throughout life  \citep{lauffenburger1996cmp,plikus2021fod,kulesa1998ncc,poujade2007cme,deng2006rmp}. During embryonic development, neural crest cells delaminate from the dorsal most aspect of the neural tube and migrate to colonise their target tissues including the gut in the case of enteric ganglia precursors and the skin in the case of melanoblasts the precursors of melanocytes \citep{spritz1994mbh}. Diseases of the neural crest are known as neurocristopathies for example failure of the enteric ganglia precursors to colonise the developing gut results in Hirschsprung’s disease \citep{bondurand2016mmh} while failure of melanoblasts to colonise the developing epidermis results in piebaldism \citep{mort2016rdm}. Therefore an in-depth understanding of cell migration is important for identifying the causes of neurocristopathies \citep{mayor2016frc,giniunaite2020mcc,thomas2008mms,wang2011asn} as well as developing new therapeutic targets to prevent metastasis in cancers \citep{mayor2016frc,deroulers2009mtc}.


Traditionally, many biological problems have been modelled using deterministic methods. However, in cell migration, randomness can play a salient role in determining a cell's trajectory and fate and hence deterministic theory may not be appropriate. Extensive research has gone into modelling cell movement as a stochastic process \citep{glazier1993sda,graner1992sbc,smith2012vma,nonaka2011mmc,osborne2017cia}. In one widely used approach, cells are modelled as agents whose positions evolve probabilistically in space and time according to a predefined set of rules. These models are commonly known as agent-based models (ABMs) or individual-based models (IBMs). The agent-based modelling paradigm can be sub-divided into off-lattice and on-lattice models, both of which have wide applicability to different problems within mathematical biology. \citet{gavagnin2018sdm} recently reviewed the most commonly used ABMs for cell movement.

In this paper, we only concern ourselves with on-lattice models of cell movement. In a lattice-based approach, the domain is divided into a series of compartments in which the cells reside. Cells take up space, preventing other cells from occupying the same space at the same time. For biological plausibility, it is often desirable that mathematical models of cell migration account for the single occupancy of sites. This realism is incorporated in an ABM via the volume-exclusion principle, which states that a cell attempting to move into a neighbouring site successfully moves only if the neighbouring site is not already occupied at the time of moving.

Models with volume exclusion at their core have been used to describe the collective migration of cells for a wide range of biological applications. \citet{mort2016rdm} used an on-lattice ABM with the exclusion principle to model the invasion of the developing epidermis by melanoblasts (the embryonic precursors of melanocytes) and investigate the basis of piebaldism in mice. Lattice-based exclusion models have also been applied to wound healing \citep{khain2007rca}, migration of breast cancer cells \citep{simpson2010mbc}, developmental processes on growing domains \citep{baker2010mmd}, cells' responses to chemotaxis \citep{charteris2014mca} and cells exhibiting pushing \citep{yates2015ipe} and pulling \citep{chappelle2019pmc} interactions in densely crowded environments.

\begin{figure}[t!]
    \centering
    \includegraphics[width=\textwidth]{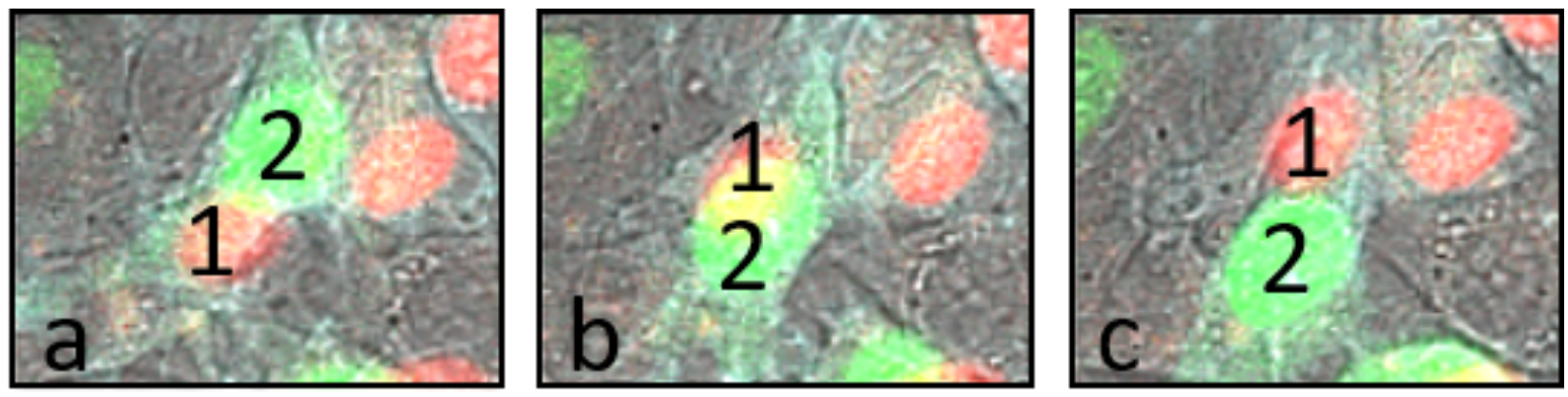}
    \caption{Two \textit{Fucci2a} labelled NIH 3T3 mouse embryonic fibroblasts swapping places with each other in culture \citep{mort2014fbc}. The colour of the cells represents their cell cycle stage (Red = G1, Green = S/G2/M) in this case making it easy to observe the swap. In (a), we show the initial placement of cells: cell 1 (shown in red) is on the bottom-left of cell 2 (shown in green). In (b) a swap starts to take place and in (c) the swap is complete and now cell 1 is on the top-right of cell 2.}
    \label{fig:cell_swapping_1}
\end{figure}


Biologically, although cells are excluded from the space occupied by other cells, their movement is not completely inhibited by them as typically assumed in volume-exclusion models. For example  melanoblasts are able to move freely between keratinocytes in the developing epidermis \citep{mort2016rdm}. Experimental data suggests that cells are often able to move past each other (passing laterally, above or below) exchanging places with one another. In Figure \ref{fig:cell_swapping_1} we show experimental images of two \textit{Fucci2a} labelled NIH-3T3 fibroblasts exhibiting the swapping behaviour. Swapping has also been observed in blood cells such as leukocytes, erythrocytes and thrombocytes \citep{Lan2014nsp} and in pattern formation for maintaining sharp boundaries between different groups of cells as part of a cell sorting mechanism \citep{dahmann2011bfm}. These examples highlight the importance of incorporating swapping into models of cell migration. To the best of our knowledge, this has not yet been explored thoroughly from a mathematical perspective.

In this paper, we develop a mathematical model to describe and analyse cell-cell swapping in two species setting. By modifying the movement rules of the traditional volume-exclusion process, we show that swapping between agents has an effect on the migration of agents at different spatial resolutions. To investigate how swapping manifests itself in the corresponding population-level model (PLM) we derive a set of partial differential equations (PDEs) describing the macroscopic dynamics of the agents. We compare numerical solutions of the PDEs with the averaged results from the ABM and comment on the agreement or discrepancy between them. We also analyse the movement of agents at an individual level and derive expressions for the individual-level diffusion coefficient.

The remainder of this paper is structured as follows. In Section \ref{sec:abm}, we develop a model that allows for swapping to take place between pairs of neighbouring agents. In Section \ref{sec:sm_pde}, we derive the macroscopic PDEs and compare
the average behaviour of the ABM to that of the PDEs. In Section \ref{sec:individual_level_analysis}, we analyse the movement of agents at an individual level and derive a relationship between the individual-level
diffusion coefficient, swapping probability and background domain density.
In Section \ref{sec:examples}, we give examples to illustrate the applications of swapping and finally in Section \ref{sec:discussion}, we conclude the paper with a summary and discussion.

\section{Cell migration model with swapping}\label{sec:swapping_model}

We begin by developing an ABM for cell movement with swapping in Section \ref{sec:abm} and we use this to investigate the effect of swapping on the mobility of agents. In particular, we look at the effect of swapping on mixing of agents in a two-species system. We derive the population-level model and compare this with the average behaviour of the ABM in Section \ref{sec:sm_pde}. We also analyse the individual-level behaviour in both single and multispecies scenarios in Section \ref{sec:individual_level_analysis}.

\subsection{On-lattice agent-based model}\label{sec:abm}

We model cell migration on a two-dimensional lattice. We discretise the domain into compartments (also known as `sites') such that there are $L_x$ compartments in the horizontal direction and $L_y$ compartments in the vertical direction. We assume that the compartments are square with side length $\mathrm{\Delta}$. Supposing that each compartment can contain no more than one agent, $\mathrm{\Delta}$ can be considered a rough proxy for a cell's diameter. A site $(i,j)$ for $i=1,...,L_x$ and $j=1,...,L_y$ can be either occupied by a type-A or a type-B agent or unoccupied. Occupancy of a site $(i,j)$ for a type-A (or type-B) agent is defined as a binary indicator, taking a value of 1 if there is a type-A (or type-B) agent at the site $(i,j)$ or 0 if the site is empty.

\begin{figure}[t!]
\begin{center}

\subfigure[]{
\includegraphics[width=0.22\textwidth]{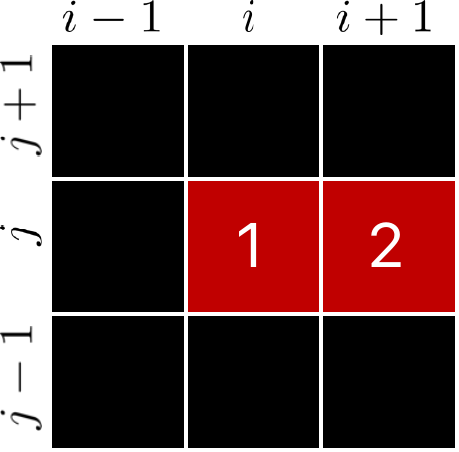}
\label{figure:lattice_sd}
}
\subfigure[]{
\includegraphics[width=0.22\textwidth]{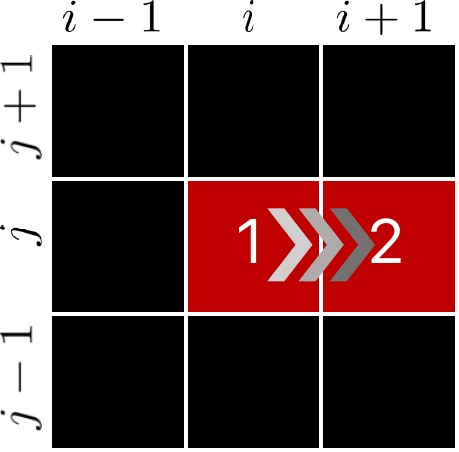}
\label{figure:lattice_swap1}
}
\subfigure[]{
\includegraphics[width=0.22\textwidth]{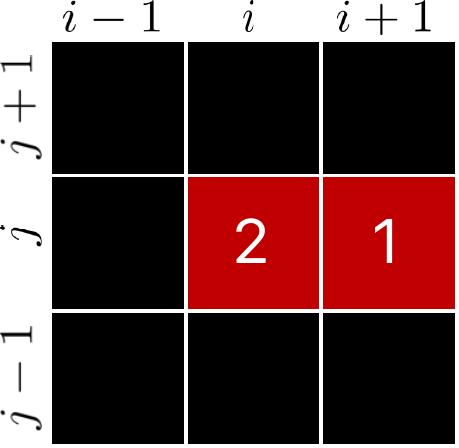}
\label{figure:lattice_swap2}
}
\end{center}
\caption{A schematic illustrating the swapping mechanism. Red sites are occupied with agents and black sites are unoccupied. The initial configuration of the lattice before is shown in \subref{figure:lattice_sd}. The agent chosen to move is at site $(i,j)$ and labelled 1. The target site is at position $(i+1,j)$ and the agent occupying the target site is labelled 2. Agent 1 attempts to move into the target site in \subref{figure:lattice_swap1}. The final configuration once the swapping move is complete is shown in \subref{figure:lattice_swap2}.}
\label{figure:swap_single_species}
\end{figure}

We initialise the lattice with species A and species B agents at densities $c_A$ and $c_B$, respectively, such that $c_A+c_B=c$ and $0 \leqslant c \leqslant 1$ where $c$ is the overall domain density. We let the positions of the agents evolve in continuous time according to the Gillespie algorithm \citep{gillespie1977ess}. Let $r_{A}$ be the rate of movement of a type-A agent and let $r_B$ be the equivalent for a type-B agent. The rates of movement are defined such that $r_A \delta t$ (and equivalently $r_B\delta t$) is the probability that a type-A (or type-B) agent attempts to move during a finite time interval of duration $\delta t$. The agent attempts to move into one of the four sites in its Von Neumann neighbourhood with equal probability. If the chosen neighbouring site is empty, the focal agent successfully moves and its position is updated. However, if another agent already occupies the site, the move is aborted \citep{gavagnin2018sdm,simpson2009dpg,yates2015ipe,chappelle2019pmc,simpson2009mse}. This blocking of the move characterises volume exclusion in our model.

Swapping works by modifying the rules of the exclusion process by allowing an exchange in the positions of a pair of neighbouring agents if the target site is already occupied. We now introduce the swapping parameter $\rho$ to denote the probability of a successful swap between a pair of neighbouring agents conditional on one of the agents attempting to move into the other's position. If $\rho=0$ then there are no swaps and we arrive back at the original exclusion process. If $\rho> 0$ then we can have different levels of swapping based on the value of $\rho$. For example, for $\rho=1$ each scenario in which a move is attempted into an occupied target site will be a successful swap and for $\rho=0.5$ half of the attempted moves into occupied sites will be successful.

\begin{figure}[t!]

\begin{center}
\subfigure[]{
\includegraphics[width=0.45\textwidth]{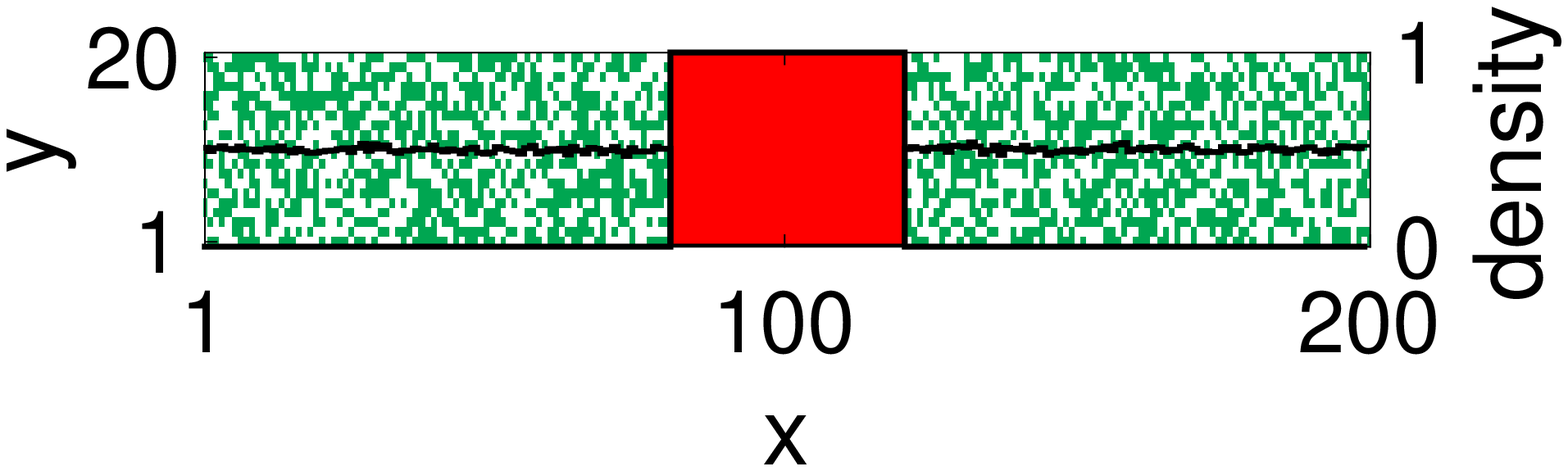}
\label{figure:two_species_rho=0_t=0}
}
\setcounter{subfigure}{3}
\subfigure[]{
\includegraphics[width=0.45\textwidth]{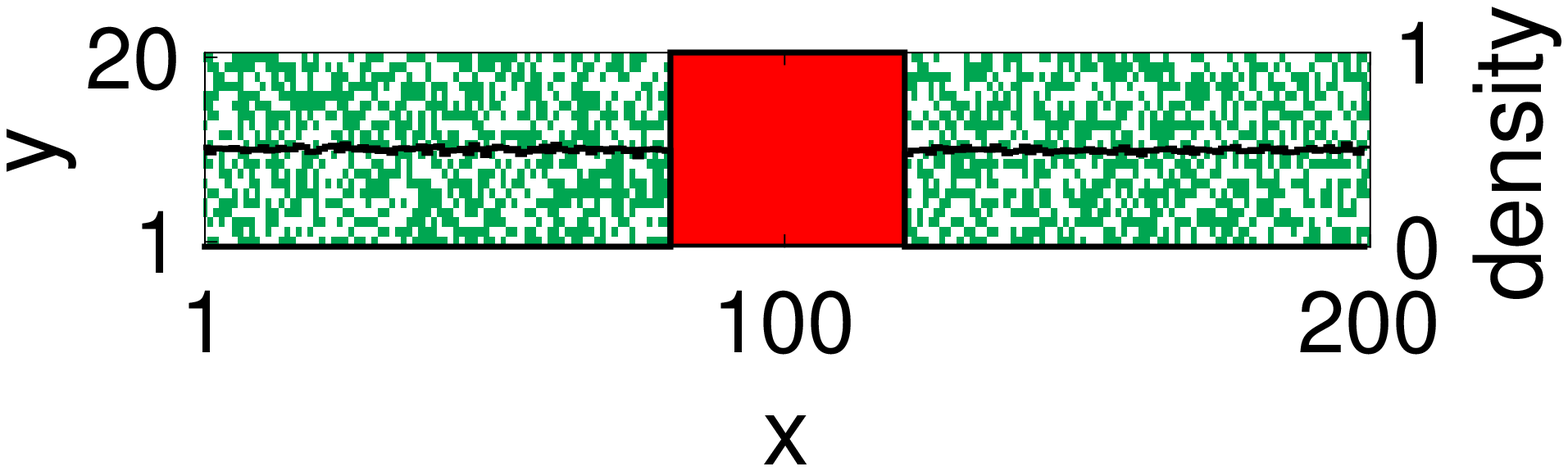}
\label{figure:two_species_rho=0.5_t=0}
}

\setcounter{subfigure}{1}
\subfigure[]{
\includegraphics[width=0.45\textwidth]{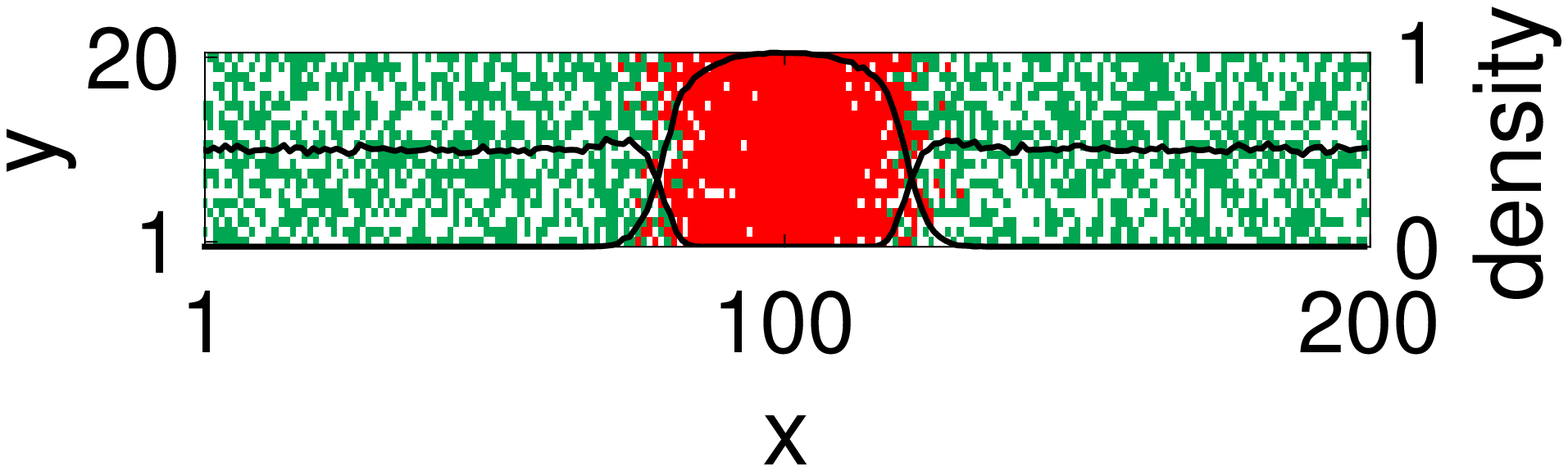}
\label{figure:two_species_rho=0_t=100}
}
\setcounter{subfigure}{4}
\subfigure[]{
\includegraphics[width=0.45\textwidth]{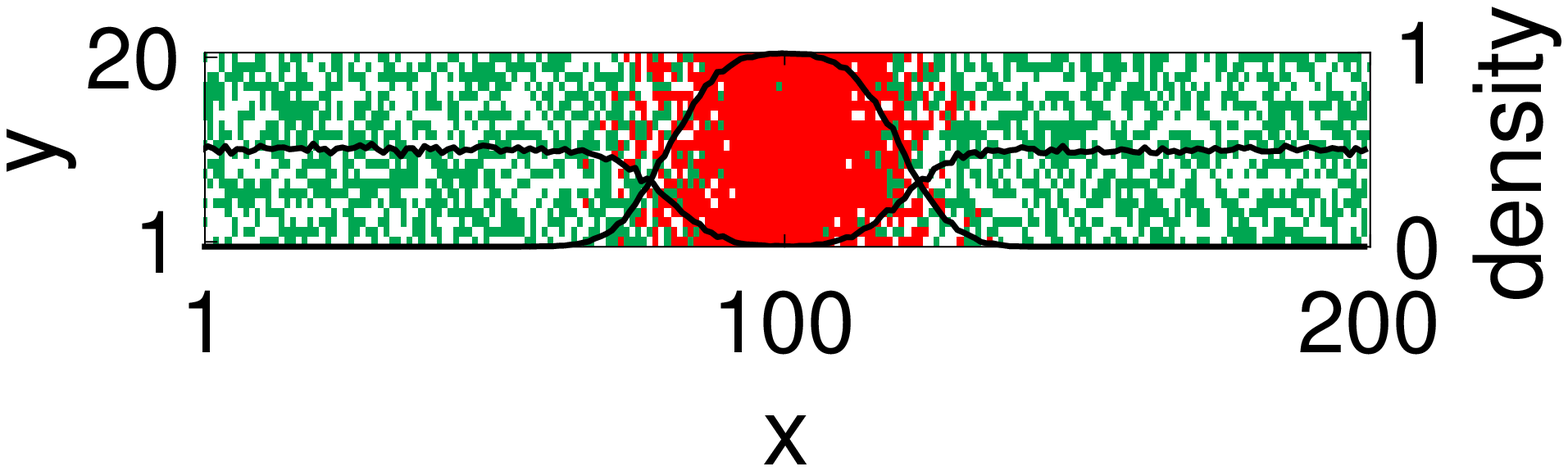}
\label{figure:two_species_rho=0.5_t=100}
}

\setcounter{subfigure}{2}
\subfigure[]{
\includegraphics[width=0.45\textwidth]{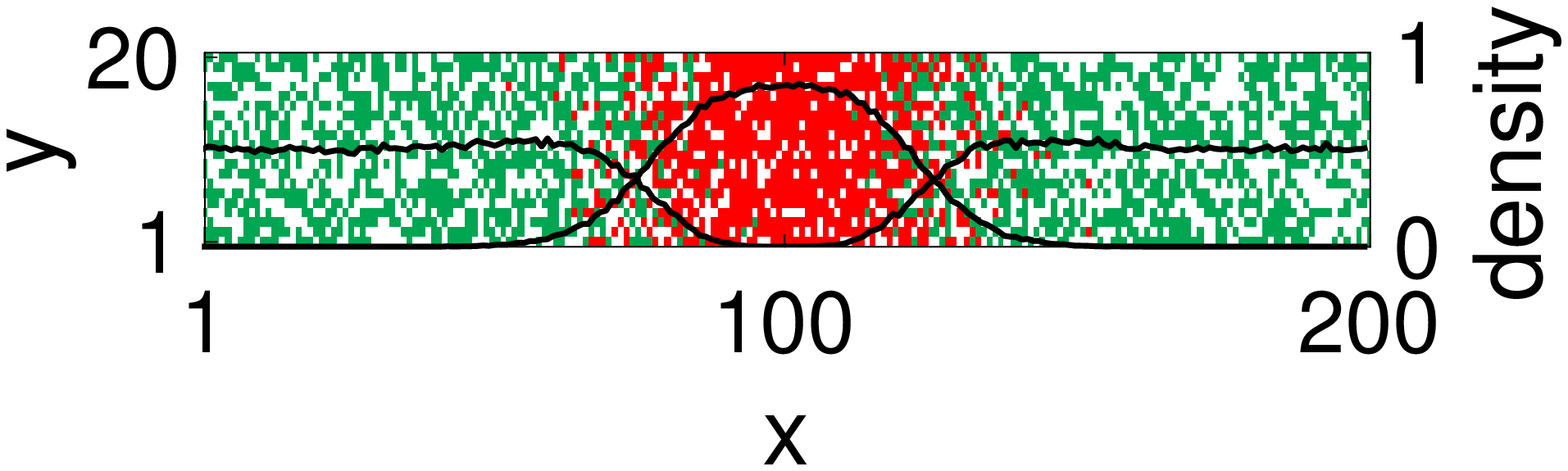}
\label{figure:two_species_rho=0_t=1000}
}
\setcounter{subfigure}{5}
\subfigure[]{
\includegraphics[width=0.45\textwidth]{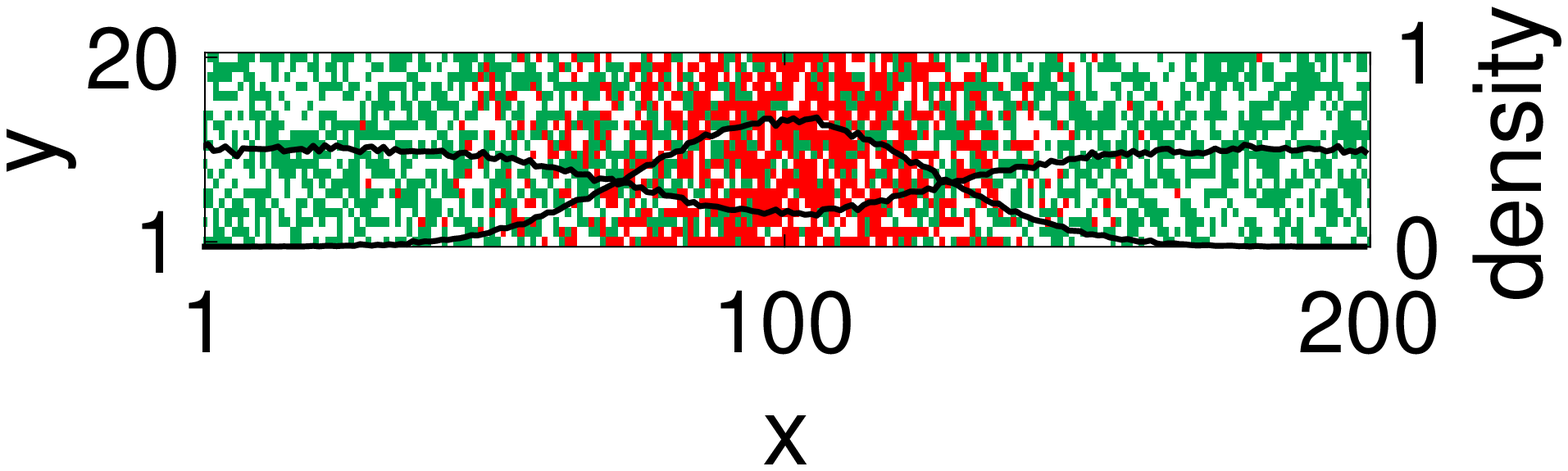}
\label{figure:two_species_rho=0.5_t=1000}
}

\subfigure[]{
\includegraphics[width=0.45\textwidth]{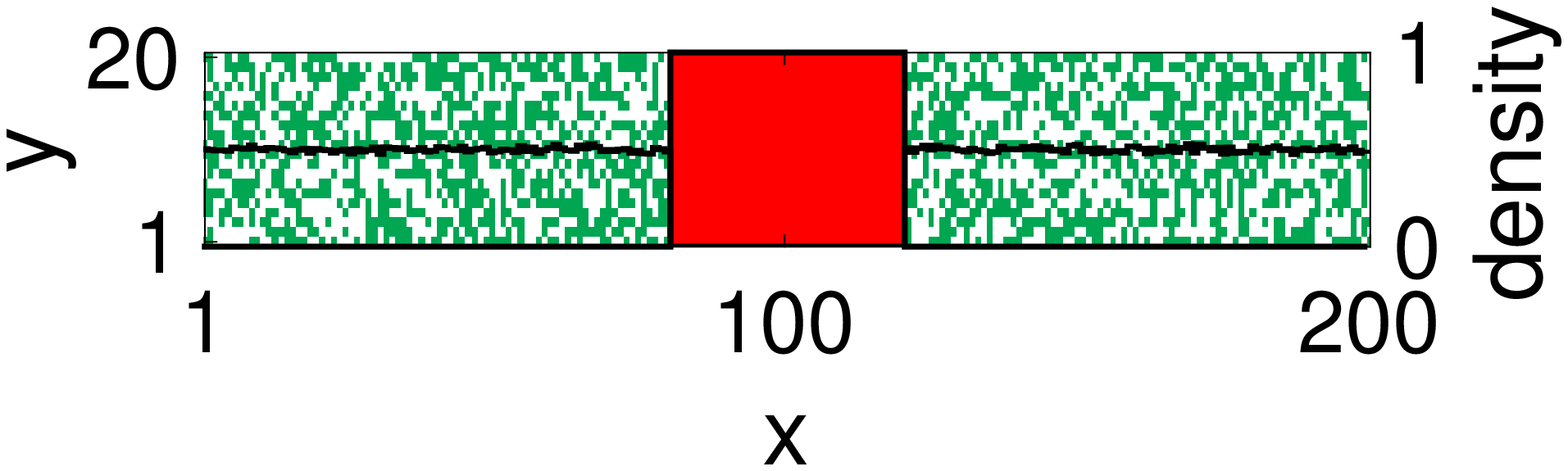}
\label{figure:two_species_rho=1_t=0}
}

\subfigure[]{
\includegraphics[width=0.45\textwidth]{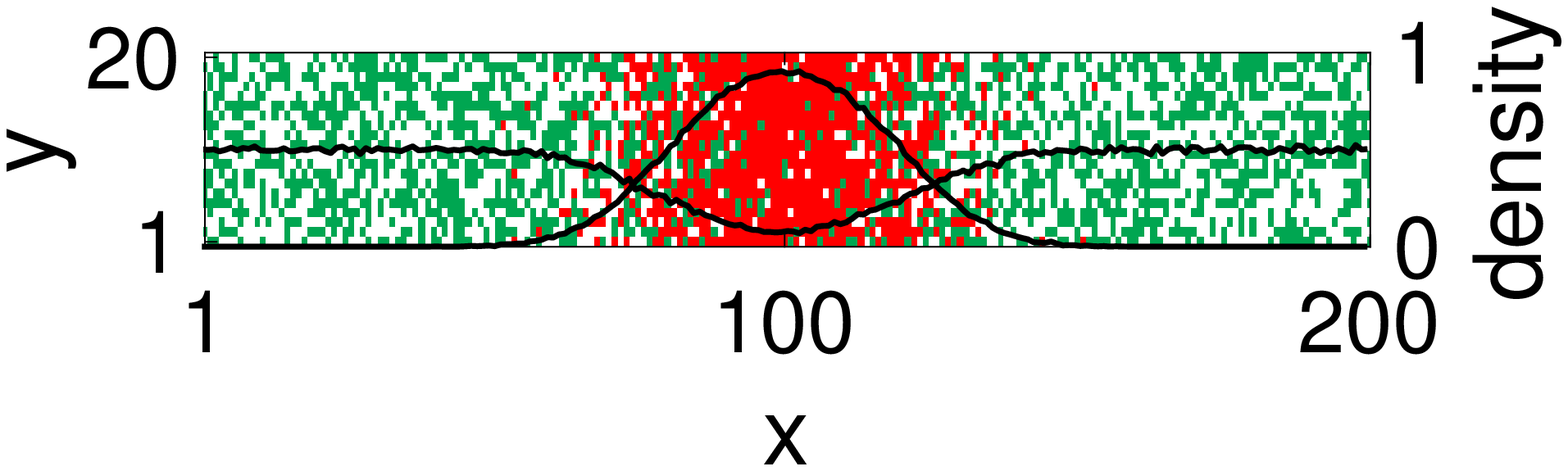}
\label{figure:two_species_rho=1_t=100}
}

\subfigure[]{
\includegraphics[width=0.45\textwidth]{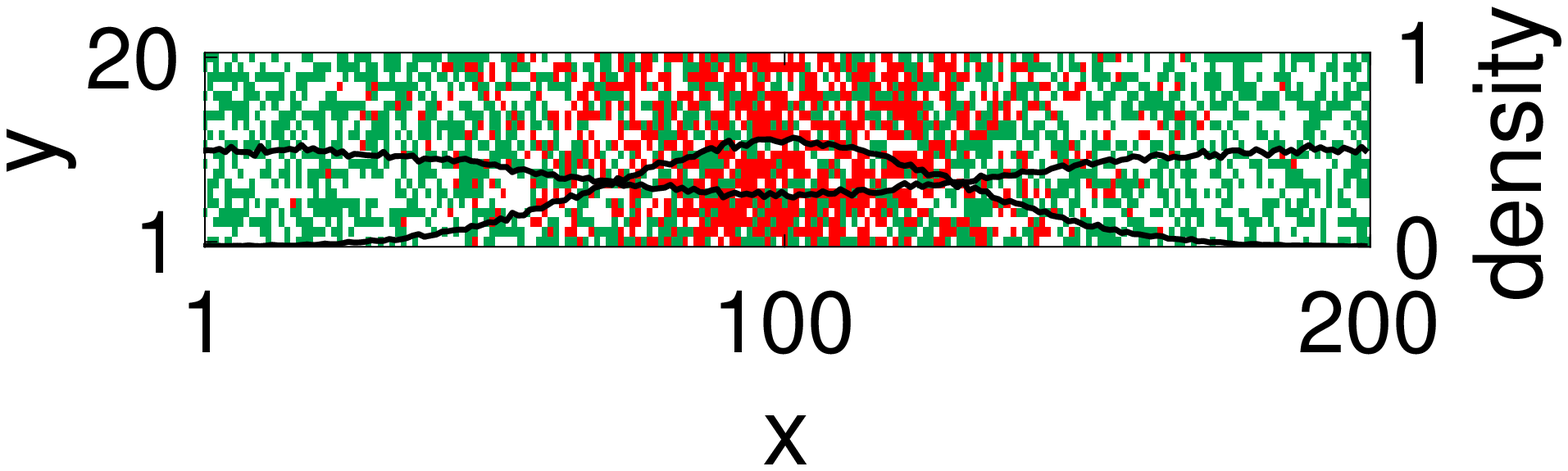}
\label{figure:two_species_rho=1_t=1000}
}

\end{center}
\caption{Snapshots of the lattice occupancy of the multi-species swapping model at $t=0,100,1000$ using $r_{A}=r_B=1$ for swapping probabilities $\rho=0$ [\subref{figure:two_species_rho=0_t=0}-\subref{figure:two_species_rho=0_t=1000}], $\rho=0.5$ [\subref{figure:two_species_rho=0.5_t=0}-\subref{figure:two_species_rho=0.5_t=1000}] and $\rho=1$ [\subref{figure:two_species_rho=1_t=0}-\subref{figure:two_species_rho=1_t=1000}]. Agents are initialised on a domain with dimensions $L_x=200$ and $L_y=20$ such that all the sites in the range $81 \leqslant x \leqslant 120$ are occupied by type-A agents (red) and the remaining sites are randomly populated with agents of type B (green) at a density of 0.5 [\subref{figure:two_species_rho=0_t=0},\subref{figure:two_species_rho=0.5_t=0},\subref{figure:two_species_rho=1_t=0}]. Further snapshots of the ABM at $t=100$ and $t=1000$ show the dispersal of agents with time. The column-averaged density of the two species over 100 runs of the ABM is also plotted (shown in black). We impose reflective boundary conditions on all four boundaries of the domain.}
\label{figure:multispecies_density}
\end{figure}

To implement swapping, we sample a random number $u$ from the uniform distribution over the unit interval $(0,1)$. If $u<\rho$, the agent at the site $(i,j)$ swaps with the agent at the target site (Figure \ref{figure:lattice_sd}-\subref{figure:lattice_swap2}), otherwise the move is aborted. Figure \ref{figure:swap_single_species} shows a successful swap between two agents labelled `1' and `2'.

In this article we present results for the two-species model only unless stated otherwise. Results for the single-species model can be found in Appendix \ref{app:appendixA}.

In Figure \ref{figure:multispecies_density} we present snapshots of lattice occupancy of a 200 by 20 grid occupied with type-A (red) and type-B (green) agents. The two species move with equal rates, $r_{A}=r_B=1$. We see that a non-zero swapping probability results in faster dispersion of the agents compared to the $\rho=0$ case. We also note that increasing the swapping probability from $\rho=0.5$ the $\rho=1$ results in faster dispersion of the agents, as expected.

In the next section, we derive the macroscopic PDEs describing the evolution of the mean lattice occupancy. By analysing the PDEs we generate further insight into the behaviours observed in Figure \ref{figure:multispecies_density}.

\subsection{Continuum model}\label{sec:sm_pde}

Let $A_{ij}^k(t)$ be the occupancy of site $(i,j)$ at time $t$ on the $k$th repeat of the ABM such that $A_{ij}^k(t)=1$ if the site is occupied by a type-A agent and 0 otherwise. Let $B_{ij}^k(t)$ be the same for a type-B agent. Then the average density of type-A and type-B agents after $K$ repeats is given by,

\begin{equation}
    \langle A_{ij}(t)\rangle=\frac{1}{K}\sum_{k=1}^K A_{ij}^k(t), \quad \text{and} \quad \langle B_{ij}(t)\rangle=\frac{1}{K}\sum_{k=1}^K B_{ij}^k(t).
\end{equation}

In what follows we will typically drop the notation for time dependence of our species densities, i.e. $\langle A_{ij}(t)\rangle = \langle A_{ij}\rangle$ and $\langle B_{ij}(t)\rangle = \langle B_{ij}\rangle$, for conciseness. By considering all the possible ways in which the site $(i,j)$ can gain or lose occupancy of either type-A or type-B agents during the time step $\delta t$, we can write down the corresponding occupancy master equations at time $t+\delta t$:

\begin{align}
    \langle A_{ij} (t+\delta t)\rangle-\langle A_{ij} \rangle&=\frac{r_{A}}{4}\delta t[(1-\langle A_{ij} \rangle-\langle B_{ij} \rangle)(\langle A_{i-1,j} \rangle+\langle A_{i+1,j} \rangle+\langle A_{i,j-1} \rangle+\langle A_{i,j+1} \rangle) \nonumber\\
    & \quad -\langle A_{ij} \rangle(4-\langle A_{i-1,j} \rangle-\langle A_{i+1,j} \rangle-\langle A_{i,j-1} \rangle-\langle A_{i,j+1} \rangle-\langle B_{i-1,j} \rangle \nonumber\\
    & \quad -\langle B_{i+1,j} \rangle -\langle B_{i,j-1} \rangle -\langle B_{i,j+1} \rangle)] \nonumber \\ 
    & \quad + \frac{(r_{A}+r_B)}{4}\rho \delta t \langle B_{ij} \rangle(\langle A_{i-1,j} \rangle   +\langle A_{i+1,j} \rangle +\langle A_{i,j-1} \rangle+\langle A_{i,j+1} \rangle)\nonumber \\
    &\quad -\frac{(r_{A}+r_B)}{4}\rho \delta t \langle A_{ij} \rangle(\langle B_{i-1,j} \rangle+\langle B_{i+1,j} \rangle+\langle B_{i,j-1} \rangle+\langle B_{i,j+1} \rangle), \label{eqn:master_eqn_M}
\end{align}

\begin{align}
    \langle B_{ij} (t+\delta t)\rangle-\langle B_{ij} \rangle&=\frac{r_B}{4}\delta t[(1-\langle B_{ij} \rangle-\langle A_{ij} \rangle)(\langle B_{i-1,j} \rangle+\langle B_{i+1,j} \rangle+\langle B_{i,j-1} \rangle+\langle B_{i,j+1} \rangle) \nonumber\\
    & \quad -\langle B_{ij} \rangle(4-\langle B_{i-1,j} \rangle-\langle B_{i+1,j} \rangle-\langle B_{i,j-1} \rangle -\langle B_{i,j+1} \rangle-\langle A_{i-1,j} \rangle \nonumber\\
    & \quad -\langle A_{i+1,j} \rangle -\langle A_{i,j-1} \rangle -\langle A_{i,j+1} \rangle)] \nonumber\\
    & \quad + \frac{(r_{A}+r_B)}{4}\rho \delta t \langle A_{ij} \rangle(\langle B_{i-1,j} \rangle+\langle B_{i+1,j} \rangle+\langle B_{i,j-1} \rangle+\langle B_{i,j+1} \rangle)\nonumber \\
    &\quad - \frac{(r_{A}+r_B)}{4}\rho \delta t \langle B_{ij} \rangle(\langle A_{i-1,j} \rangle + \langle A_{i+1,j} \rangle+\langle A_{i,j-1} \rangle+\langle A_{i,j+1} \rangle). \label{eqn:master_eqn_X}
\end{align}

For illustration, we describe the terms in Equation \eqref{eqn:master_eqn_M}. The terms in Equation \eqref{eqn:master_eqn_X} carry similar interpretation. A site $(i,j)$ can gain occupancy of type A in one of the following three ways:

\begin{enumerate}
    \item The site $(i,j)$ is unoccupied and a type-A agent moves in from a neighbouring site (line 1 in Equation \eqref{eqn:master_eqn_M}).
    
    \item The site $(i,j)$ is occupied by a type-B agent, which initiates and completes a swap with a type-A agent at a neighbouring site (line 3 in Equation \eqref{eqn:master_eqn_M}).
    
    \item The site $(i,j)$ is occupied by a type-B agent and a type-A agent at a neighbouring site initiates a swap to exchange positions with the type-B agent in the site $(i,j)$ (also line 3 in Equation \eqref{eqn:master_eqn_M}).
\end{enumerate}

In all three cases, a type-A agent moves into the site $(i,j)$. Similarly, there are three ways for a type-A agent to move \emph{out} of the site $(i,j)$ leading to a loss in the corresponding occupancy:

\begin{enumerate}
    \item The site is occupied by a type-A agent which jumps out to an unoccupied neighbouring site, leaving the site $(i,j)$ empty (line 2 in Equation \eqref{eqn:master_eqn_M}).
    
    \item The site $(i,j)$ is occupied by a type-A agent which initiates a swap with a type-B agent in its neighbourhood.
    
    \item The site $(i,j)$ is occupied by a type-A agent and a type-B agent in the neighbouring site initiates a swap to exchange positions with the agent in site $(i,j)$ (line 4 in Equation \eqref{eqn:master_eqn_M}).
\end{enumerate}

In all three cases, a type-A agent moves out of the site $(i,j)$.

To obtain the continuum model, we Taylor expand the appropriate terms on the RHS of Equations \eqref{eqn:master_eqn_M} and \eqref{eqn:master_eqn_X} around the site $(i,j)$ keeping terms of up to second order. By letting $\mathrm{\Delta} \to 0$ and $\delta t \to 0$ such that $\mathrm{\Delta}^2/\delta t$ is held constant, we arrive at the coupled PDEs,

\begin{align}
    \D A t &= \nabla \cdot [D_1(B) \nabla A + D_2(A)\nabla B], \label{eqn:pde_M}\\
    \D B t &= \nabla \cdot [D_3(A)\nabla B + D_4(B)\nabla A] \label{eqn:pde_X},
\end{align}
where,
\begin{equation}
    D_1(B)=D_A(1-B)+\rho(D_A+D_B)B, \qquad D_2(A)=(D_A -\rho(D_A+D_B))A,
\end{equation}

\begin{equation}
     D_3(A)=D_B(1-A)+\rho(D_A+D_B)A, \qquad D_4(B)=(D_B - \rho(D_A+D_B))B.
\end{equation}
Here,
\begin{equation*}
    D_A=\lim_{\mathrm{\Delta} \to 0} \frac{r_{A} \mathrm{\Delta}^2}{4}, \quad \text{and} \quad D_B=\lim_{\mathrm{\Delta} \to 0} \frac{r_B \mathrm{\Delta}^2}{4},
\end{equation*}
are the macroscopic diffusion coefficients corresponding to species A and B, respectively. Setting $\rho=0$ in equations \eqref{eqn:pde_M} and \eqref{eqn:pde_X} leads to,

\begin{align}
    \D A t &= D_A \nabla \cdot [(1-B)\nabla A + A\nabla B], \label{eqn:two_species_A_simpson_mss}\\
    \D B t &= D_B \nabla \cdot [(1-A)\nabla B + B\nabla A], \label{eqn:two_species_B_simpson_mss}
\end{align}
which are the macroscopic equations for the two-species volume-exclusion process \citep{simpson2009mse}.

\begin{figure}[h!]
\begin{center}

\subfigure[]{
\includegraphics[width=0.31\textwidth]{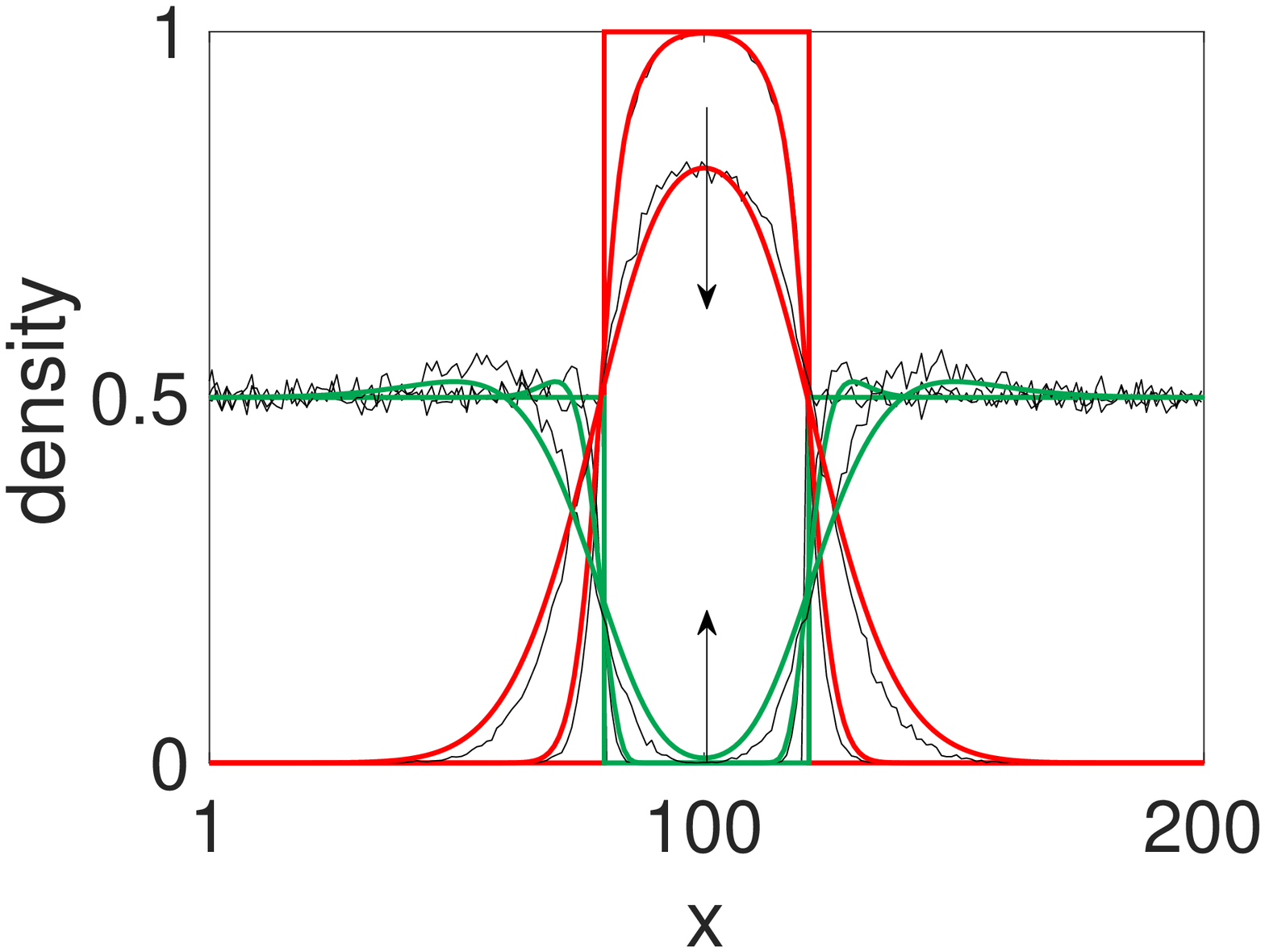}
\label{figure:IC=1_P_s=0}
}
\subfigure[]{
\includegraphics[width=0.31\textwidth]{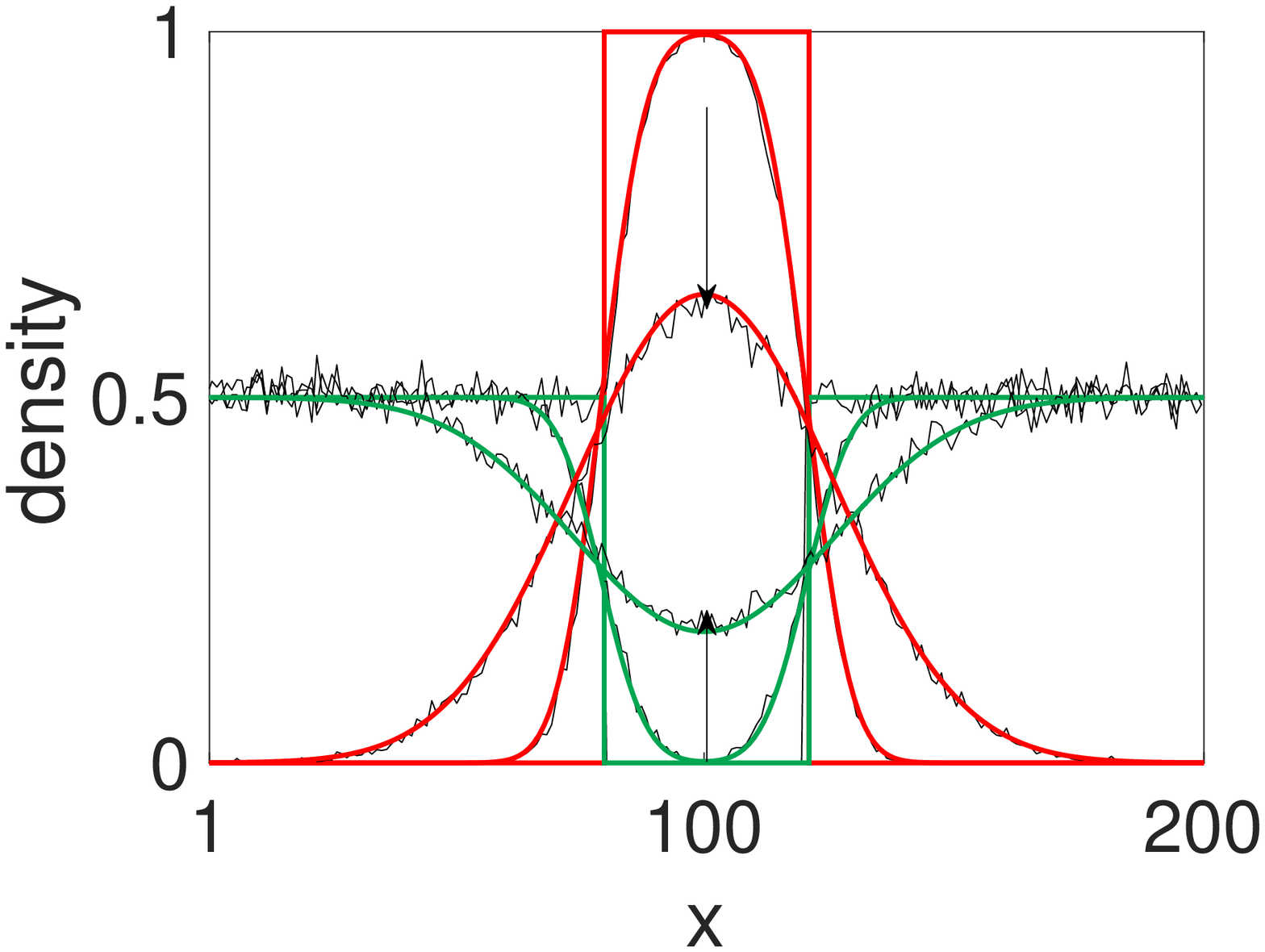}
\label{figure:IC=1_P_s=0.5}
}
\subfigure[]{
\includegraphics[width=0.31\textwidth]{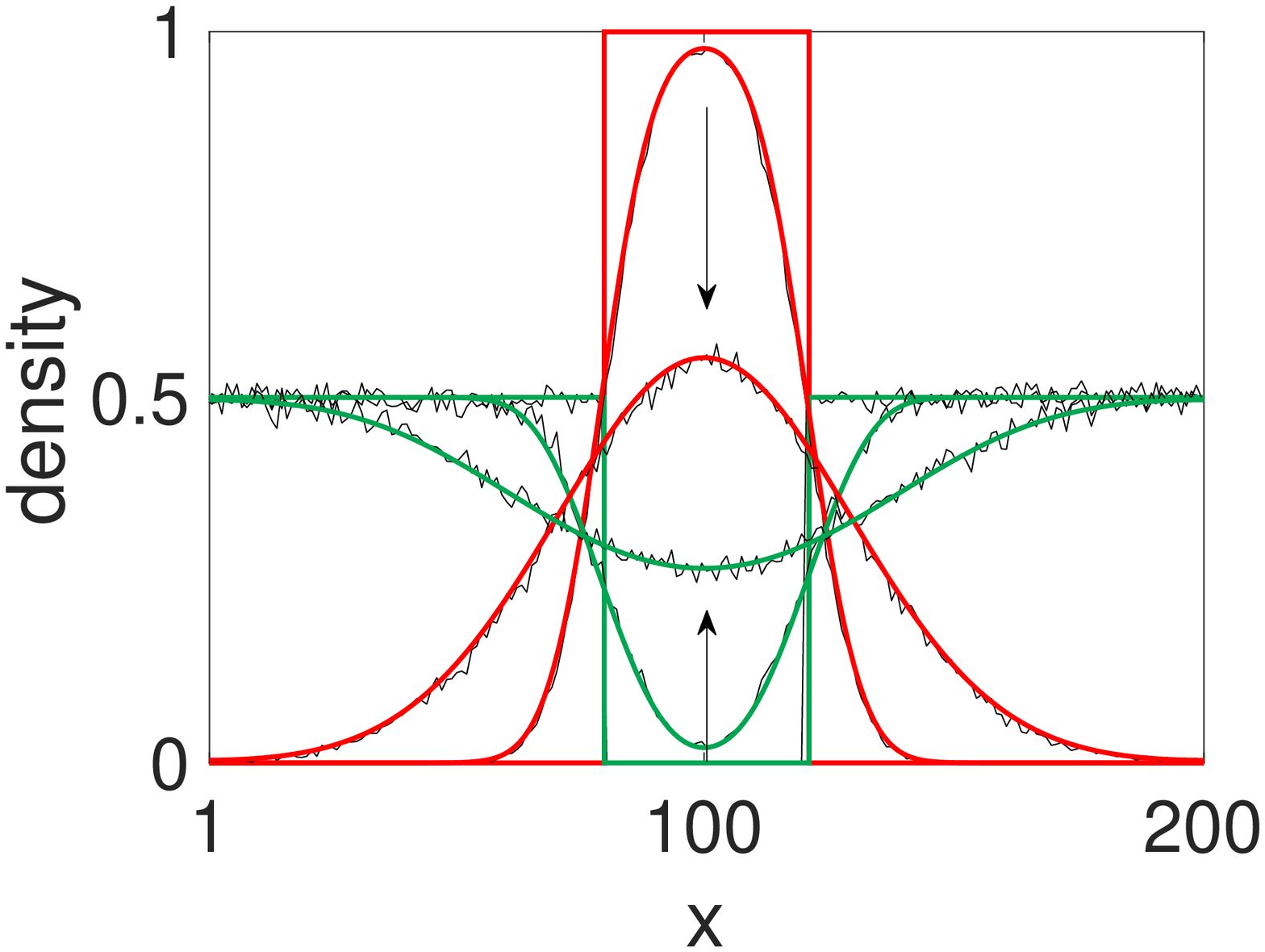}
\label{figure:IC=1_P_s=1}
}

\subfigure[]{
\includegraphics[width=0.31\textwidth]{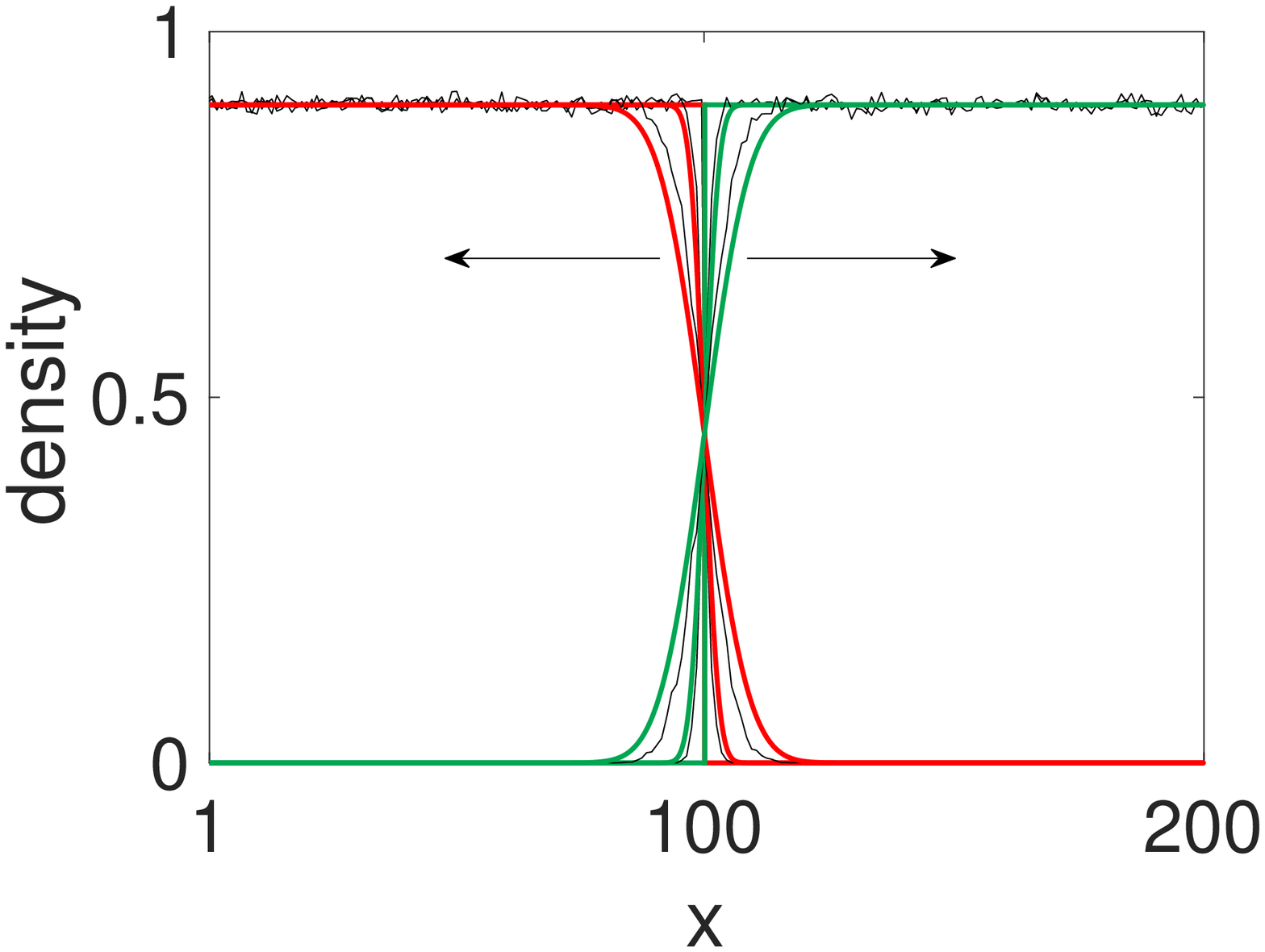}
\label{figure:IC=2_P_s=0}
}
\subfigure[]{
\includegraphics[width=0.31\textwidth]{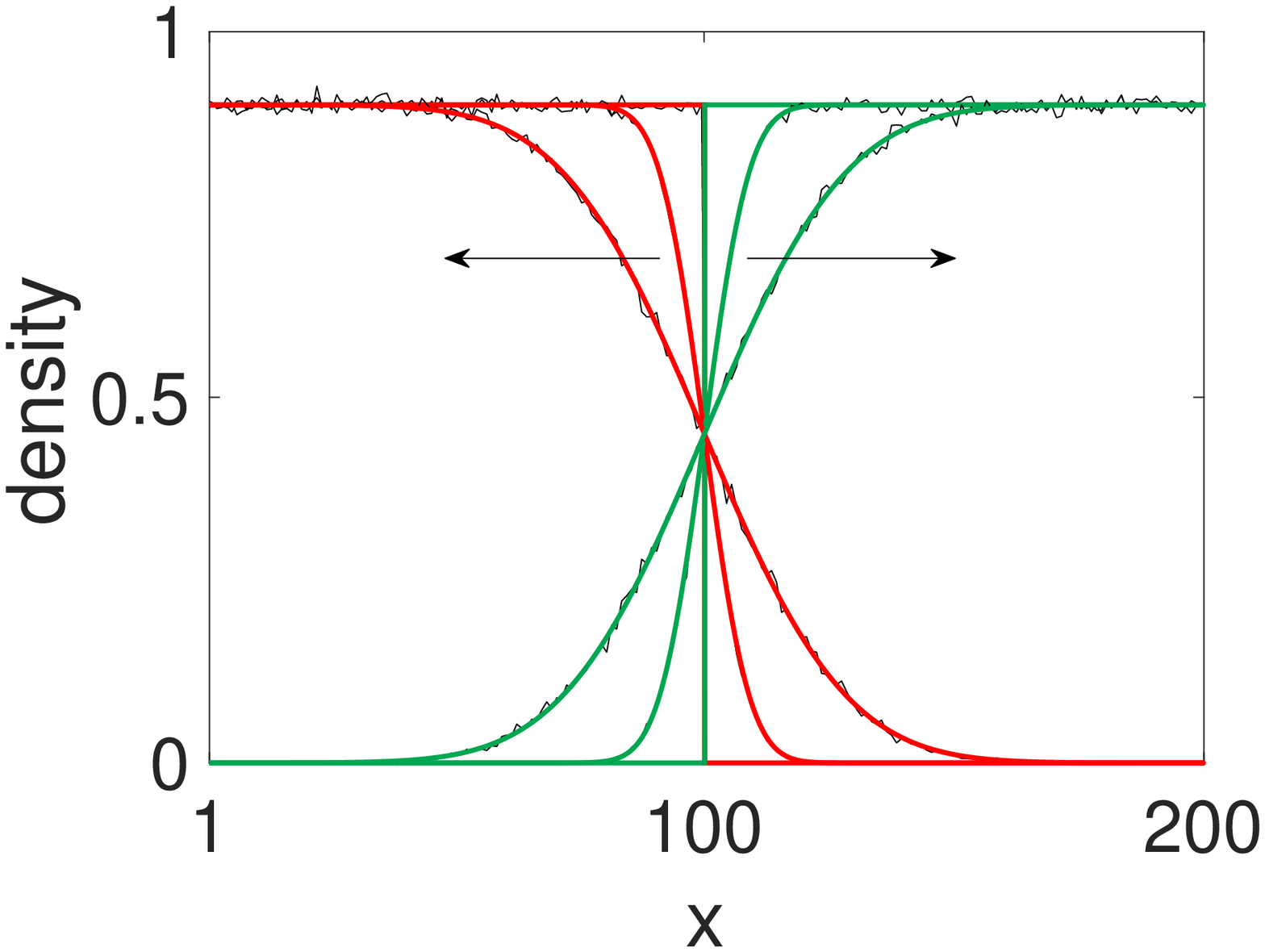}
\label{figure:IC=2_P_s=0.5}
}
\subfigure[]{
\includegraphics[width=0.31\textwidth]{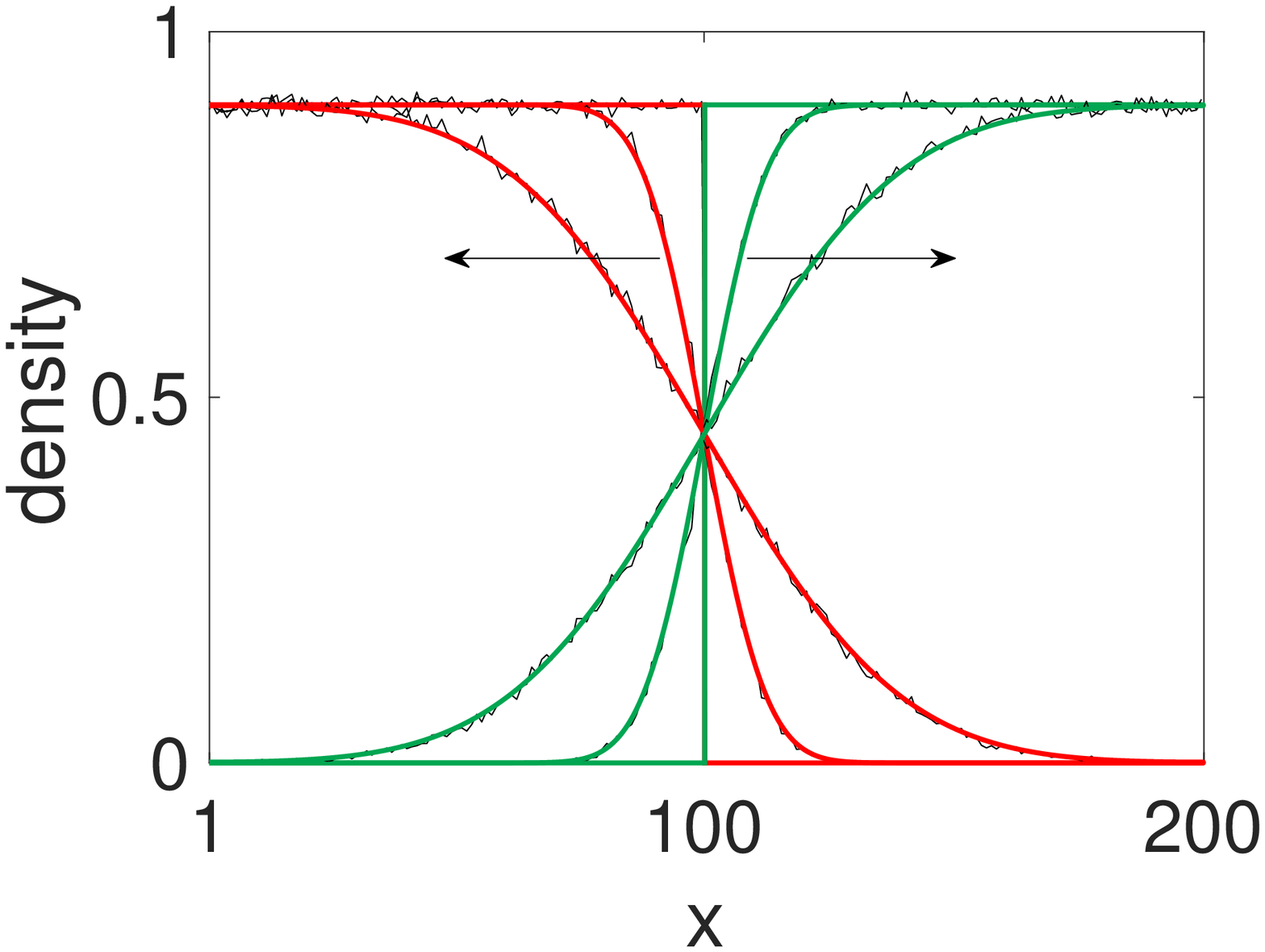}
\label{figure:IC=2_P_s=1}
}

\subfigure[]{
\includegraphics[width=0.31\textwidth]{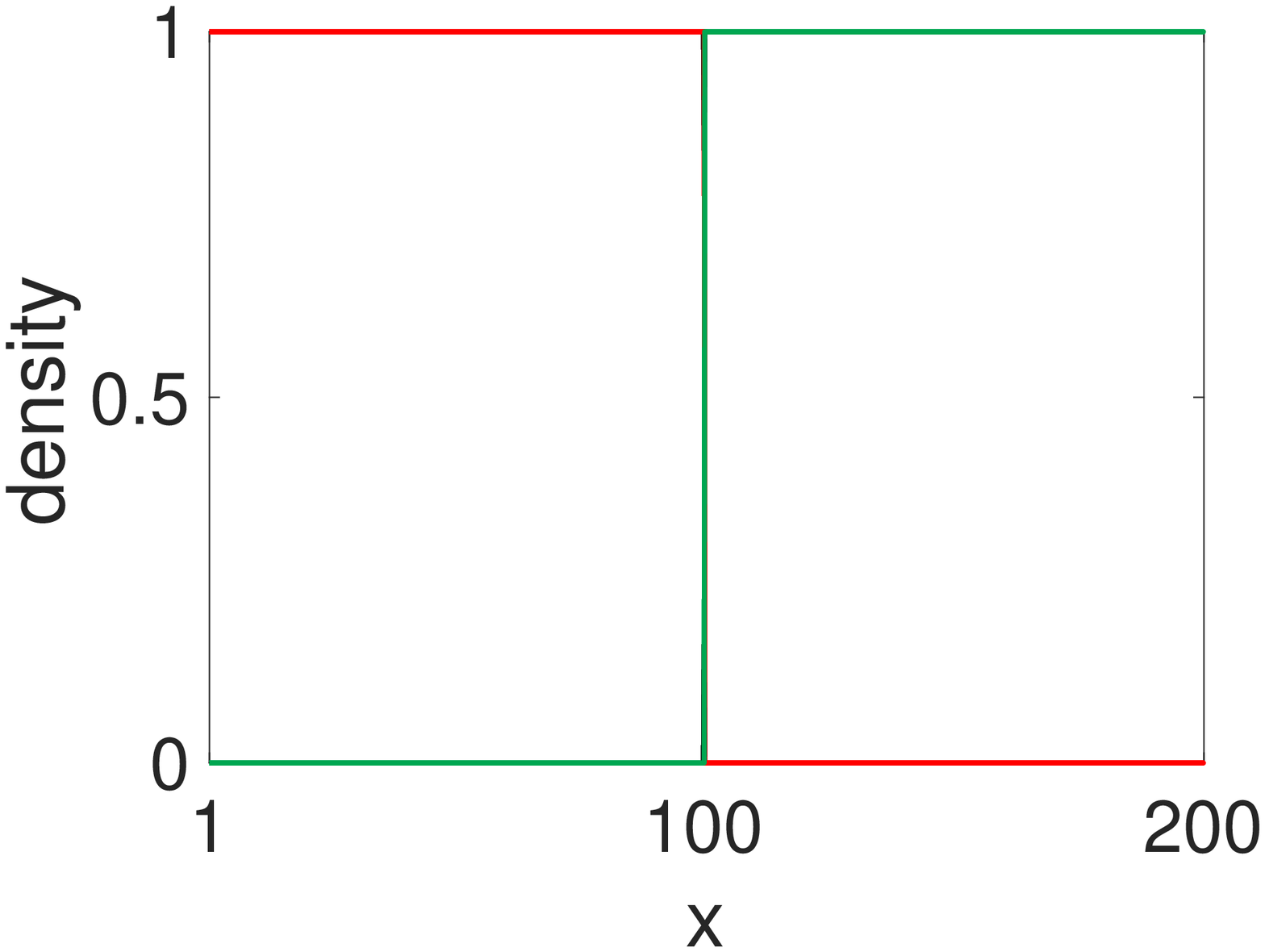}
\label{figure:IC=4_P_s=0}
}
\subfigure[]{
\includegraphics[width=0.31\textwidth]{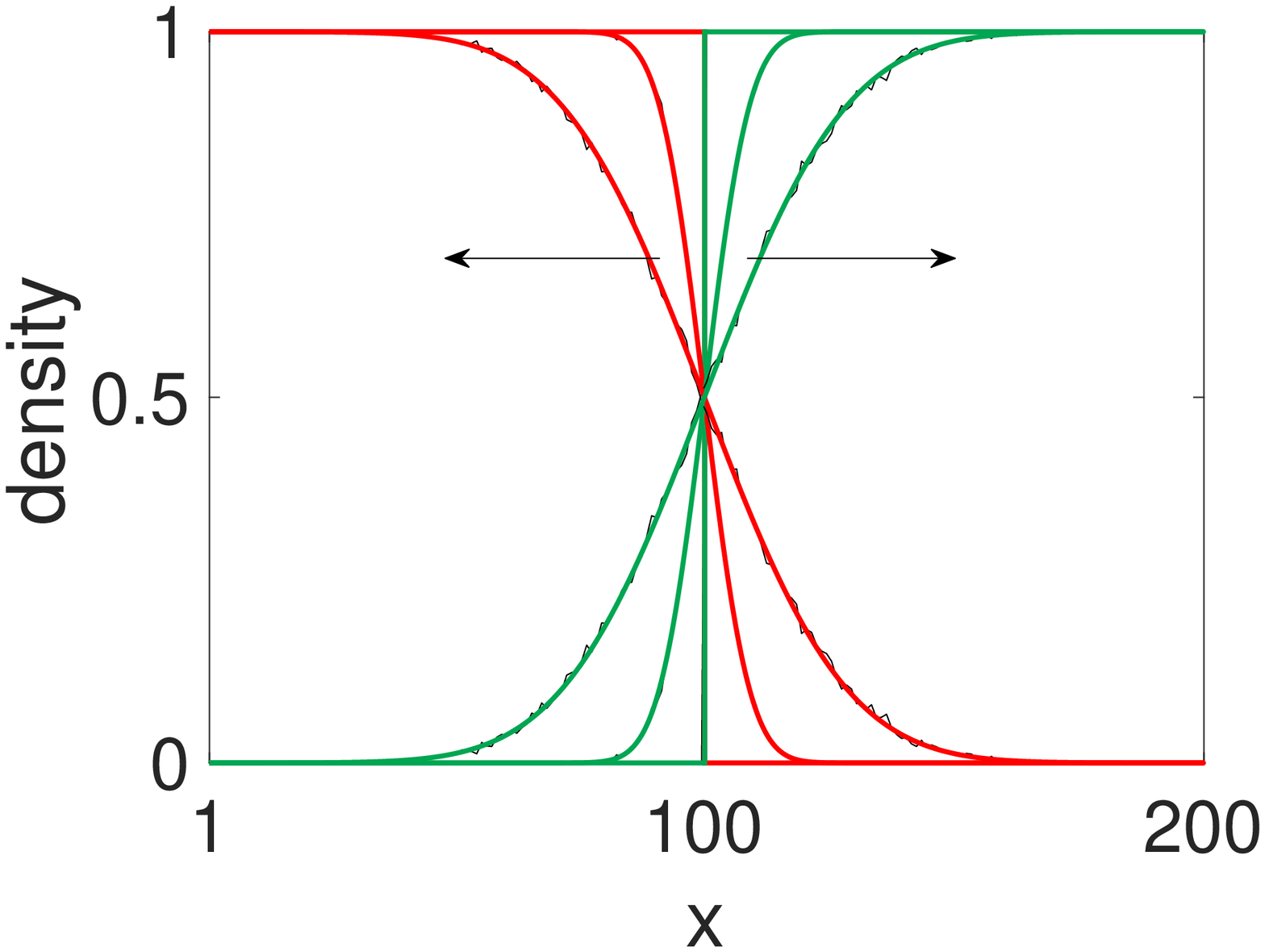}
\label{figure:IC=4_P_s=0.5}
}
\subfigure[]{
\includegraphics[width=0.31\textwidth]{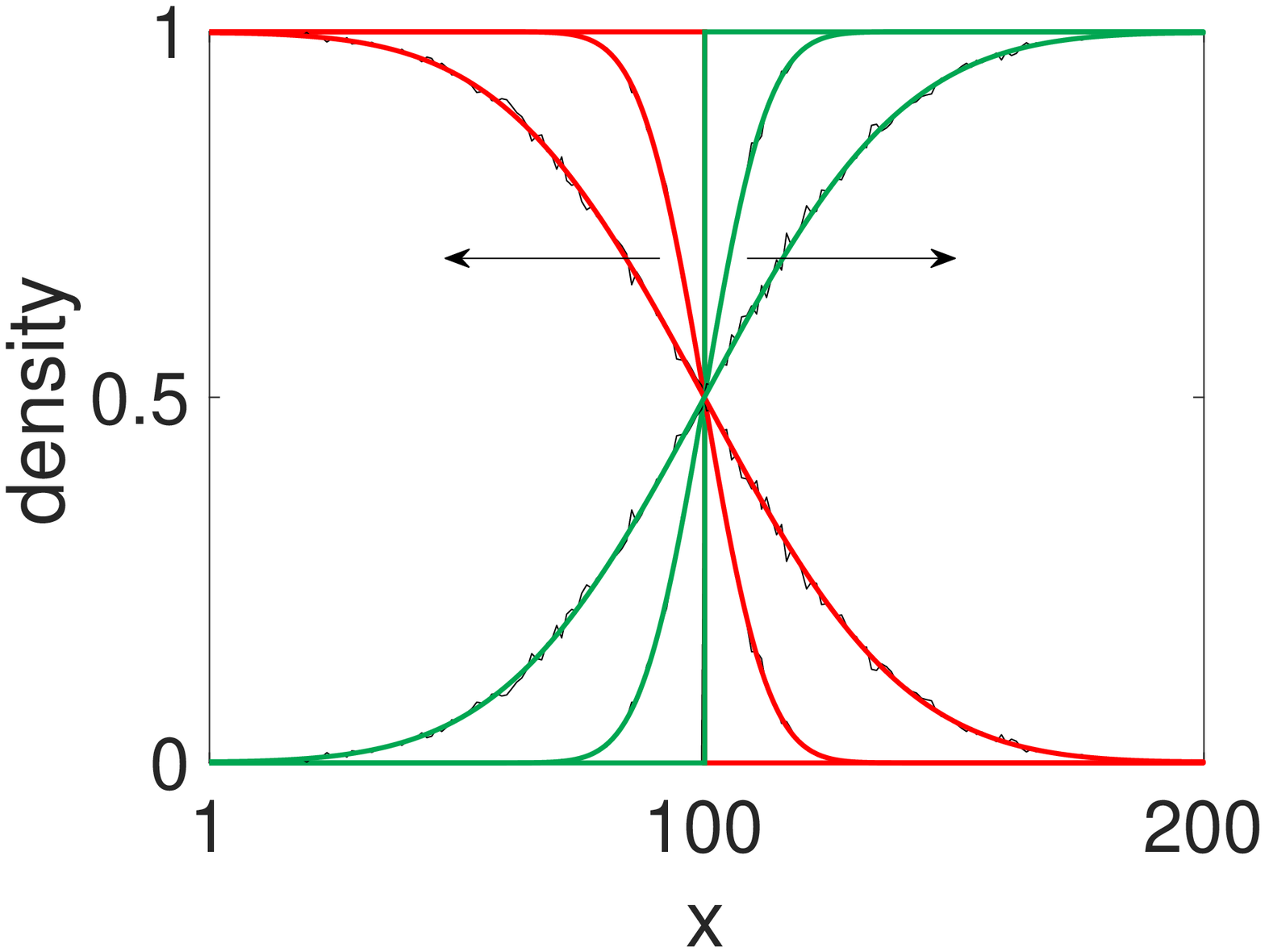}
\label{figure:IC=4_P_s=1}
}
\end{center}
\caption{A comparison between the numerical solution of the PDEs \eqref{eqn:pde_M} and \eqref{eqn:pde_X} and the averaged behaviour of the ABM for the migration process with swapping with different swapping probabilities: $\rho=0$ for [\subref{figure:IC=1_P_s=0}, \subref{figure:IC=2_P_s=0}, \subref{figure:IC=4_P_s=0}]; $\rho=0.5$ for [\subref{figure:IC=1_P_s=0.5}, \subref{figure:IC=2_P_s=0.5}, \subref{figure:IC=4_P_s=0.5}] and $\rho=1$ for [\subref{figure:IC=1_P_s=1}, \subref{figure:IC=2_P_s=1}, \subref{figure:IC=4_P_s=1}]. The initial conditions for [\subref{figure:IC=1_P_s=0}-\subref{figure:IC=1_P_s=1}] are the same as Figure \ref{figure:multispecies_density}. For [\subref{figure:IC=2_P_s=0}-\subref{figure:IC=2_P_s=1}] we initialised the region $1 \leqslant x \leqslant 100$ with agents of type A at a density of 0.9 and the remaining sites with agents of type B also at a density of 0.9. In [\subref{figure:IC=4_P_s=0}-\subref{figure:IC=4_P_s=1}] the lattice was initialised such that all the sites in the region $1 \leqslant x \leqslant 100$ are occupied by agents of type A and all the sites in the region $101 \leqslant x \leqslant 200$ are occupied by agents of type B. We present solutions at $t=0$, $t=100$ and $t=1000$ in all cases. The averaged column densities $\bar{A}_i$ and $\bar{B}_i$ are shown in black and the approximate PDE solution trajectories are shown in colour (red for agents of type A and green for agents of type B). The black arrows show the direction of increasing time.}
\label{figure:density_compar}
\end{figure}

In Figure \ref{figure:density_compar}, we compare the column-averaged density of the ABM given by,

\begin{equation*}
    \bar{A}_i=\frac{1}{L_y}\sum_{j=1}^{L_y} A_{ij}, \quad \bar{B}_i=\frac{1}{L_y}\sum_{j=1}^{L_y} B_{ij},
\end{equation*}

to the numerical solution of the one-dimensional analogue of Equations \eqref{eqn:pde_M} and \eqref{eqn:pde_X} with reflective boundary conditions by averaging the PDEs over the $y$ direction\footnote{Note that in Figure \ref{figure:IC=4_P_s=0} in the fully occupied domain the agreement is trivially perfect since volume exclusion prohibits departure from the initial condition.}. In the $\rho=0$ case (Figure \ref{figure:density_compar}, first column), the PDE solutions and the ABM do not agree well as evidenced by the disparity between the two profiles. This discrepancy can also be seen in \citet{simpson2009mse} where the authors devise an ABM for multi-species exclusion processes ($\rho=0$ case in our model) and compare their ABM with the corresponding continuum model. We remark that one reason for the discrepancy is that in crowded environments where the movement of agents is frequently inhibited by other agents, lattice occupancies cannot be considered independent of each other and spatial correlation are not dissipated efficiently \citep{simpson2013emi,markham2013isc,markham2013smi}. Independence of lattice sites is a key assumption that is typically made when deriving the continuum models such as the one above \citep{simpson2007sic,simpson2009mse,simpson2010mbc,simpson2009dpg,gavagnin2018sdm,chappelle2019pmc,yates2015ipe}. 


For non-zero swapping probabilities, the agreement is significantly improved between the deterministic and stochastic model (Figure \ref{figure:density_compar}, second and third columns). Swapping helps to break down the spatial correlations improving the agreement between the PLM and the ABM. We also see that in the zero swapping case, crowding of the green agents behind the red agents leads to profiles for which the maximum density at non-zero time (shown $t=100$ and $t=1000$ in Figure \ref{figure:IC=1_P_s=0}) is higher than the initial maximum density ($t=0$). The reason for this is that for a multispecies migration process with cross-diffusion terms there is no maximum principle for the individual species \citep{rahman2016dmc,jungel2015bmc,simpson2009mse}. We note that the enhanced diffusion which swapping engenders eliminates this effect. However, this does not necessarily mean that a maximum principle now holds for the systems under consideration. Investigating this further is beyond the scope of this article.

\section{Individual-level analysis}\label{sec:individual_level_analysis}

In this section, we analyse the movement of agents at the individual level in the single-species and two-species case to assess how swapping affects the movement of agents.

\subsection{Single-species individual-level analysis}\label{sec:ss_individual_level_analysis}

The single-species swapping discrete and continuum models are given in Appendix \ref{app:appendixA}. Here, we present the individual-level analysis for the single-species model. Our aim is to quantify the movement of individually tagged agents by their individual-level diffusion coefficient. For the analysis that we present next, we neglect any long range temporal correlations in the agents' movement. We first derive the individual-level time-uncorrelated diffusion coefficient analytically and then we compare it with its ABM approximation.

Let $P_{ij}(t)=P_{ij}$ denote the probability that a focal agent is at position $(i,j)$ at time $t$. Defining $\delta t$ as an infinitesimally small change in time, we can write down the probability of the agent being at position $(i,j)$ at time $t+\delta t$ as,

\begin{multline}\label{eqn:prob_master_eqn}
    P_{ij}(t+\delta t)=\frac{r}{4}P_{i-1,j}[(1-c)+2c\rho]\delta t+\frac{r}{4}P_{i+1,j}[(1-c)+2c\rho]\delta t+\frac{r}{4}P_{i,j-1}[(1-c)+2c\rho]\delta t\\+\frac{r}{4}P_{i,j+1}[(1-c)+2c\rho]\delta t+P_{ij}[(1-r\delta t)+rc(1-\rho)\delta t-r c\rho\delta t] + o(\delta t).
\end{multline}

A lattice site is occupied with probability $c$ and empty with probability $1-c$. The first four terms in Equation \eqref{eqn:prob_master_eqn} are obtained by considering an agent at each of the four lattice sites in the neighbourhood the site $(i,j)$ that has attempted to move into site $(i,j)$ with probability $r/4 \hspace{4pt} \delta  t$. If the site $(i,j)$ is vacant, the agent jumps from the neighbouring site to site $(i,j)$. Otherwise, the agents at the neighbouring site and site $(i,j)$ swap their positions with probability $\rho$. The 2 in $2\rho c$ corresponds to the two ways in which the position of the agent at site $(i,j)$ can change due to a swap: either the agent at site $(i,j)$ initiates and successfully completes the swap with the agent at the neighbouring site or the agent at the neighbouring site initiates and successfully swaps with the agent at site $(i,j)$. The last term in Equation \eqref{eqn:prob_master_eqn} gives the probability that the agent already occupying position $(i,j)$ does not attempt to move during the time interval $[t,t+\delta t]$ with probability $(1-r \delta  t)$ and the probability that the agent attempts to swap with an agent at an occupied neighbouring site but fails to complete the swap with probability $rc(1-\rho)\delta t$ and lastly the probability that a neighbouring agent successfully swaps with the agent at site $(i,j)$ with probability $rc\rho\delta t$.

By rearranging, dividing both sides of Equation \eqref{eqn:prob_master_eqn} by $\delta  t$ and taking the limit as $\delta  t \to 0$ leads to the system of ODEs given by,

\begin{equation}{\label{eqn:ss_tagged_master_eqn}}
    \frac{\mathrm{d} P_{ij}}{\mathrm{d} t}=\frac{r}{4}[(1-c)+2c\rho](P_{i-1,j}+P_{i+1,j}+P_{i,j-1}+P_{i,j+1})-r P_{i,j}[(1-c)+2c\rho],
\end{equation}
which describe the time evolution of the probability of finding an agent at position $(i,j)$ at time $t$. From Equation \eqref{eqn:ss_tagged_master_eqn} it can be shown that the expected net displacement of an agent is zero. This gives us no information about the statistical fluctuations in the movement of an agent and therefore we use the variance to quantify the net displacement \citep{codling2008rwm,simpson2009dpg}. The equations describing the time-evolution of the variances $\langle i(t)^2\rangle$ and $\langle j(t)^2\rangle$ are given by,

\begin{equation}\label{eqn:mth_moment_i}
	\frac{\mathrm{d} \langle i^2 \rangle}{\mathrm{d} t} = \frac{\mathrm{d} }{\mathrm{d}  t}\left(\sum_{i=1}^{L_x} i^2 P_{ij} \right) = \frac{r}{2}[(1-c)+2c\rho],
\end{equation}
and,
\begin{equation}\label{eqn:mth_moment_j}
    \frac{\mathrm{d} \langle j^2 \rangle}{\mathrm{d} t} = \frac{\mathrm{d} }{\mathrm{d}  t}\left(\sum_{j=1}^{L_y} j^2 P_{ij} \right) = \frac{r}{2}[(1-c)+2c\rho].
\end{equation}



Under the initial condition that at time $t=0$, $\langle i^2 \rangle=\langle j^2 \rangle=0$, Equations \eqref{eqn:mth_moment_i} and \eqref{eqn:mth_moment_j} solve to give,

\begin{equation}
    \langle i(t)^2 \rangle = \langle j(t)^2 \rangle = \frac{r}{2}[(1-c)+2c\rho]t.
\end{equation}

The time-uncorrelated individual-level diffusion coefficient can be retrieved as,

\begin{equation}\label{eqn:D_theory}
    D^\star(c,\rho) = \frac{r}{4}[(1-c)+2c\rho].
\end{equation}

As briefly discussed above, the master equation neglects temporal correlations in an agent's position. Consequently, the diffusion coefficient derived from the master equation will necessarily be inaccurate and fail to represent the true dynamics of the agents in the system. As a result, we refer to the expression given in Equation \eqref{eqn:D_theory} as the time-uncorrelated individual-level diffusion coefficient.


To approximate the $D^\star$ using the ABM, we initialise a $150$ by $150$ lattice and randomly seed it with agents at different background densities $\mathbf{c}=[0,0.25,0.5,0.75,1]$, where $c=0$ corresponds to one agent with no agents in the background to interact with and $c=1$ corresponds to a fully populated lattice. We let the positions of the agents evolve according to the single-species version of the ABM described in Section \ref{sec:abm} using the movement rate $r=1$ for a range of swapping probabilities $\boldsymbol\rho=[0, 0.25, 0.5, 0.75, 1]$. The positions $(X,Y)$ of agents whose initial position is in the region defined by the central square $[51,100]\times [51,100]$ are recorded over a regular time grid $t_\text{rec}=[0,\Delta t, 2\Delta t,...,T \Delta t]$ to create a track where $\Delta t$ is the recording step and $T$ is the track length. Each track traces the path of an agent over time. We impose periodic boundary conditions on the domain. The simulation ends at $T_{final}=1000$ or when a tracked agent hits a boundary, whichever happens first.

We analyse individual movement of agents using the sum of squared displacement (SSD) \citep{simpson2009dpg,simpson2007sic},

\begin{equation}
    S_t^{(x)}=\sum_{j=1}^{t'} (X(t+j\Delta t) - X(t+(j-1)\Delta t))^2,
\end{equation}
and,

\begin{equation}
    S_t^{(y)}=\sum_{j=1}^{t'} (Y(t+j\Delta t) - Y(t+(j-1)\Delta t))^2, \qquad t'=1,...,T
\end{equation}

By only taking the difference between successive positions, the SSD neglects temporal correlations in an agent's position which is consistent with our master equation in Equation \eqref{eqn:prob_master_eqn}. For each value of $c$ and $\rho$ we average the SSD over an ensemble of tracks and fit a linear model of the form $\hat S_t^{x}=a_xt$ and $\hat S_t^{y}=a_yt$ in each orthogonal direction $x$ and $y$, respectively. The ABM approximation of the time-uncorrelated individual-level diffusion coefficient can be extracted from the gradient of these linear equations,

\begin{equation}
    \hat D^\star(c,\rho)=\frac{a_x+a_y}{2d},
\end{equation}

where $d=2$ is the dimension dimension of the lattice. We put a `hat' symbol over $D^\star$ to differentiate it from the exact time-uncorrelated diffusion coefficient $D^\star$ in Equation \eqref{eqn:D_theory}.

\begin{figure}[t!]
\begin{center}

\subfigure[]{
\includegraphics[width=0.47\textwidth]{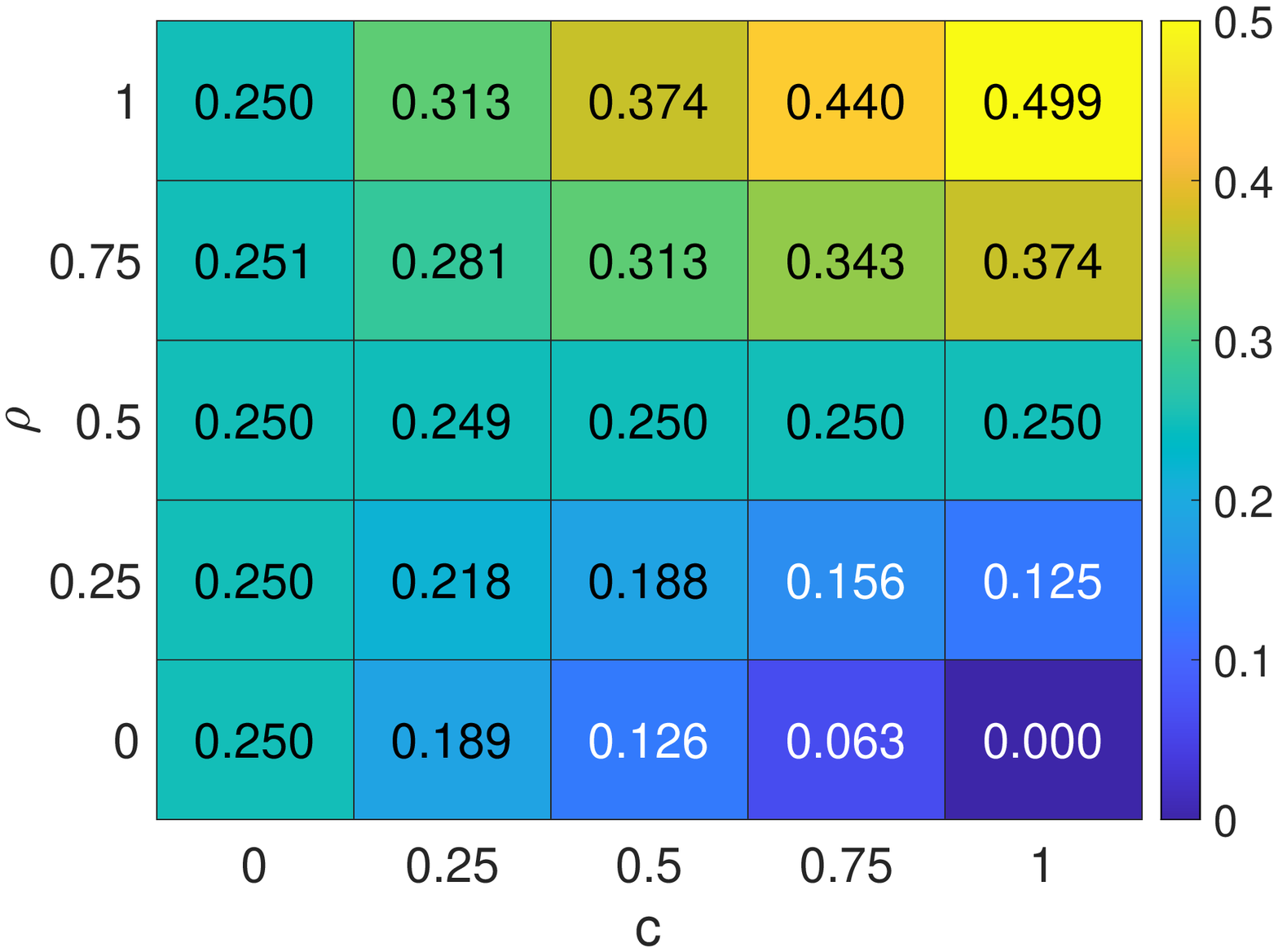}
\label{figure:heatmap_simul}
}
\subfigure[]{
\includegraphics[width=0.47\textwidth]{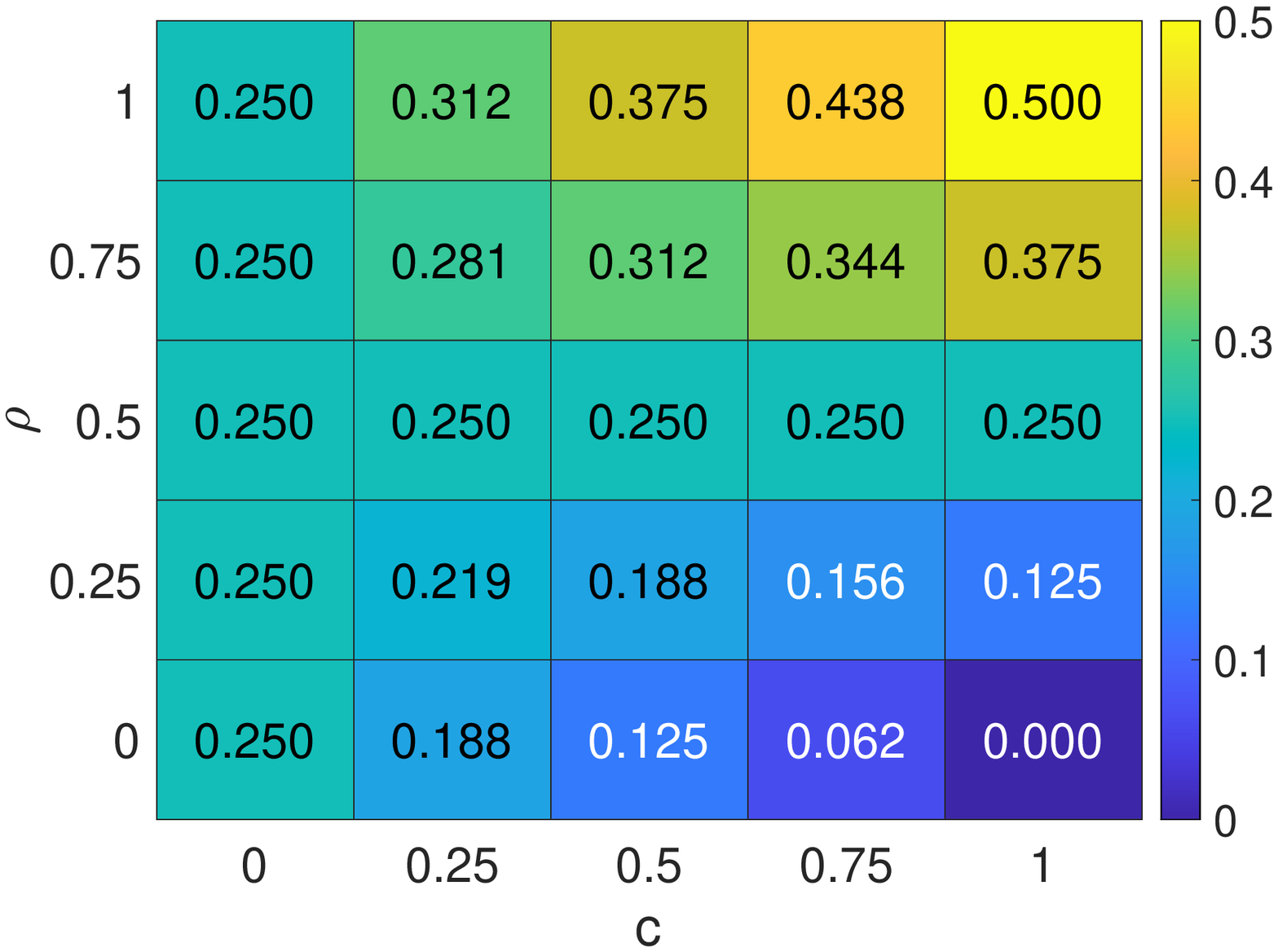}
\label{figure:heatmap_theory}
}

\subfigure[]{
\includegraphics[width=0.5\textwidth]{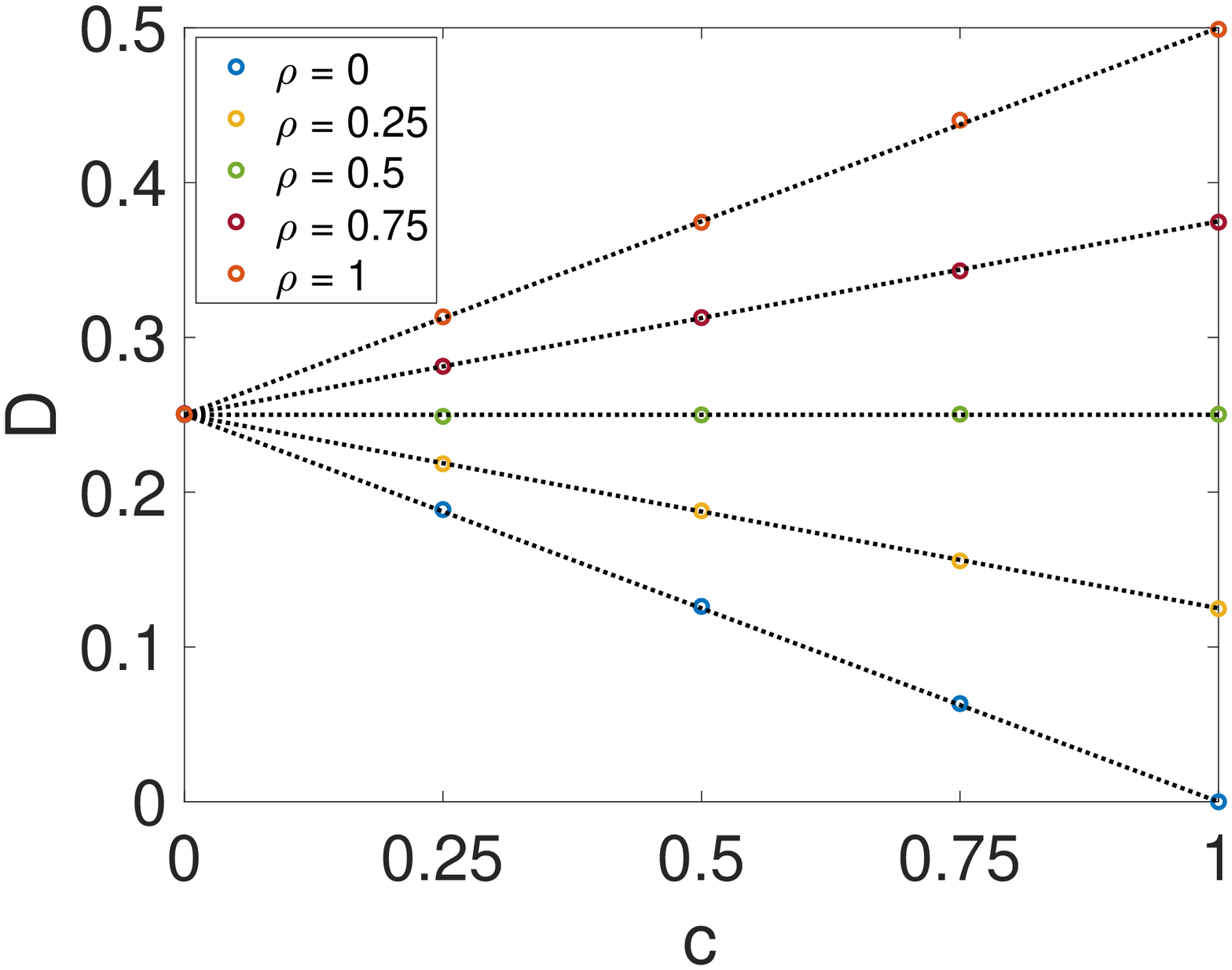}
\label{figure:D_vs_c_vs_rho}
}

\end{center}
\caption{Heat map showing the relationship between the time-uncorrelated individual-level diffusion coefficient, $D^\star$, the domain density, $c$, and the swapping probability, $\rho$. For \subref{figure:heatmap_simul}, we initialised a 150 by 150 periodic domain with density $c \in \mathbf{c}=[0,0.25,0.5.0.75,1]$ and for the values of swapping probability $\boldsymbol \rho = [0,0.25,0.5,0.75,1]$ we let the positions of the agents evolve according to the single-species ABM in Section \ref{app:appendixA} with $r=1$. We tracked all agents in the region defined by the square $[51,100]\times[51,100]$. Theoretical heatmap in \subref{figure:heatmap_theory} was obtained using Equation \eqref{eqn:D_theory}. In \subref{figure:D_vs_c_vs_rho} we show the linear relationship between $D^\star$, the domain density, $c$ and the swapping probability, $\rho$. The circles represent ABM approximations of $\hat D^\star$ whereas the dotted black lines represent the exact value $D^\star$ given in Equation \eqref{eqn:D_theory}.}
\label{figure:heatmaps}
\end{figure}

In Figure \ref{figure:heatmaps}, we compare the ABM approximation $\hat D^\star$ (shown in \subref{figure:heatmap_simul}) with the derived expression $D^\star$ (shown in \subref{figure:heatmap_theory}) as a pair of heat maps for the range of values of $c$ and $\rho$ defined earlier. We also show a line plot that nicely depicts the linear relationship between $D^\star$ and $c$ for different value of $\rho$ (shown in \subref{figure:D_vs_c_vs_rho}). 
We can see that there is excellent agreement between the analytical expression and the simulated results. We note that in the case of zero background density ($c=0$) $D^\star$ is always 0.25 regardless of the swapping probability since for a single agent with no other agents to interact with, the movement of the agent can neither be inhibited by volume exclusion nor enhanced by swapping. We also note that for swapping probability $\rho=0.5$, the value of $D^\star$ is always 0.25 irrespective of the density since half the number of times a focal agent attempts to move into an occupied site the moves will be rejected. If this were the only impact of swapping, we would expect $D^\star$ of the focal agent to be reduced. However, just as often as a focal agent attempts to move into an occupied neighbour's position, an agent occupying a neighbouring site tries to move into the focal agent's position -- achieving this successfully with probability $\rho=0.5$. Assuming a well-mixed scenario, this exactly compensates for the number of aborted moves the focal agent makes, meaning movement is as if the focal agent were on an unoccupied domain irrespective of density.
For $c=1$ and $\rho=1$ we note that $D^\star=0.5$ (i.e. twice as large compared to an agent moving on a domain with zero background density) as every attempted move by the focal agent is executed successfully and the focal agent is also moved equally often by neighbouring agents swapping into its position.

\subsection{Two-species individual-level analysis}\label{sec:ms_individual_level_analysis}

In this section, we perform the individual-level analysis for a two-species system as set out for the single-species case in Section \ref{sec:ss_individual_level_analysis}.

Let $P^{A}_{ij}(t)=P^A_{ij}$ be the probability that a focal agent of type-A occupies the position $(i,j)$ and let $P^B_{ij}(t)=P^B_{ij}$ be the equivalent for a type-B focal agent. Recalling $\delta t$ as a small change in time we can write down the master equations for species A and B at time $t+\delta t$,

\begin{align}
\frac{d P^{A}_{ij}}{dt}&= \left(\frac{r_{A}}{4}(1-c)+\frac{r_{A}}{2}c_A\rho+\frac{(r_{A}+r_B)}{4}c_B\rho \right) (P^{A}_{i-1,j}+P^{A}_{i+1,j}+P^{A}_{i,j-1}+P^{A}_{i,j+1}) \nonumber\\
&\quad + (-r_{A}(1-c) - 2r_{A}c_A\rho - (r_{A}+r_{B})c_B\rho)P^{A}_{ij}, \label{eqn:ms_tagged_M}\\
\frac{d P^{B}_{ij}}{dt}&= \left(\frac{r_B}{4}(1-c)+\frac{r_B}{2}c_B\rho+\frac{(r_A+r_{B})}{4}c_A\rho \right) (P^{B}_{i-1,j}+P^{B}_{i+1,j}+P^{B}_{i,j-1}+P^{B}_{i,j+1}) \nonumber\\
&\quad + (-r_B(1-c) - 2r_Bc_B\rho - (r_{A} + r_B)c_A\rho)P^{A}_{ij}.\label{eqn:ms_tagged_X}
\end{align}

A full derivation of Equations \eqref{eqn:ms_tagged_M} and \eqref{eqn:ms_tagged_X} can be found in Appendix \ref{app:two_species} accompanying this article.

The individual-level time-uncorrelated diffusion coefficients $D_A^\star$ and $D_B^\star$ for the species A and B, respectively, are given by,

\begin{align}
    D_A^\star&=\frac{1}{4}(r_A(1-c)+(r_Ac+r_Ac_A+r_Bc_B)\rho), \label{eqn:ms_diffusion_coefficient_DM}\\
    D_B^\star&=\frac{1}{4}(r_B(1-c)+(r_Bc+r_Ac_A+r_Bc_B)\rho).\label{eqn:ms_diffusion_coefficient_DX}
\end{align}

In order to investigate the effect of swapping on a two-species system and to verify our theoretical results, we simulate the two-species model, track a set of tagged agents and analyse their movement using the SSD as done in Section \ref{sec:ss_individual_level_analysis}.

\begin{figure}[h!]
\begin{center}

\subfigure[]{
\includegraphics[width=0.475\textwidth]{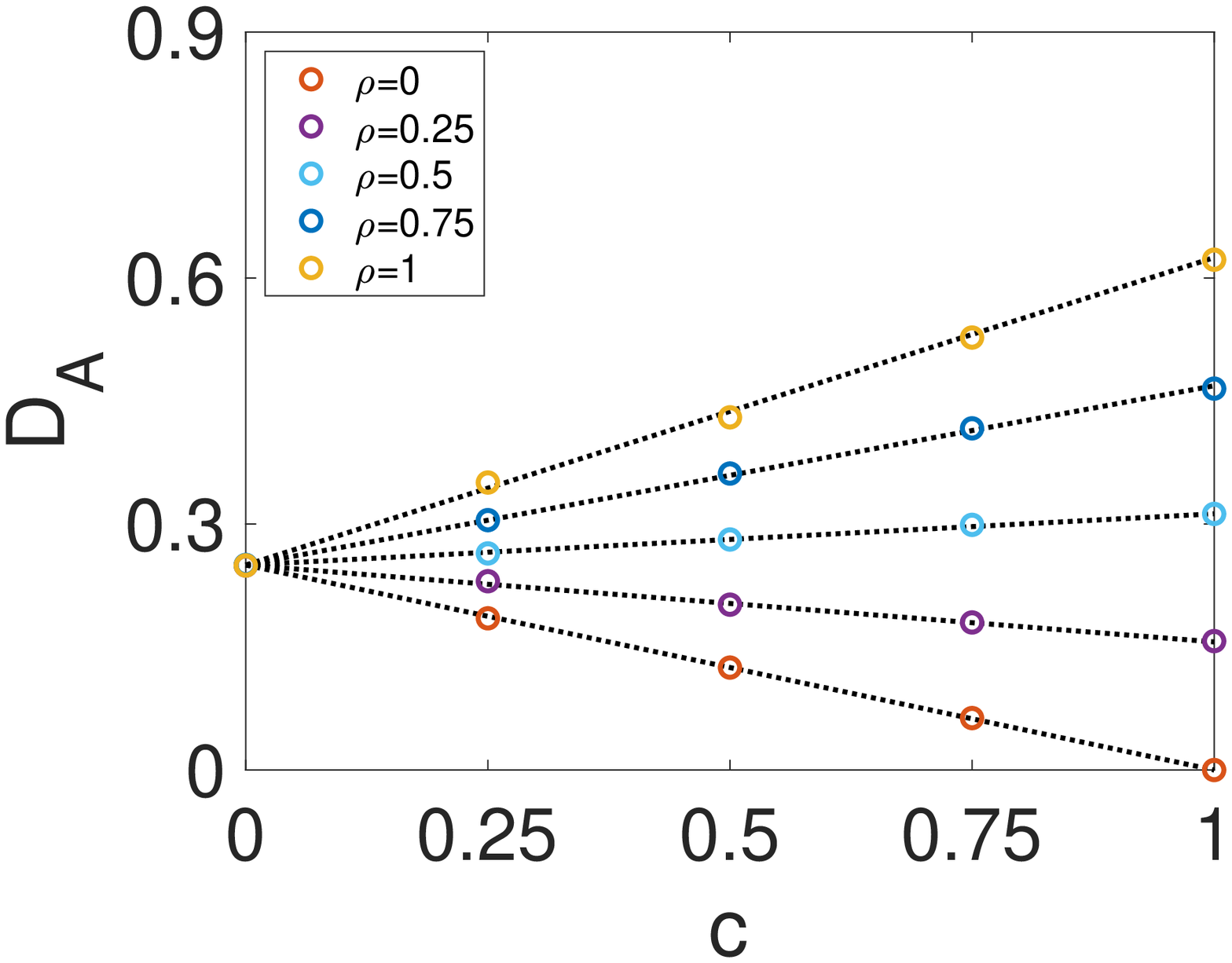}
\label{figure:D1_c=0.5_rm=1}
}
\subfigure[]{
\includegraphics[width=0.475\textwidth]{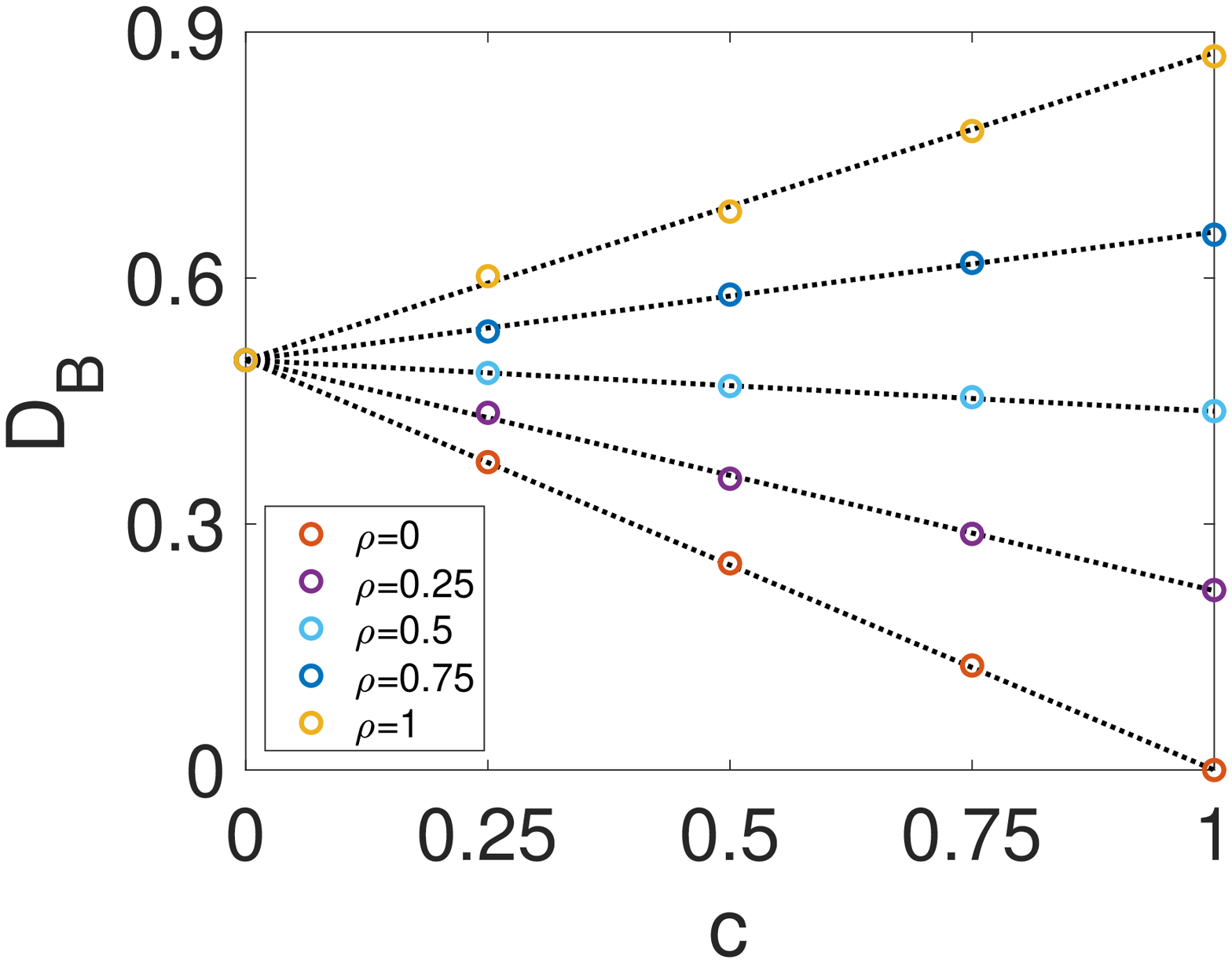}
\label{figure:D2_c=0.5_rm=1}
}
\end{center}
\caption{Time-uncorrelated diffusion coefficient of tagged type-A \subref{figure:D1_c=0.5_rm=1} and type-B agents \subref{figure:D2_c=0.5_rm=1} plotted against background densities $c=[0,0.25,0.5,0.75,1]$ for different swapping probabilities $\rho=[0,0.25,0.5,0.75,1]$ with fixed movement rate $r_A=1$ and $r_B=2$. Both species are present in equal proportions (i.e. $c_A=c_B=0.5c$). The circles represent the ABM approximations $\hat D_A^\star$ and $\hat D_B^\star$ and the dotted lines represent the theoretical values $D_A^\star$ and $D_B^\star$ from Equations \eqref{eqn:ms_diffusion_coefficient_DM} and \eqref{eqn:ms_diffusion_coefficient_DX}.}
\label{figure:ms_tagged_agent_1}
\end{figure}

\begin{figure}[h!]
\begin{center}

\subfigure[]{
\includegraphics[width=0.475\textwidth]{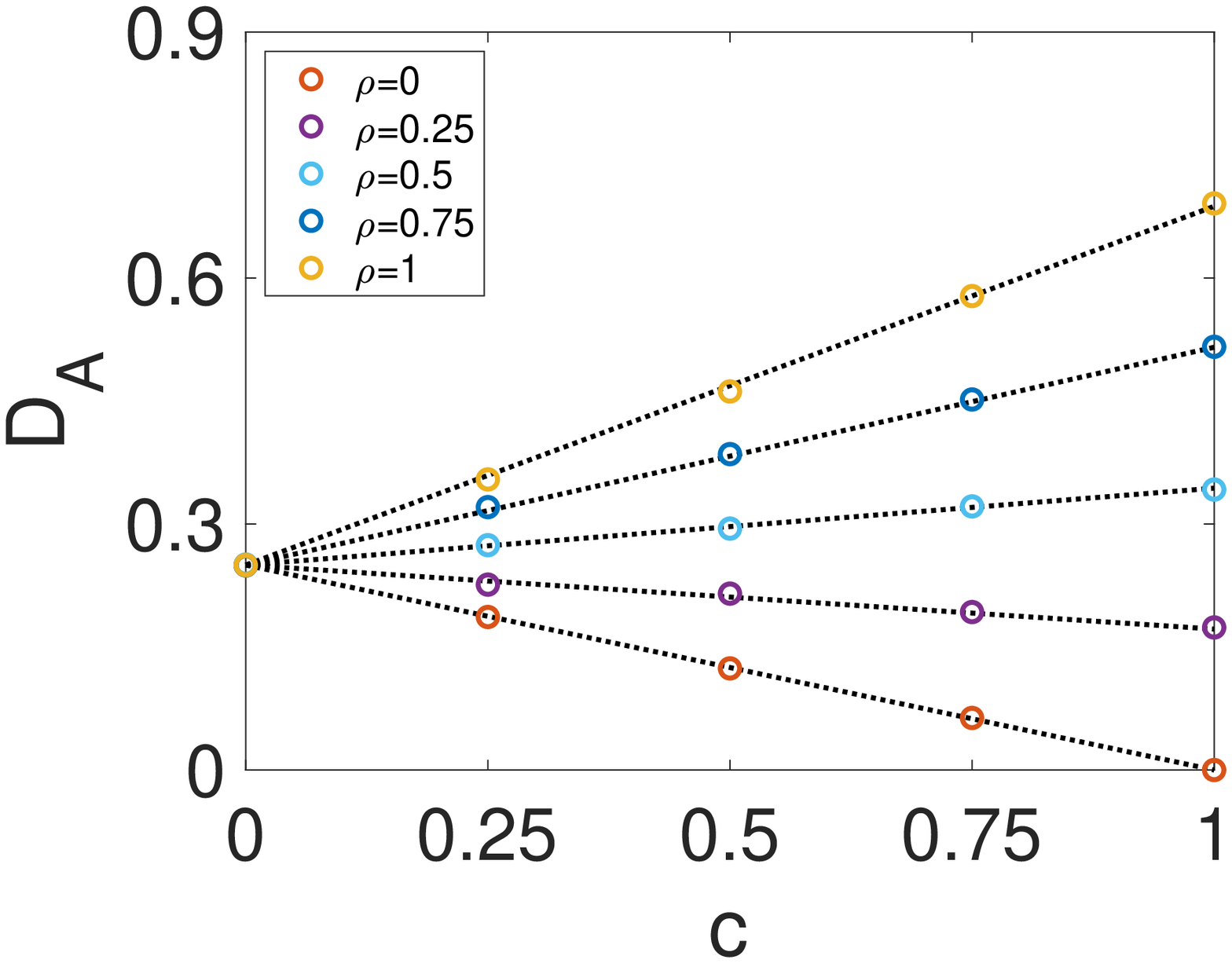}
\label{figure:D1_cm=0.25_c=0.5_rm=1}
}
\subfigure[]{
\includegraphics[width=0.475\textwidth]{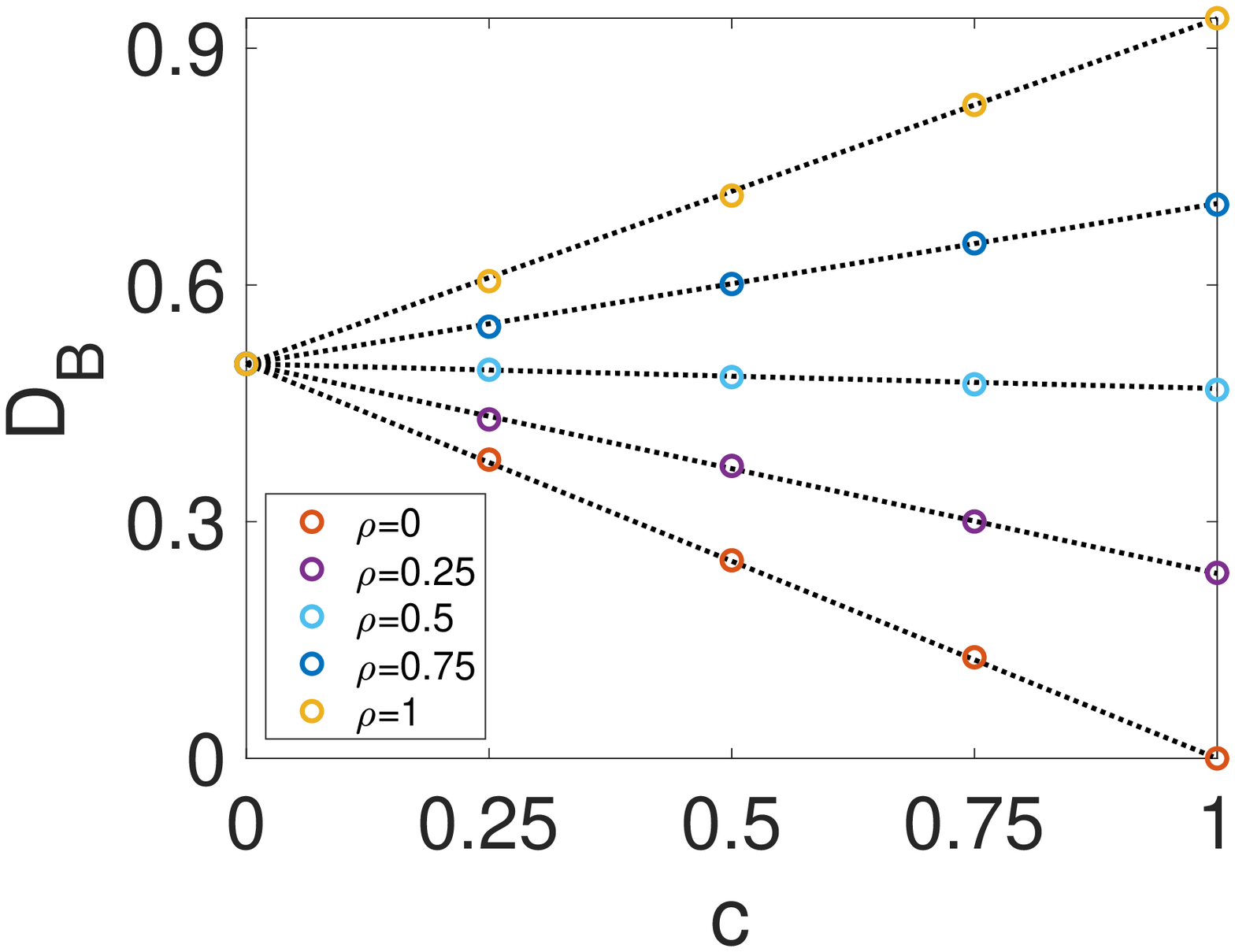}
\label{figure:D2_cm=0.25_c=0.5_rm=1}
}
\end{center}
\caption{Time-uncorrelated diffusion coefficient of tagged type-A \subref{figure:D1_c=0.5_rm=1} and type-B agents \subref{figure:D2_c=0.5_rm=1} plotted against background densities $c=[0,0.25,0.5,0.75,1]$ for different swapping probabilities $\rho=[0,0.25,0.5,0.75,1]$ with fixed movement rate $r_A=1$ and $r_B=2$. Type-A agents are present at a density $c_A=0.25c$ and type-B agents are at $c_B=0.75c$. The Left panel \subref{figure:D1_c=0.5_rm=1} shows $D_A^\star$ and the right-hand panel \subref{figure:D2_c=0.5_rm=1} shows $D_B^\star$.}
\label{figure:ms_tagged_agent_2}
\end{figure}

In Figure \ref{figure:ms_tagged_agent_1} we show the time-uncorrelated diffusion coefficient of species A and species B plotted against background density $c=[0,0.25,0.50,0.75,1]$ for different values of swapping probability $\rho=[0,0.25$, $0.50,0.75,1]$ for the specific case in which $r_A=1$ and $r_B=2$. In this figure, both species are present in equal proportions on the domain. We see good agreement between the ABM approximations and the theoretical values. We also see that, as expected and as noted in the single-species system, swapping speeds up the movement of the agents in the multi-species setting compared to the pure volume-excluded scenario ($\rho=0$). Furthermore, since species B has a higher movement rate than species A, species B diffuses faster than species A, apart from the $c=1$ and $\rho=0$ trivial case in which $D_A^\star$ and $D_B^\star$ are both 0 since the agents have nowhere to go.

In Figure \ref{figure:ms_tagged_agent_2} we present a similar comparison as in Figure \ref{figure:ms_tagged_agent_1} but this time the two species are not present in equal proportions. In this particular case, we consider a scenario in which type-A agents make up 25\% of the total population and type-B agents 75\%. Again, we see good agreement between the ABM and theoretical values and that swapping speeds up the movement of agents.

Comparing Figure \ref{figure:ms_tagged_agent_1} to Figure \ref{figure:ms_tagged_agent_2} we see that for the $\rho=0$ case $D_A^\star$ and $D_B^\star$ in Figure \ref{figure:ms_tagged_agent_1} are equal to their respective values in Figure \ref{figure:ms_tagged_agent_2} and hence unaffected by changes in the densities $c_A$ and $c_B$. Analytically, this can be observed by setting $\rho=0$ in Equations \eqref{eqn:ms_diffusion_coefficient_DM} and \eqref{eqn:ms_diffusion_coefficient_DX}. In this case, $D_A^\star$ and $D_B^\star$ depend on the overall background density and not the species proportions. For $\rho>0$, in the instance where $r_B > r_A$ both species show faster movement in Figure \ref{figure:ms_tagged_agent_2} than in Figure \ref{figure:ms_tagged_agent_1}. Again, this can be checked by referring to the expressions for $D_A^\star$ and $D_B^\star$. Changing the proportions of the species (while keeping $r_A$, $r_B$ and $c$ constant) the term $(r_Ac_A+r_Bc_B)\rho$ is sensitive to changes in the proportions in both $D_A^\star$ and $D_B^\star$. When $r_B>r_A$ and we increase the proportion of the faster moving species, $c_B$, this results in more swapping events which enhance the movement of both species. On the other hand, if we were to increase the proportion of species A (while simultaneously decreasing the proportion of species B to keep $c$ constant) we would observe reduced movement for both species since an increase in the proportion of the slower moving species (and a decrease in the proportion of faster moving species) would result in fewer swaps.

\section{Illustrative examples}\label{sec:examples}

In this section, we show examples of the situations in which swapping has important applications. In Section \ref{sec:example_prolif}, we build the swapping mechanism into a cell migration model with proliferation and in Section \ref{sec:example_adhesion}, we show how the swapping mechanism in conjunction with cell-cell adhesion can facilitate spontaneous pattern formation in densely crowded environments.

\subsection{Swapping model with cell proliferation}\label{sec:example_prolif}

We look at the role of swapping in cell migration with proliferation. For this example, we concern ourselves with the two-species cell migration model. The movement kinetics of the agents are the same as the swapping model described in Section \ref{sec:abm} but in addition to migrating, agents can attempt to proliferate, placing a daughter at a randomly chosen neighbouring site if the site is empty, otherwise the division event is aborted. The proliferation rates per unit time for the two species A and B are denoted by $r_{p}^A$ and $r_{p}^B$, respectively.

We initialise the domain with $L_x=100$ sites in the horizontal direction and $L_y=20$ sites in the vertical direction. We fill all the sites in the range $41 \leqslant x \leqslant 60$ with agents of type A and all the remaining sites with type-B agents at a density of 0.5. The movement rates of agents are set to $r_A=r_B=1$ and the proliferation rates as $r_{p}^A=0.01$ and $r_{p}^B=0$, i.e. only the agents of type A divide and the number of type-B agents are held constant. We let the system evolve according to the specified ABM.

\begin{figure}[t!]
\begin{center}

\subfigure[]{
\includegraphics[width=0.30\textwidth]{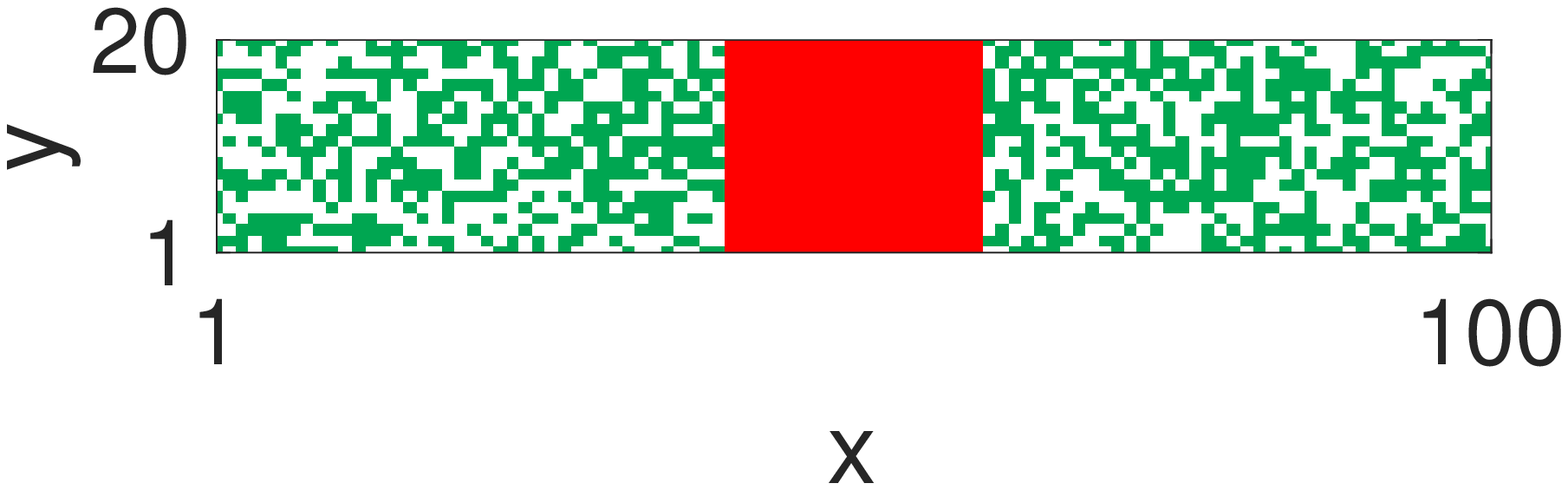}
\label{figure:prolif_agents_Ps=0_T=0}
}
\setcounter{subfigure}{3}
\subfigure[]{
\includegraphics[width=0.30\textwidth]{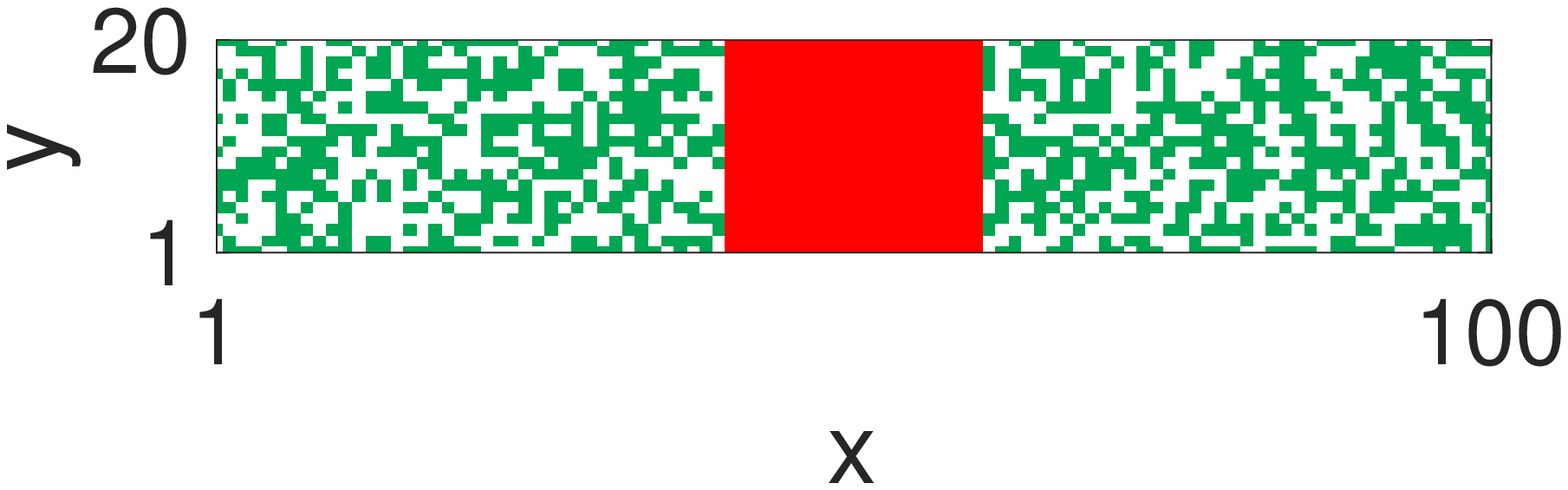}
\label{figure:prolif_agents_Ps=0.5_T=0}
}
\setcounter{subfigure}{6}
\subfigure[]{
\includegraphics[width=0.30\textwidth]{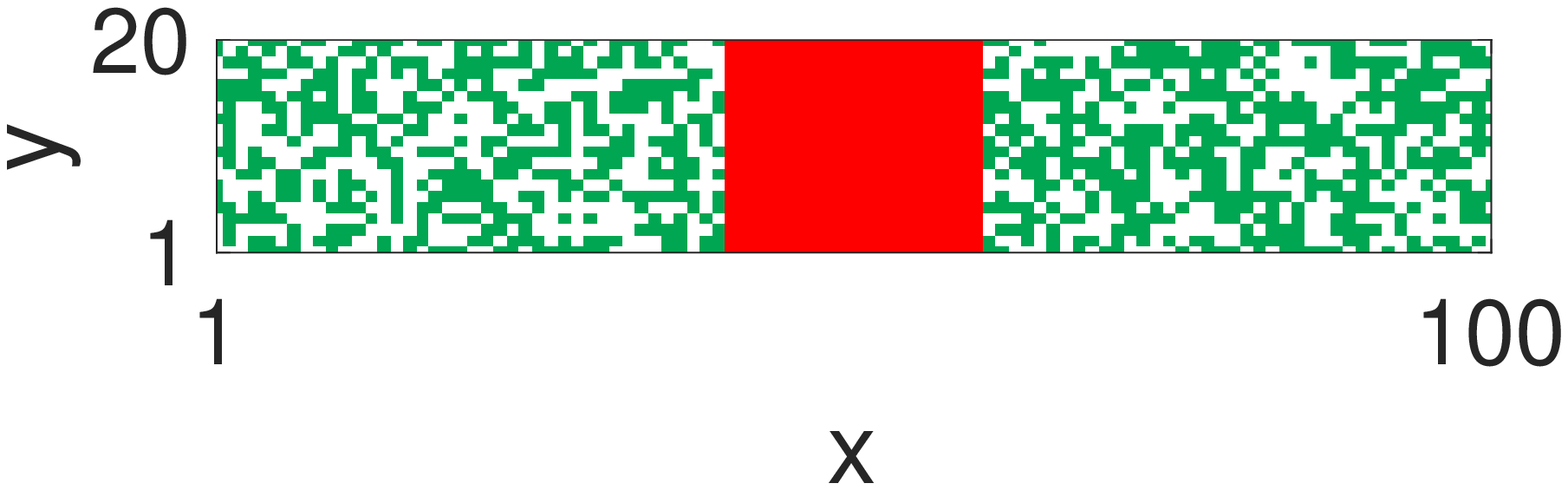}
\label{figure:prolif_agents_Ps=1_T=0}
}

\setcounter{subfigure}{1}
\subfigure[]{
\includegraphics[width=0.30\textwidth]{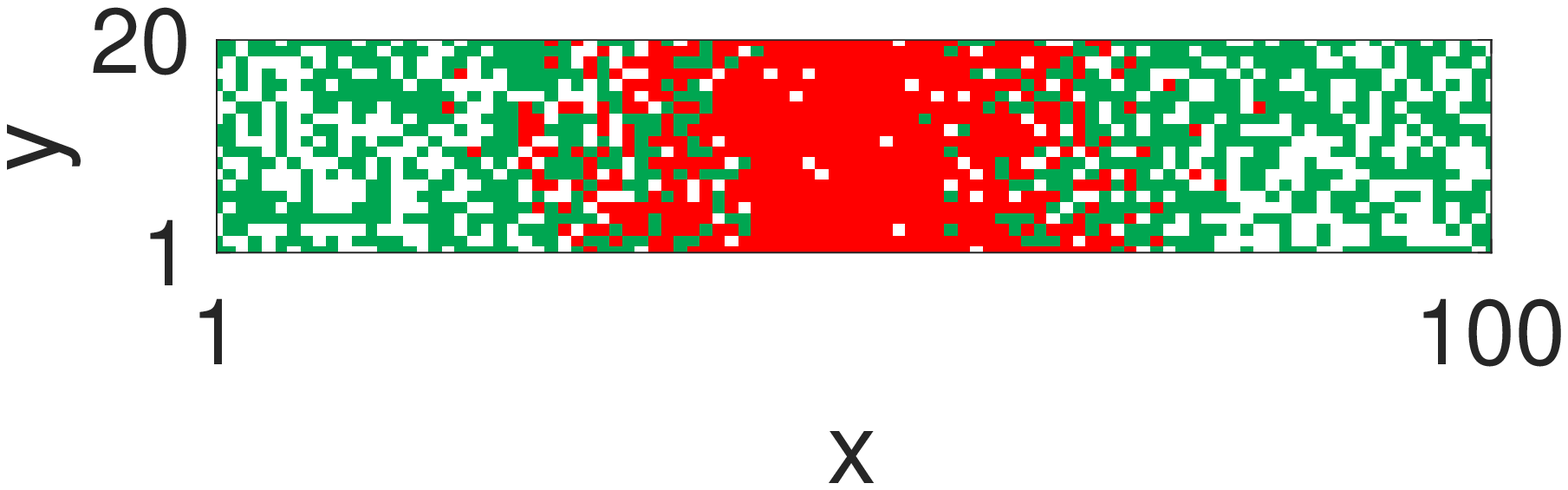}
\label{figure:prolif_agents_Ps=0_T=500}
}
\setcounter{subfigure}{4}
\subfigure[]{
\includegraphics[width=0.30\textwidth]{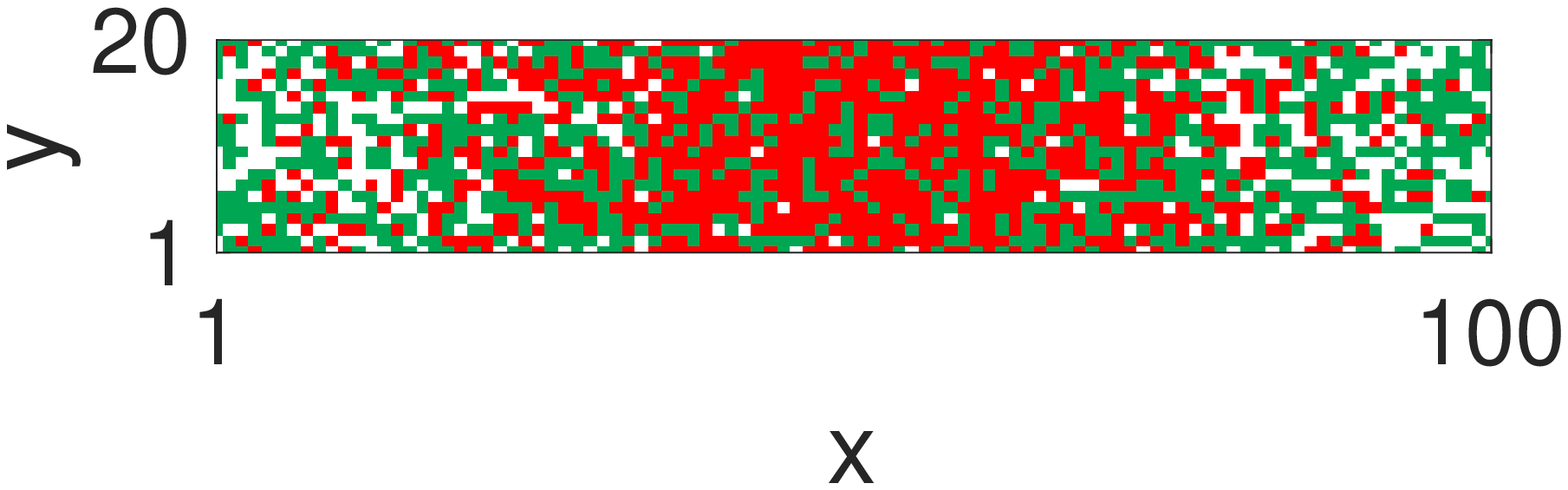}
\label{figure:prolif_agents_Ps=0.5_T=500}
}
\setcounter{subfigure}{7}
\subfigure[]{
\includegraphics[width=0.30\textwidth]{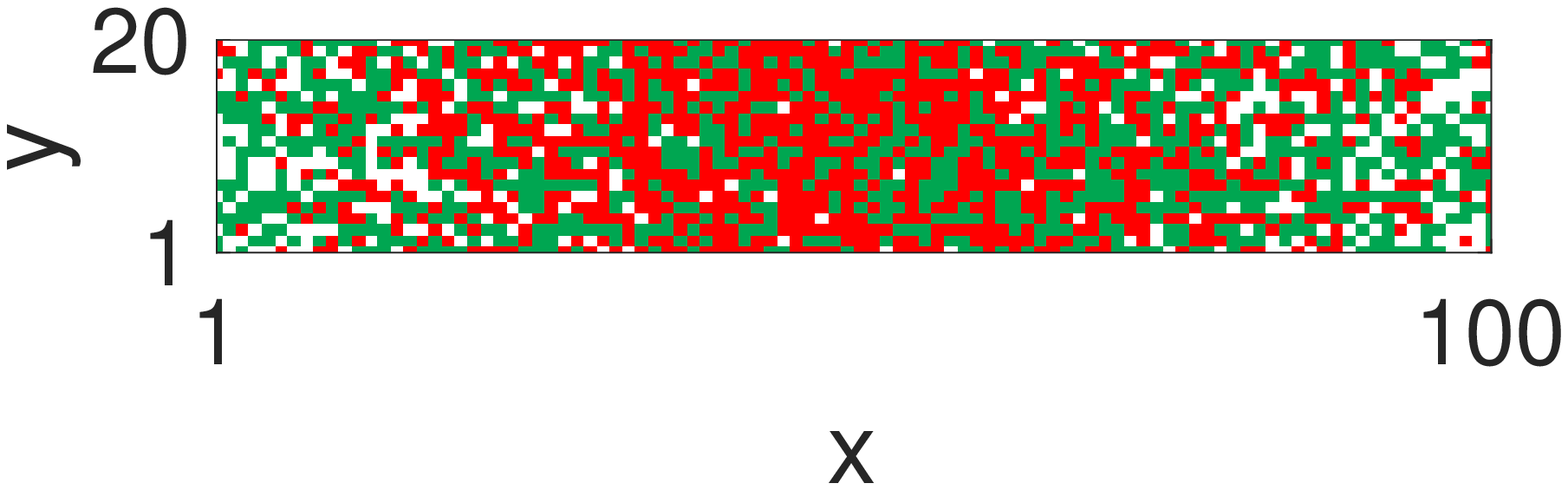}
\label{figure:prolif_agents_Ps=1_T=500}
}

\setcounter{subfigure}{2}
\subfigure[]{
\includegraphics[width=0.30\textwidth]{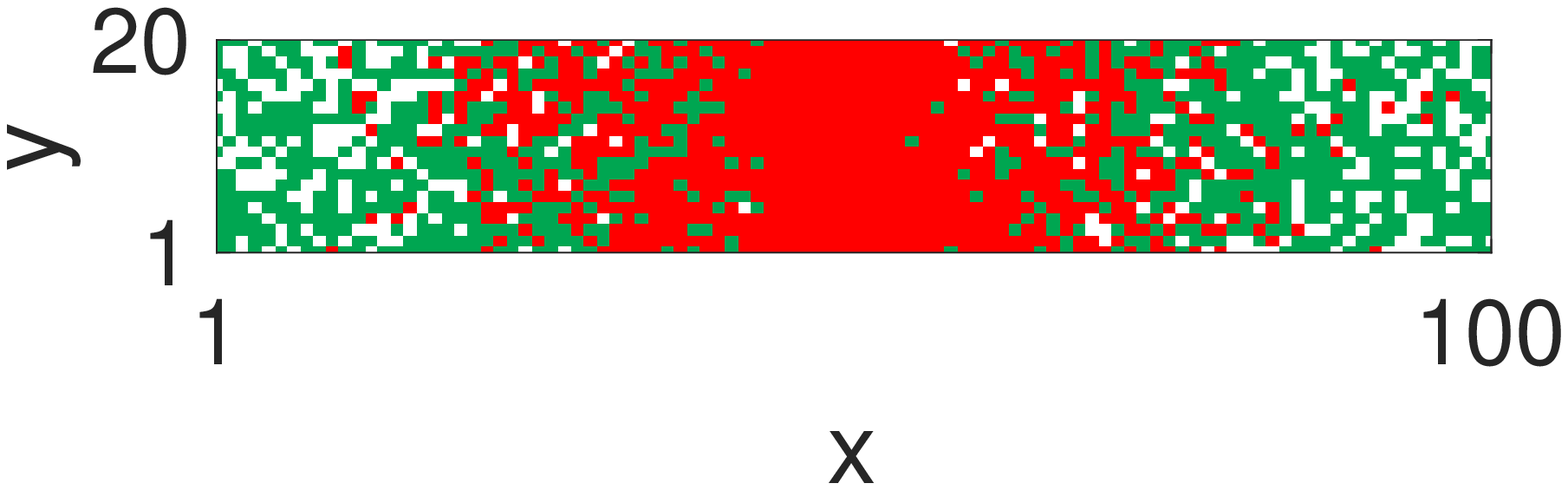}
\label{figure:prolif_agents_Ps=0_T=1000}
}
\setcounter{subfigure}{5}
\subfigure[]{
\includegraphics[width=0.30\textwidth]{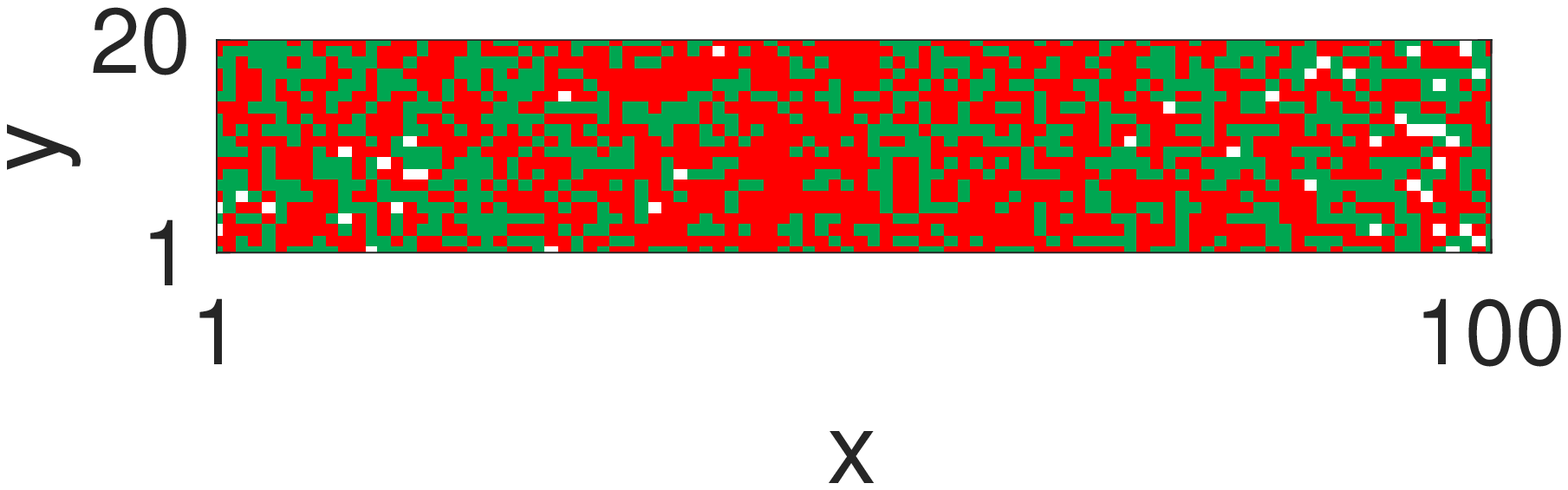}
\label{figure:prolif_agents_Ps=0.5_T=1000}
}
\setcounter{subfigure}{8}
\subfigure[]{
\includegraphics[width=0.30\textwidth]{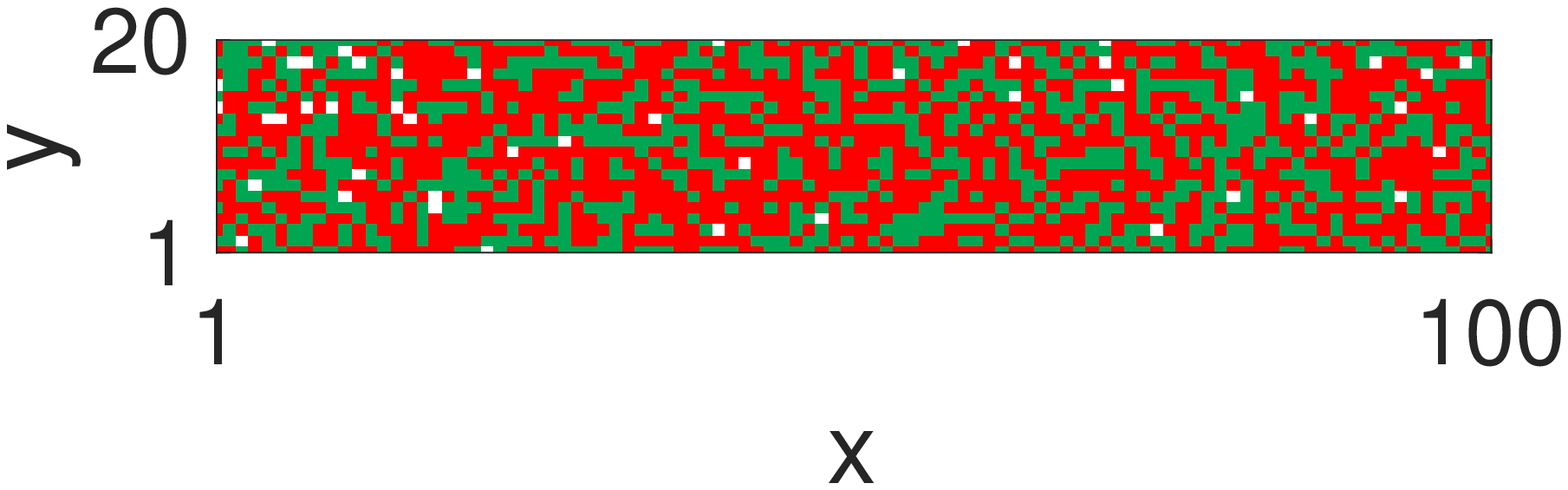}
\label{figure:prolif_agents_Ps=1_T=1000}
}

\end{center}
\caption{Snapshots of the lattice occupancy with swapping probability $\rho=0$ in [\subref{figure:prolif_agents_Ps=0_T=0}-\subref{figure:prolif_agents_Ps=0_T=1000}], $\rho=0.5$ in [\subref{figure:prolif_agents_Ps=0.5_T=0}-\subref{figure:prolif_agents_Ps=0.5_T=1000}] and $\rho=1$ in [\subref{figure:prolif_agents_Ps=1_T=0}-\subref{figure:prolif_agents_Ps=1_T=1000}] at $t=0,500,1000$ for the cell migration process with swapping and proliferation. We initialise the domain as a $20$ by $100$ lattice where all the sites in the horizontal range $41$ to $60$ are occupied by the agents of type A (red) and the remaining sites are inhabited by agents of type B at a density of $0.5$. Both species diffuse at equal rates $r_A=r_B=1$. Rates of proliferation are given by $r_{p}^A=0.01$ and $r_{p}^B=0$.}
\label{figure:prolif_snapshots}
\end{figure}

\begin{figure}[t!]
    \centering
    \includegraphics[width=0.6\textwidth]{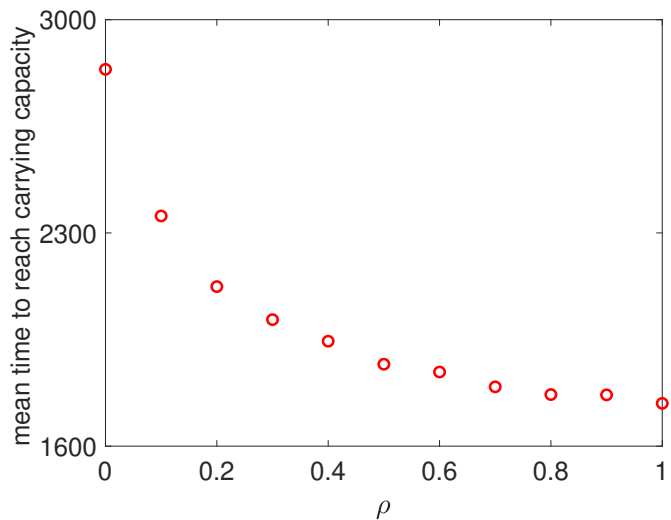}
    \caption{Time to reach the carrying capacity. The red circles show the mean time for the number of agents to reach the carrying capacity of the domain for a range of swapping probabilities (averaged over 100 repeats).}
    \label{fig:time_to_reach_K}
\end{figure}

In Figure \ref{figure:prolif_snapshots} we provide snapshots of the evolving lattice occupancy for $\rho=0, 0.5, 1$ (columns 1, 2 and 3, respectively) at $t=0, 500, 1000$ (rows 1, 2 and 3, respectively). We see, at the same time points, that cells are unsurprisingly more well-mixed in the case of non-zero swapping (second and third columns in Figure \ref{figure:prolif_snapshots}) compared to the zero-swapping situation (first column in Figure \ref{figure:prolif_snapshots}). We also see faster colonisation of the domain overall in the non-zero swapping cases than without swapping. This is because swapping allows the proliferating red agents to disperse more quickly into less dense regions, which in turn increases the probability of a successful division events for these agents. Without swapping, it takes longer for proliferative red agents to find the space to proliferate into. This trend of decreasing colonisation time with increasing swapping probability is reinforced in Figure \ref{fig:time_to_reach_K} where we see that the time to reach the domain's carrying capacity is a decreasing function of the swapping probability, $\rho$.

\begin{figure}[t!]
\begin{center}

\subfigure[]{
\includegraphics[width=0.30\textwidth]{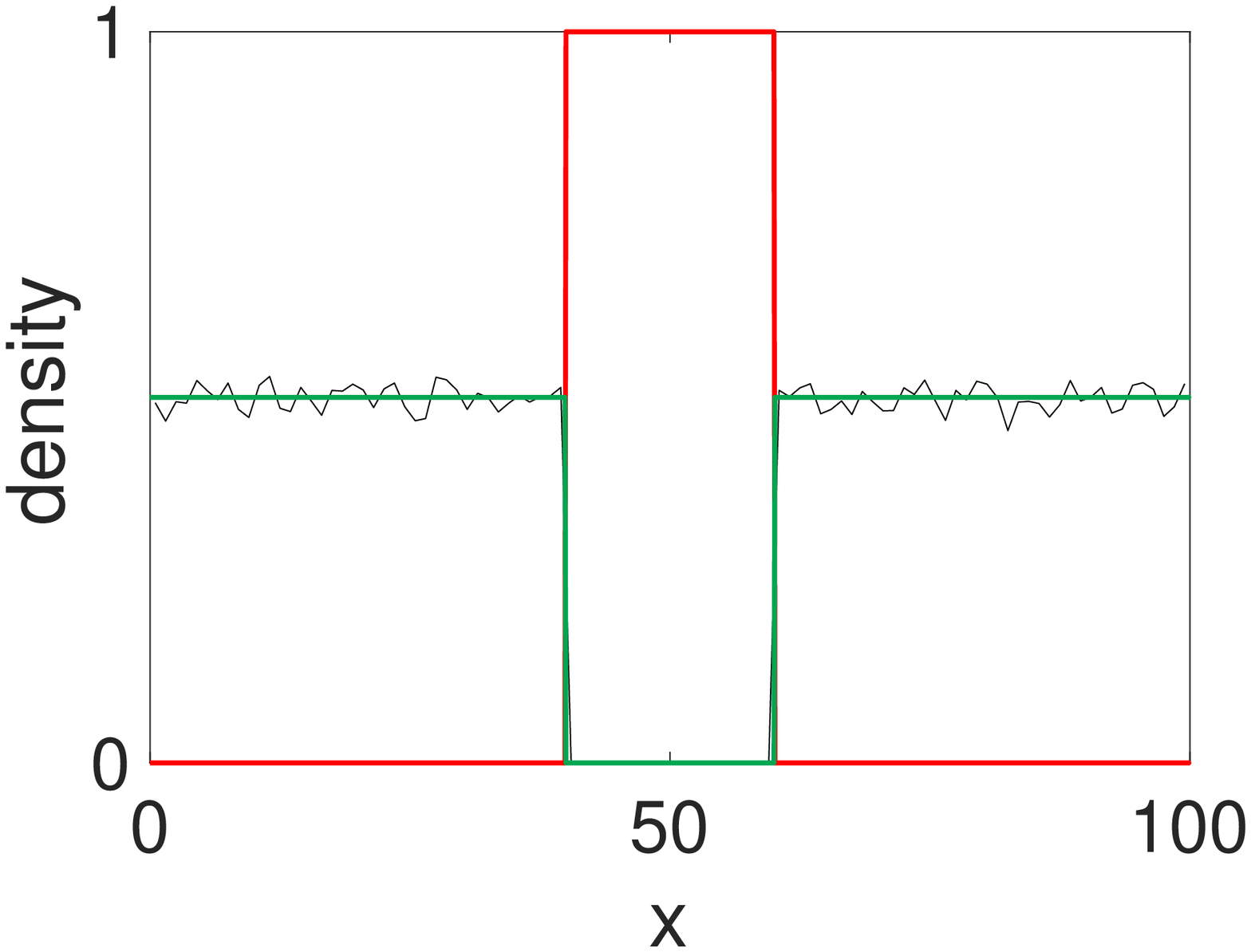}
\label{figure:prolif_agents_density_rho=0_T=0}
}
\setcounter{subfigure}{3}
\subfigure[]{
\includegraphics[width=0.30\textwidth]{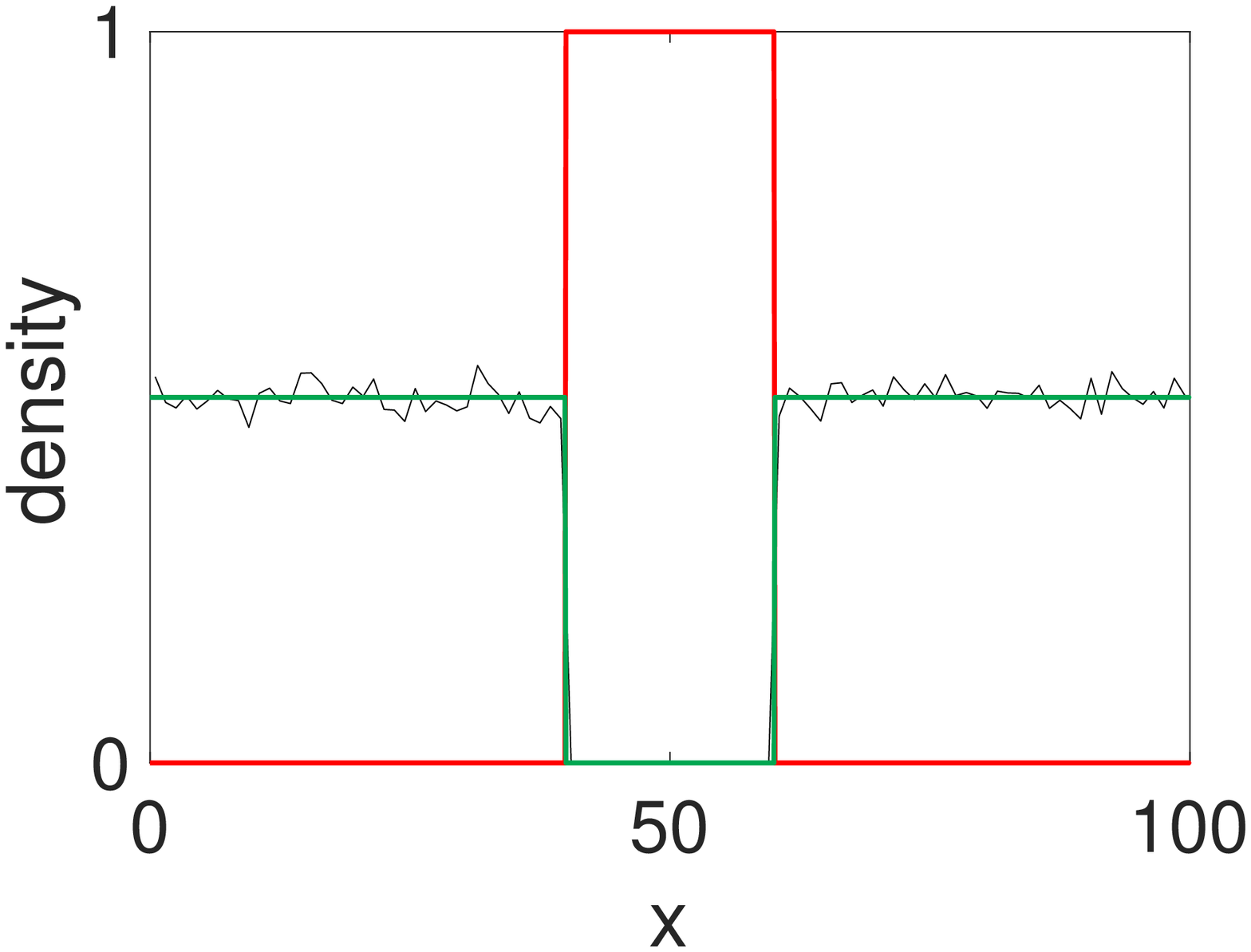}
\label{figure:prolif_agents_density_rho=0.5_T=0}
}
\setcounter{subfigure}{6}
\subfigure[]{
\includegraphics[width=0.30\textwidth]{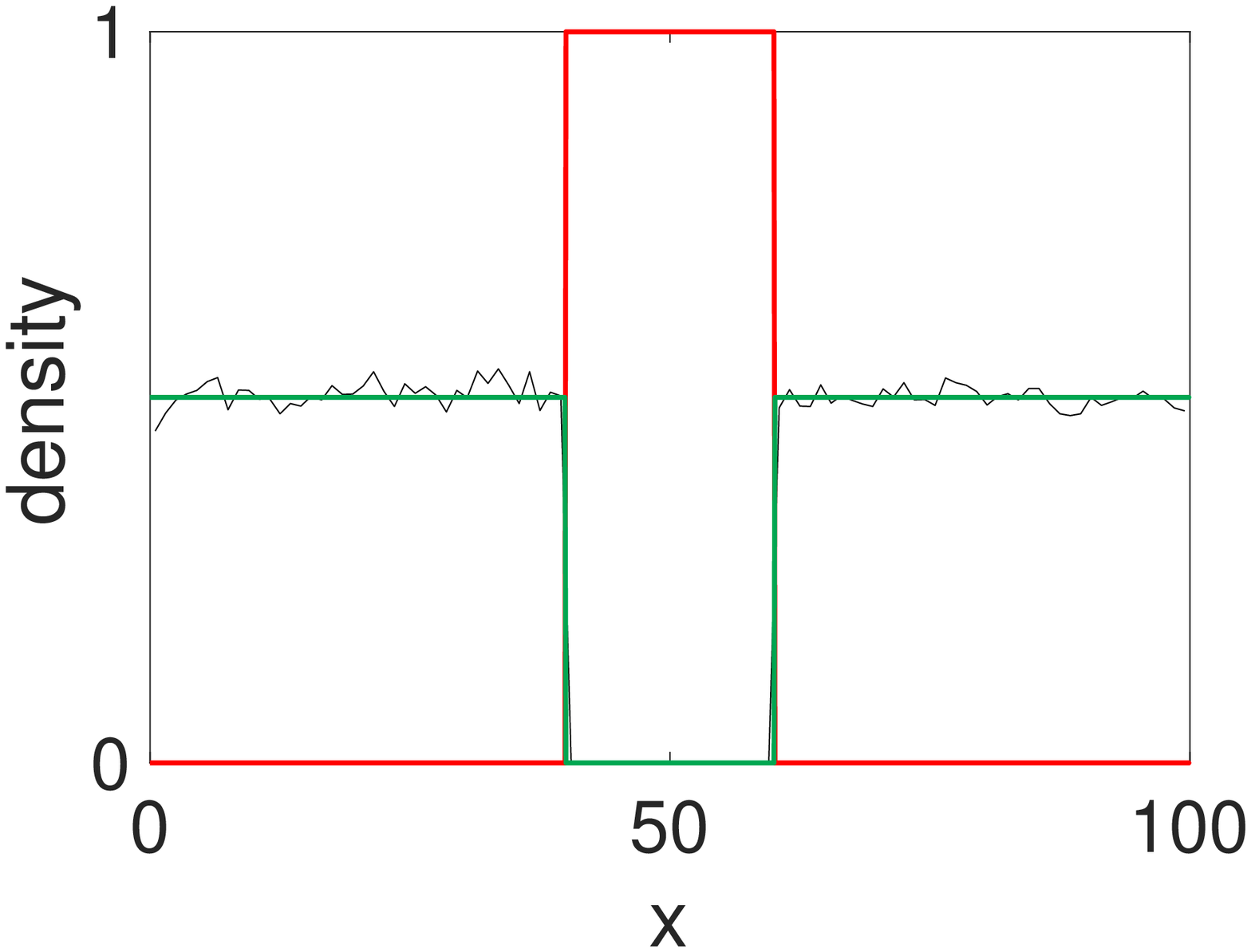}
\label{figure:prolif_agents_density_rho=1_T=0}
}

\setcounter{subfigure}{1}
\subfigure[]{
\includegraphics[width=0.30\textwidth]{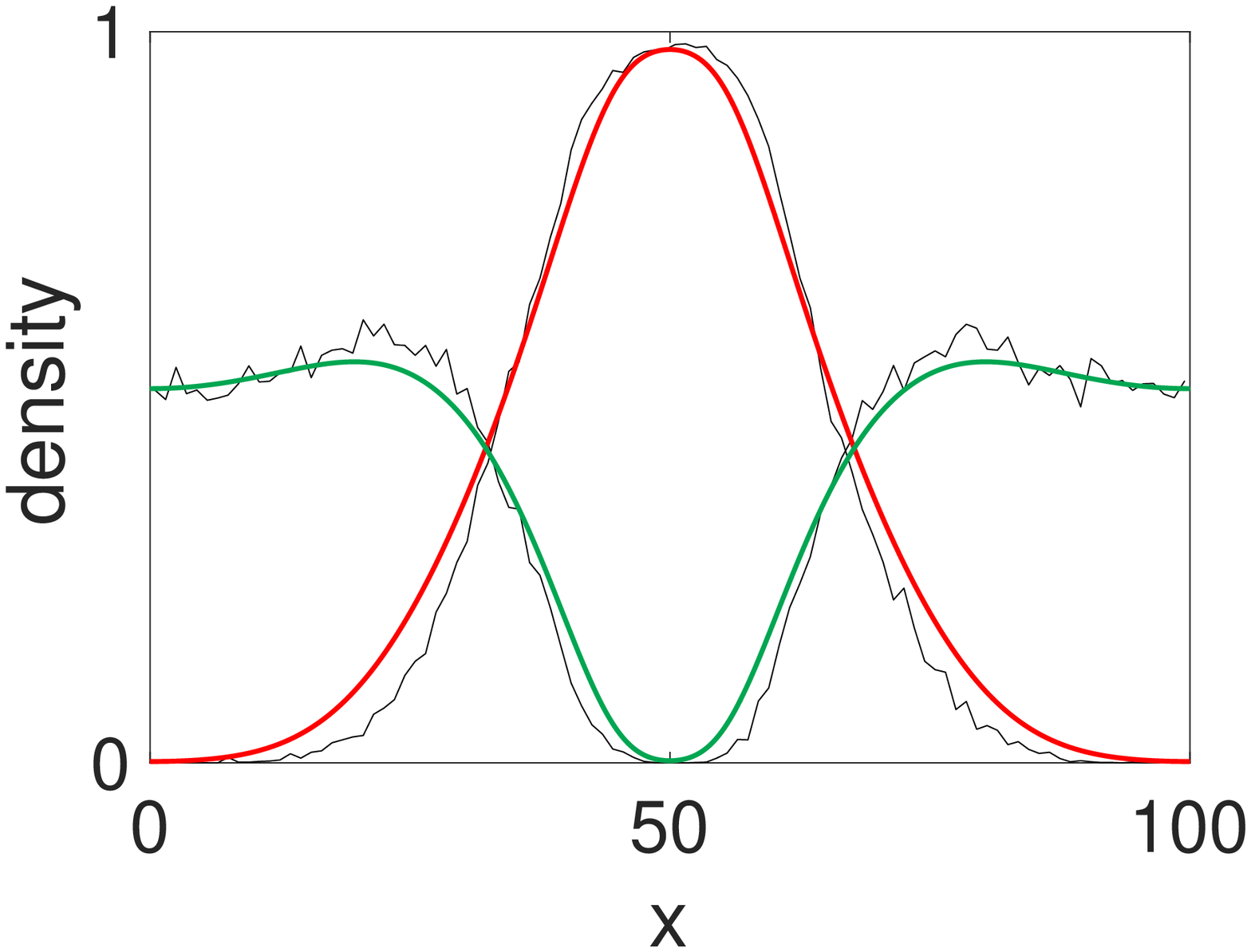}
\label{figure:prolif_agents_density_rho=0_T=500}
}
\setcounter{subfigure}{4}
\subfigure[]{
\includegraphics[width=0.31\textwidth]{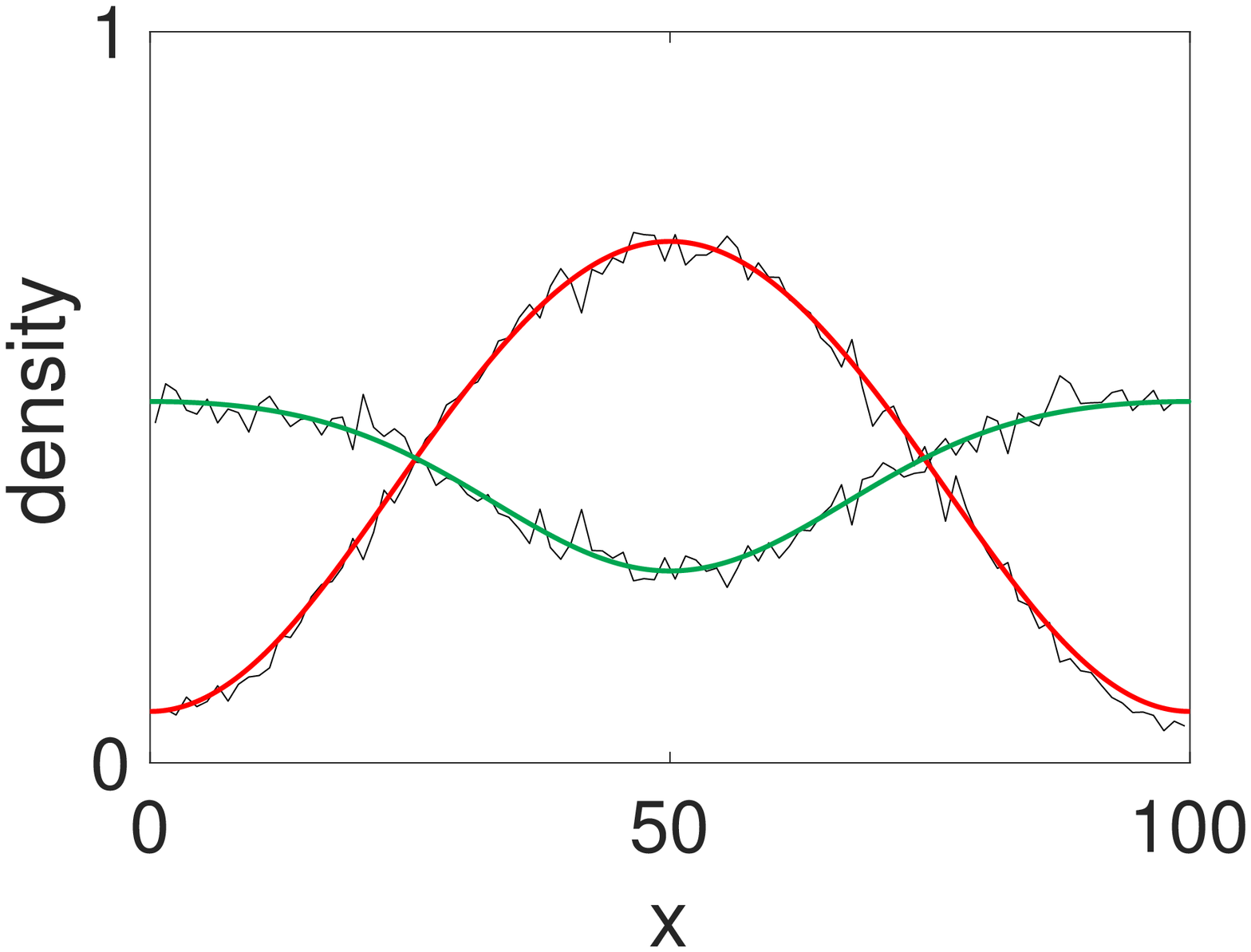}
\label{figure:prolif_agents_density_rho=0.5_T=500}
}
\setcounter{subfigure}{7}
\subfigure[]{
\includegraphics[width=0.30\textwidth]{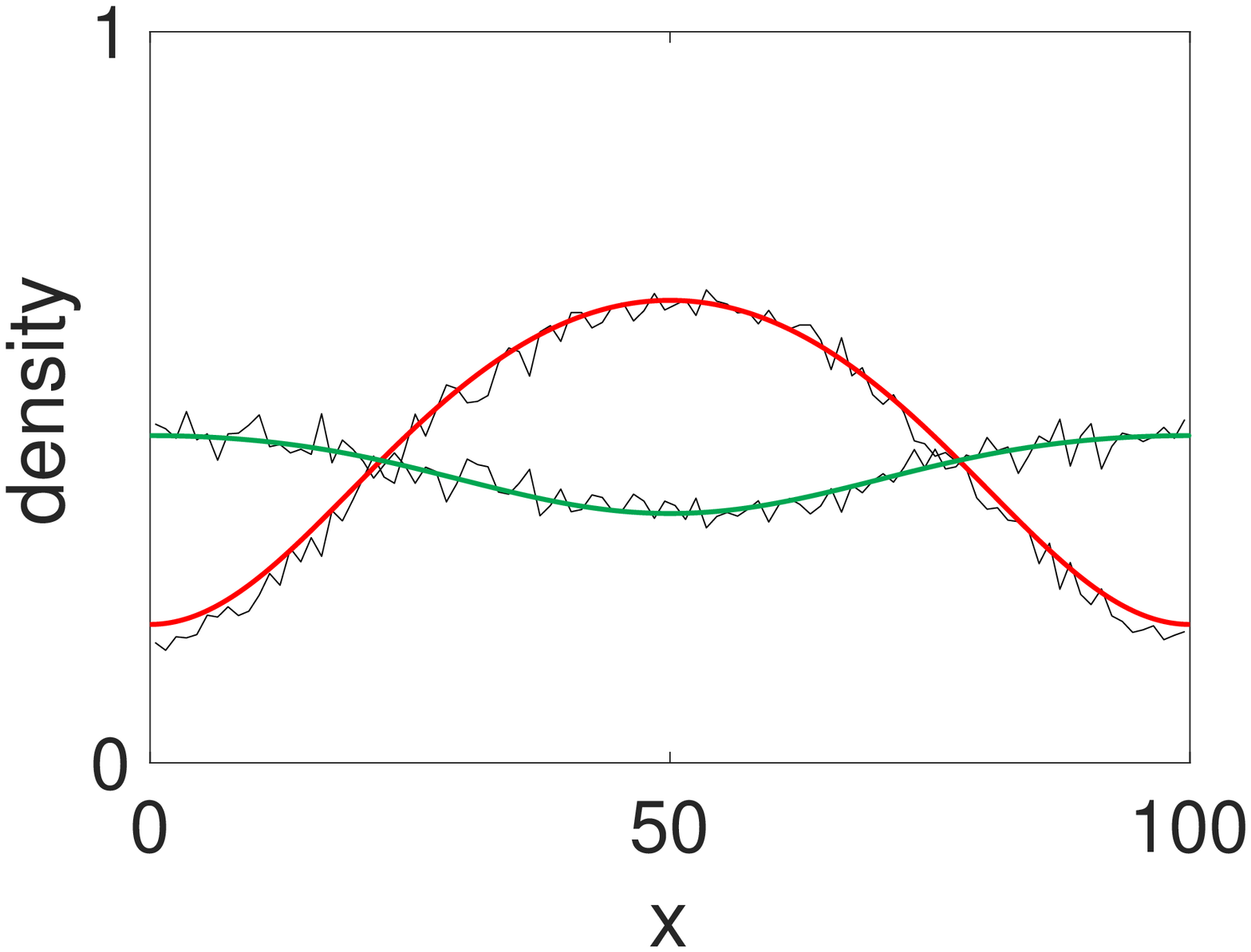}
\label{figure:prolif_agents_density_rho=1_T=500}
}
\setcounter{subfigure}{2}
\subfigure[]{
\includegraphics[width=0.30\textwidth]{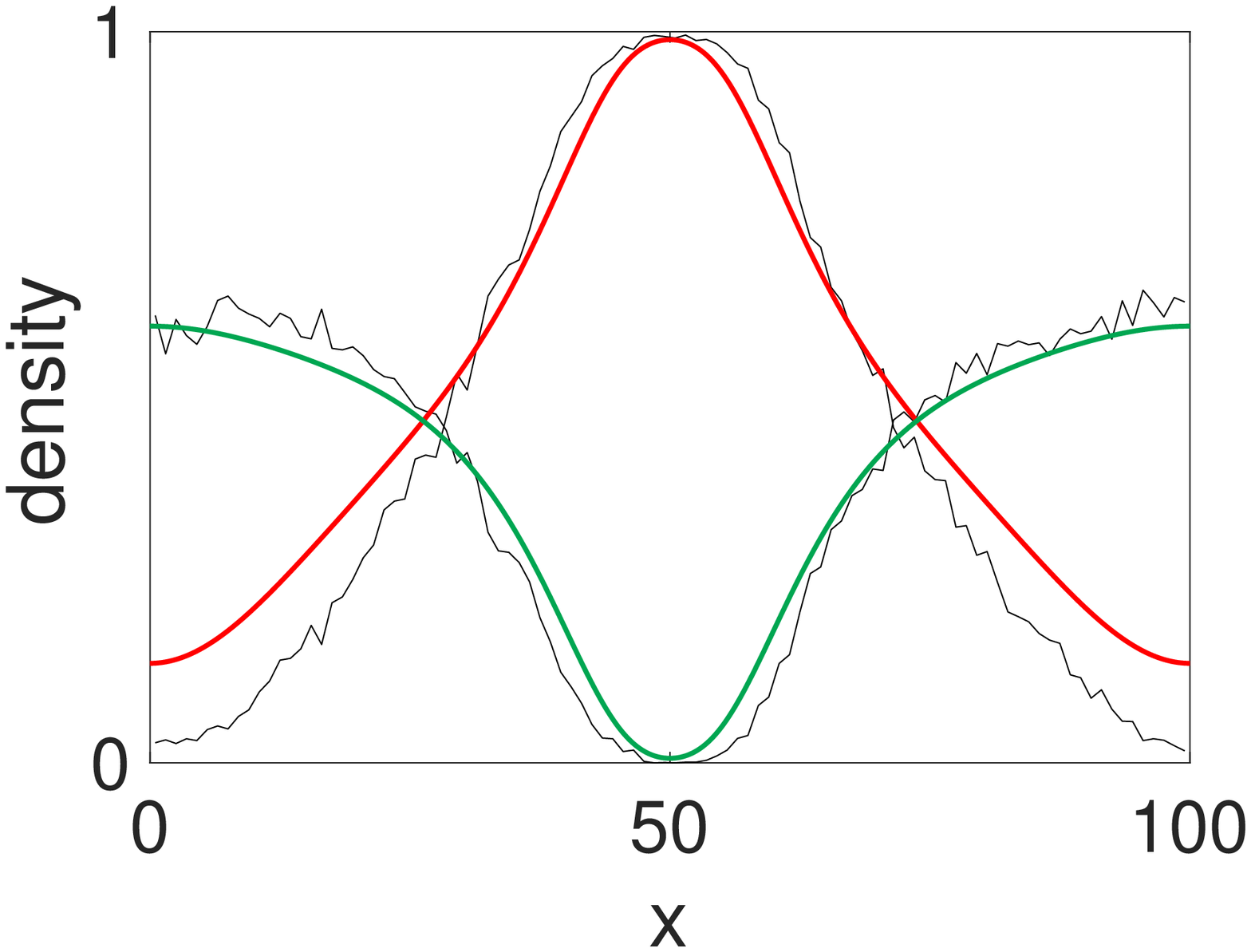}
\label{figure:prolif_agents_density_rho=0_T=1000}
}
\setcounter{subfigure}{5}
\subfigure[]{
\includegraphics[width=0.30\textwidth]{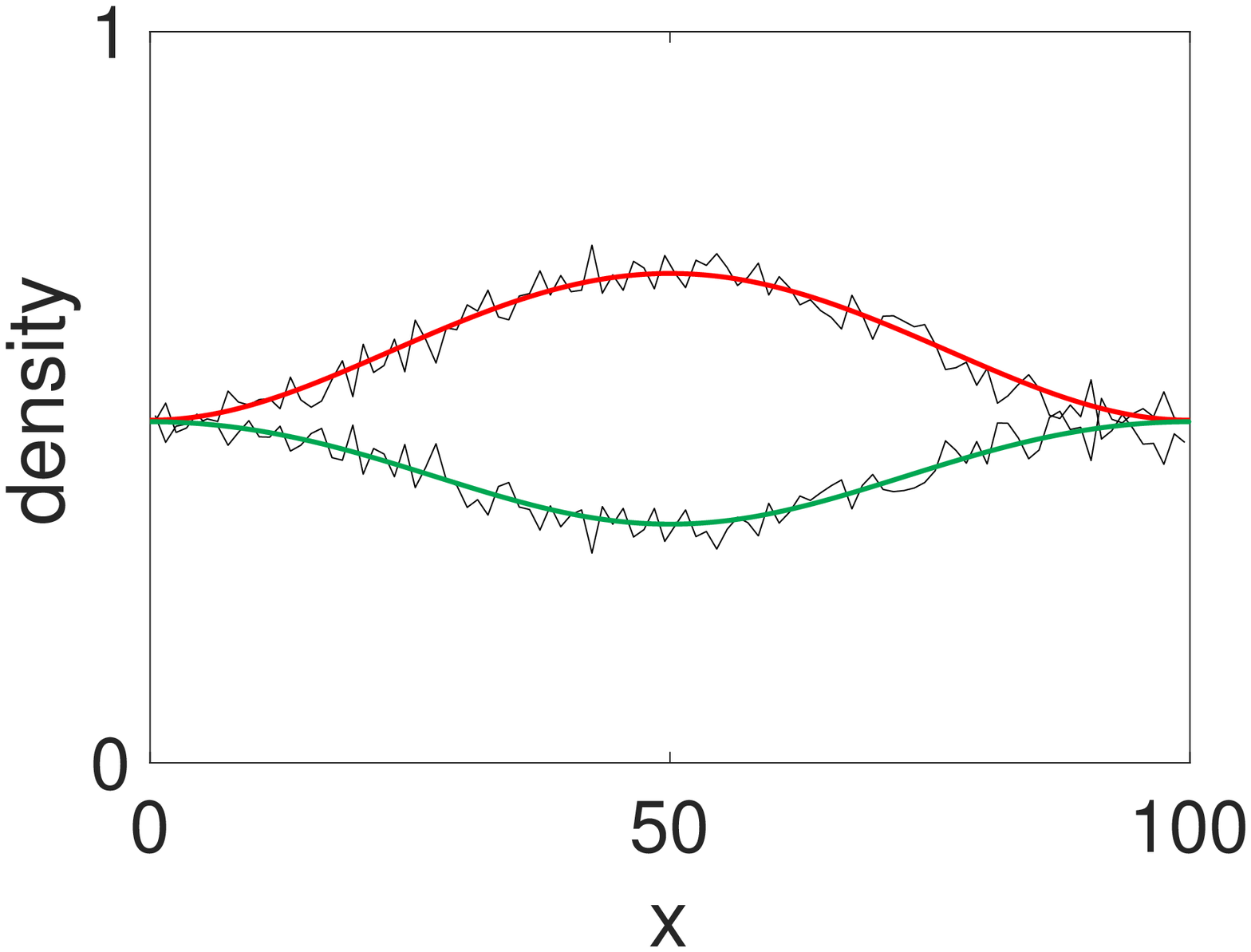}
\label{figure:prolif_agents_density_rho=0.5_T=1000}
}
\setcounter{subfigure}{8}
\subfigure[]{
\includegraphics[width=0.30\textwidth]{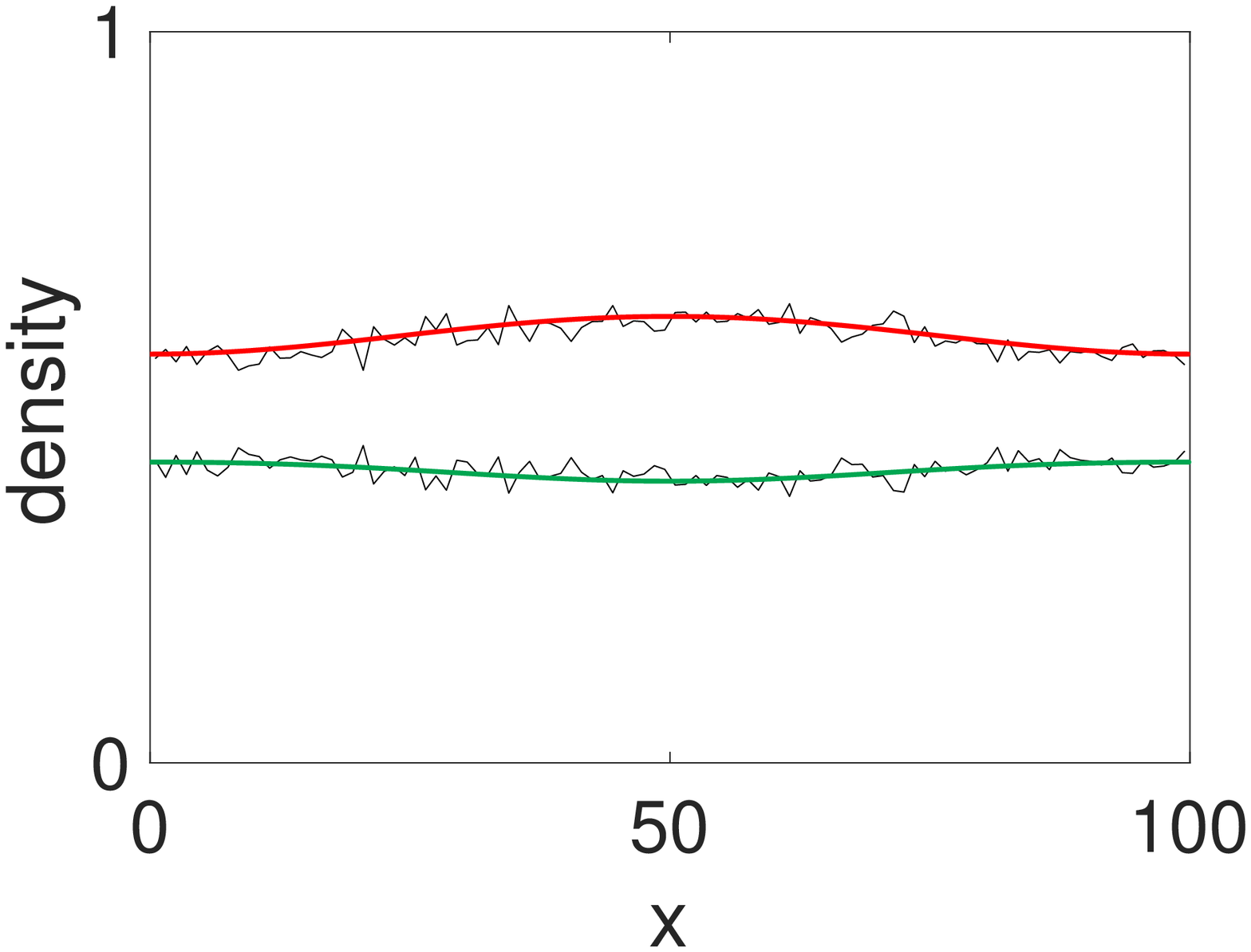}
\label{figure:prolif_agents_density_rho=1_T=1000}
}

\end{center}
\caption{Density profiles for cell migration process with swapping and proliferation. All parameters and initial conditions are the same as Figure \ref{figure:prolif_snapshots}. Here, we present the column densities at $t=0,500,1000$ for $\rho=0$ in [\subref{figure:prolif_agents_density_rho=0_T=0}-\subref{figure:prolif_agents_density_rho=0_T=1000}], $\rho=0.5$ in [\subref{figure:prolif_agents_density_rho=0.5_T=0}-\subref{figure:prolif_agents_density_rho=0.5_T=1000}] and $\rho=1$ in [\subref{figure:prolif_agents_density_rho=1_T=0}-\subref{figure:prolif_agents_density_rho=1_T=1000}] averaged over 100 repeats of the ABM described above. We also plot the corresponding mean-field PDE solutions with $r_p^A=0.01$ in red for species $A$ and in green for species $B$.}
\label{figure:prolif_density_plots}
\end{figure}

The mean-field PDEs describing the approximate population-level dynamics of the agent are given by,

\begin{align}
    \D A t &= \nabla \cdot [D_1(B) \nabla A + D_2(A)\nabla B]+ r_p^A A\left(1-(A+B)\right), \label{eqn:prolif_PDE_M}\\
    \D B t &= \nabla \cdot [D_3(A)\nabla B + D_4(B)\nabla A] \label{eqn:prolif_PDE_X},
\end{align}
where $D_1, D_2, D_3$ and $D_4$ are as defined previously. Notice that proliferation of type-A agents gives rise to an additive source term in Equation \eqref{eqn:prolif_PDE_M}. The derivation of this source is standard and can be found in \citet{plank2012mcc,simpson2010cip}, for example.


In Figure \ref{figure:prolif_density_plots} we compare the average column density of the ABM with the numerical solution of the one-dimensional analogue of the mean-field PDEs obtained by averaging over the $y$ direction with $r_p^A=0.01$. We chose this value for the proliferation rate as we wanted to keep the ratio $r_A/r_p^A \ll 1$ \citep{fadai2019aed,simpson2010cip}. There are two reasons for this: firstly, a modelling choice to prevent agents clustering into proliferation-induced patches and secondly, it is biological realism that given the parameters of the model we could expect real biological cells will attempt proliferation events less frequently than movement events. We see good agreement between the two profiles for non-zero swapping probability. However, when the swapping probability is set to 0, we see discrepancies arising that are amplified as time increases. Recall that the disparity between the PDE and ABM profiles can be also observed in Figure \ref{figure:multispecies_density} due to spatial correlations that are not accounted for by the mean-field PDEs. The addition of proliferation into the model increases the spatial correlations between site occupancies, leading to greater disparity. One way to partially rectify this problem is by modelling the higher order moments in the PDE description \citep{markham2013isc,markham2013smi,simpson2013emi} however for the purposes of this work, we simply note that allowing swapping breaks up the correlations more effectively than in its absence leading to better agreement between the ABM and the population-level densities.

\subsection{Swapping model with cell-cell adhesion}\label{sec:example_adhesion}

Another interesting application of swapping is the formation of patterns in densely crowded environments. In this section, we use a cell-cell adhesion model with swapping to investigate how biologically plausible patterns can form starting from a randomly seeded domain. Our model is based on the similar cell-cell adhesion model studied previously by \citep{simpson2010mbc,charteris2014mca,khain2007rca} who consider adhesion between identical agents. Here, we extend the model to incorporate two types of agents with swapping to facilitate the movement events.

For the purpose of this paper, we assume adhesion between two species, A and B, on a fully populated domain. For a simple exclusion-based ABM the agents on the fully populated domain would not successfully move at all. This is since the exclusion principle forbids cells from occupying the already occupied lattice sites. In our model, the movement of the agents and the formation of patterns will be facilitated by the swapping mechanism. In an on-lattice adhesion model, agents can adhere to other agents in their neighbourhood, making them less likely to successfully complete the movement event. As well as the number of agents in the neighbourhood, the strength of adhesion determines how likely an agent is to successfully move. In a simple model with species A and B, $0\leqslant p\leqslant 1$ characterises the strength of adhesion between two type-A agents and $0\leqslant q\leqslant 1$ the strength of adhesion between two type-B agents.

We assume cell movement in a densely crowded domain where the underlying lattice is fully populated with type-A and B agents, their positions chosen uniformly at random. When an agent is chosen to move into an occupied neighbouring site, we check the feasibility of swapping by sampling a random number $u$ from the standard uniform distribution and comparing it with the swapping probability, $\rho$. A swapping move breaks existing interactions between the two swapping agents and their neighbours and makes new connections following a successful swap. Since the movement of the focal agent and the target agent depends on their respective neighbours, the success of a swapping move depends on whether the respective neighbouring sites are occupied by type-A or type-B agents. Suppose the focal agent is at site $(i,j)$ and let $Z_{ij}=\{(i-1,j),(i+1,j),(i,j-1),(i,j+1) \}$ be the set containing the positions its neighbouring sites on the two-dimensional lattice. If the focal agent is a type-A agent then the probability of it breaking existing connections with its neighbours is given by,

\begin{equation}\label{eqn:p_move_agent_A}
p_\text{break}^\text{agent}=(1-p)^{\mathrm{\Sigma}_{z\in Z} A_z}.
\end{equation}

However, if the focal agent is a type-B agent then,

\begin{equation}\label{eqn:p_move_agent_B}
p_\text{break}^\text{agent}=(1-q)^{\mathrm{\Sigma}_{z\in Z} B_z}.
\end{equation}

Here, $A_z$ are binary taking a value of unity if the site with position $z\in Z_{ij}$ is occupied by a type-A agent or 0 otherwise. Therefore, $\sum_{z\in Z} A_z$ is the sum of occupancies of the sites in $Z_{ij}$ that are occupied by type-A agents. Likewise, $\sum_{z\in Z} B_z$ is the sum of occupancies of the sites in $Z_{ij}$ that are occupied by type-B agents. Similarly, for a type-A agent occupying the target site the probability that it breaks existing connections with its neighbours is given by,

\begin{equation}{\label{eqn:q_move_targ_B}}
p_\text{break}^\text{targ}=(1-p)^{\mathrm{\Sigma}_{y\in Y_z} A_y}.
\end{equation}

However, if the agent occupying the target site is a type-B agent then,

\begin{equation}{\label{eqn:p_move_targ_B}}
p_\text{break}^\text{targ}=(1-q)^{\mathrm{\Sigma}_{y \in Y_z} B_y}.
\end{equation}

Here, the set $Y_z$ contains the positions of sites in the neighbourhood of the target site, $z$. Therefore, $\sum_{y\in Y_z} A_y$ is the sum of occupancies of all the sites in $Y_z$ that are occupied by a type-A agent and $\sum_{y\in Y_z} B_y$ denotes the sum of occupancies of all the sites in $Y_z$ that are occupied by a type-B agent.

The probability of a successful swap given a movement event has been attempted is therefore a product of the swapping probability, the probability of the focal agent at site $(i,j)$ breaking links with its neighbours in order to move out and the probability of the agent at the target site breaking links with its neighbours in order to move in, i.e.,

\begin{equation}
    p_\text{swap} = \rho p_\text{break}^\text{agent} p_\text{break}^\text{targ}.
\end{equation}

\begin{figure}[t!]
\begin{center}

\subfigure[]{
\includegraphics[width=0.3\textwidth]{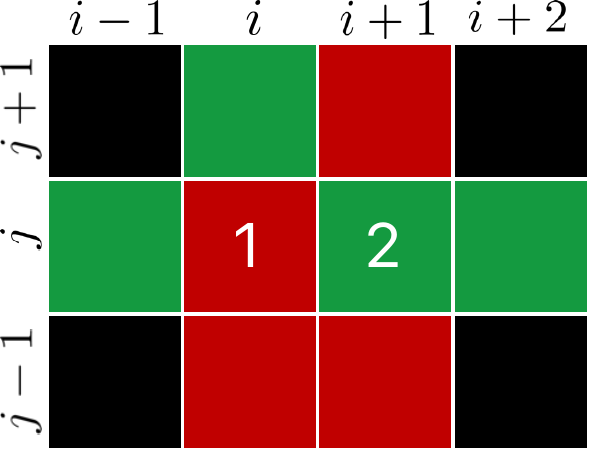}
\label{figure:adh_lattice_1}
}
\subfigure[]{
\includegraphics[width=0.3\textwidth]{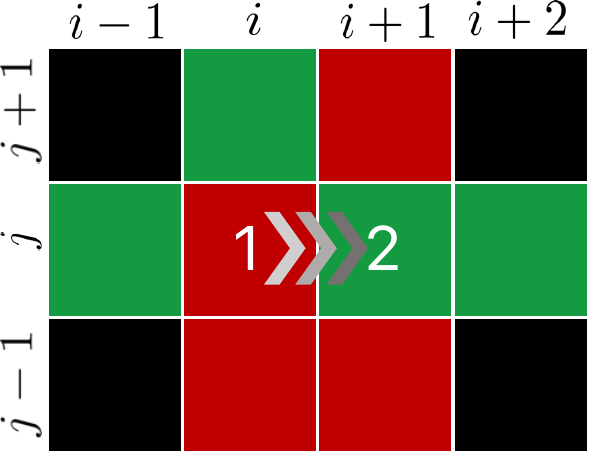}
\label{figure:adh_lattice_2}
}
\subfigure[]{
\includegraphics[width=0.3\textwidth]{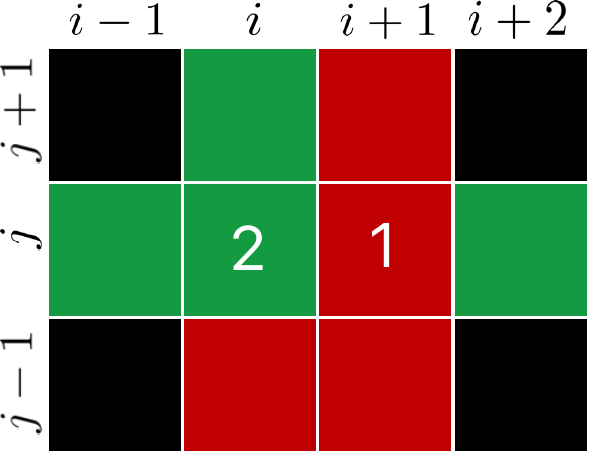}
\label{figure:adh_lattice_3}
}
\end{center}
\caption{A schematic illustrating swapping in the cell-cell adhesion model. A red site represents a type-A agent and a green site represent a type-B agent. Agent 1 at site $(i,j)$ attempts to swap with agent 2 at site $(i+1,j)$. We have coloured in black the neighbouring sites that are unimportant in this context (i.e. occupancy of these sites does not affect the probability of a successful swap happening between agent 1 and agent 2). The initial configuration of the lattice is shown in \subref{figure:adh_lattice_1}. Agent 1 attempts to swap with agent 2 \subref{figure:adh_lattice_2} and the state of the lattice after the successful swap is shown in \subref{figure:adh_lattice_3}.}
\label{figure:adh_lattice_all}
\end{figure}

Here we are considering that the link between the two swapping agents has to be broken in order for the swap to take place. As an example, consider the case in which the focal agent at site $(i,j)$ attempts to jump into the neighbouring site $(i+1,j)$ (Figure \ref{figure:adh_lattice_all}). The agent occupying the focal site is a type-A agent, shown in red and labelled as 1 and the target site $(i+1,j)$ is occupied by a type-B agent, shown in green and labelled as 2. Since the focal agent is red, it is adhesive to other red agents in its Von Neumann neighbourhood. There is only one site (with position $(i,j-1)$) neighbouring agent 1 that is occupied by a red agent and therefore, $\sum_{z\in Z_{ij}} A_z = 1$. Since agent 2 (target site) is a green agent, it is adhesive to the other green agents in its Von Neumann neighbourhood. There is only one green agent in the neighbourhood (position $(i+2,j)$) and hence, $\sum_{y\in Y_{i+1,j}} B_y = 1$. For an arbitrary $\rho$, $p$ and $q$, the probability of swap in this situation given a neighbouring site to move into has already been chosen can be written as $p_\text{swap}=\rho(1-p)(1-q)$.

\begin{figure}[h!]
\begin{center}

\subfigure[]{
\includegraphics[width=0.31\textwidth]{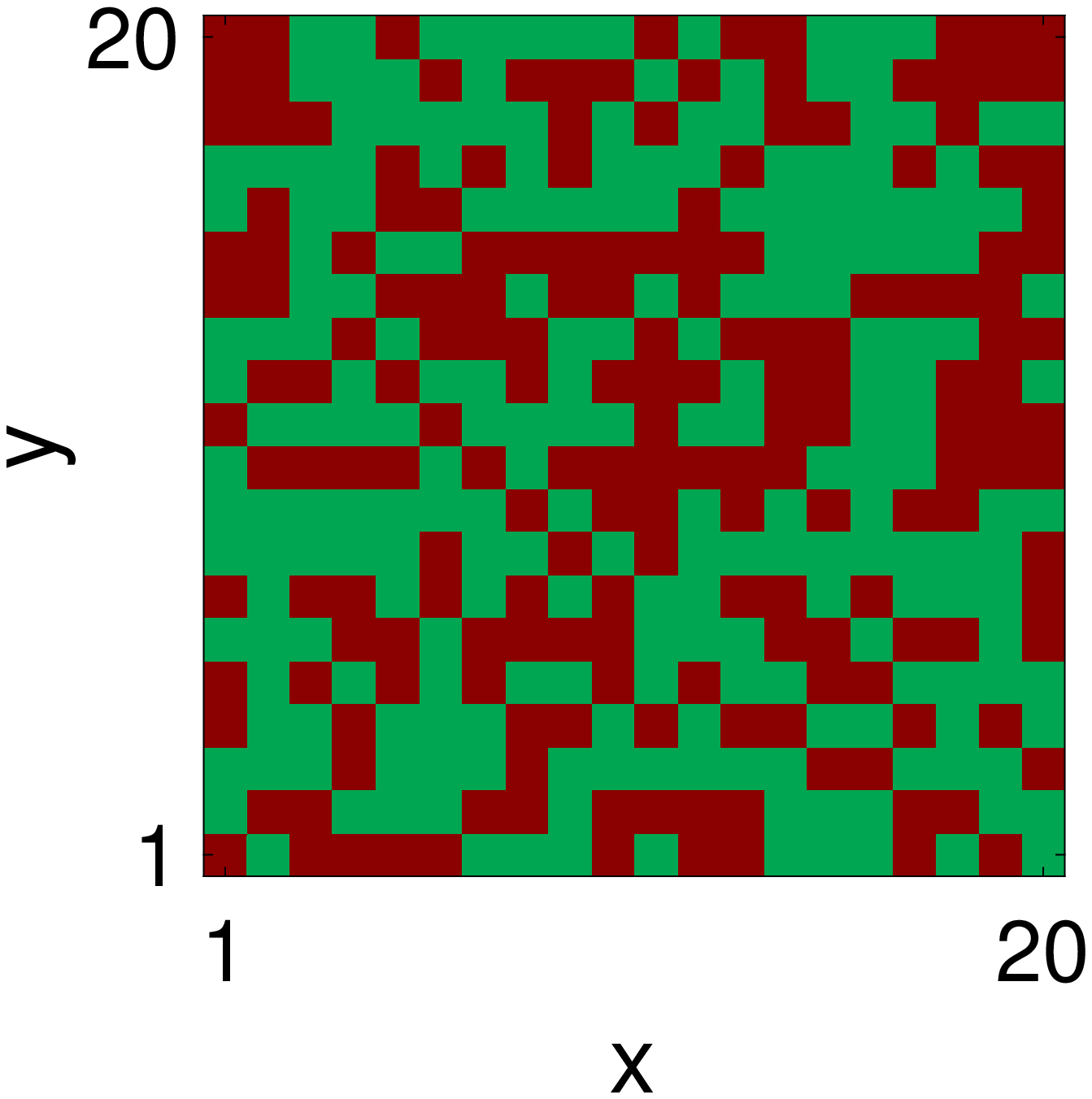}
\label{figure:pattern_t=0}
}
\subfigure[]{
\includegraphics[width=0.31\textwidth]{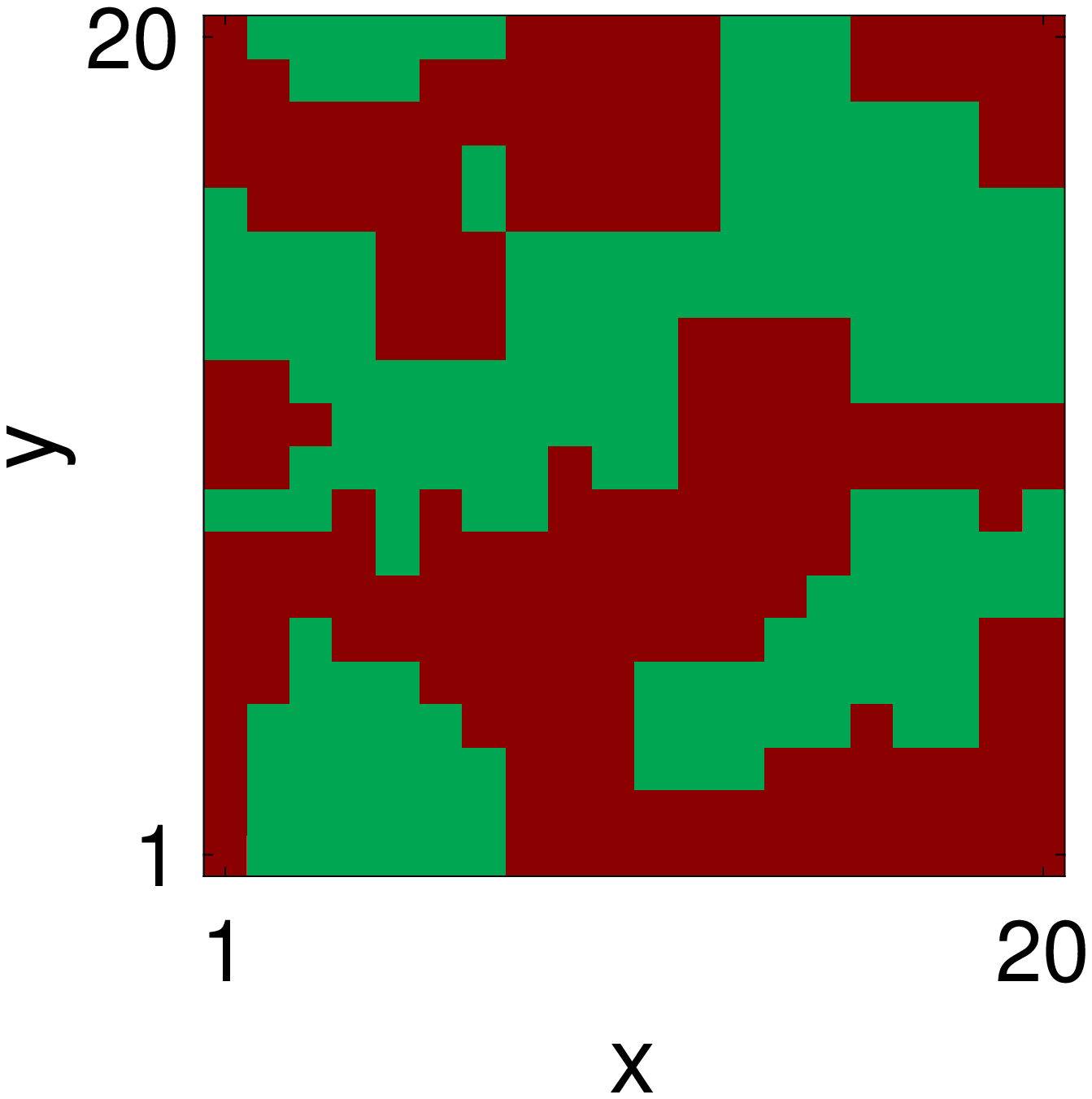}
\label{figure:pattern_b}
}
\subfigure[]{
\includegraphics[width=0.31\textwidth]{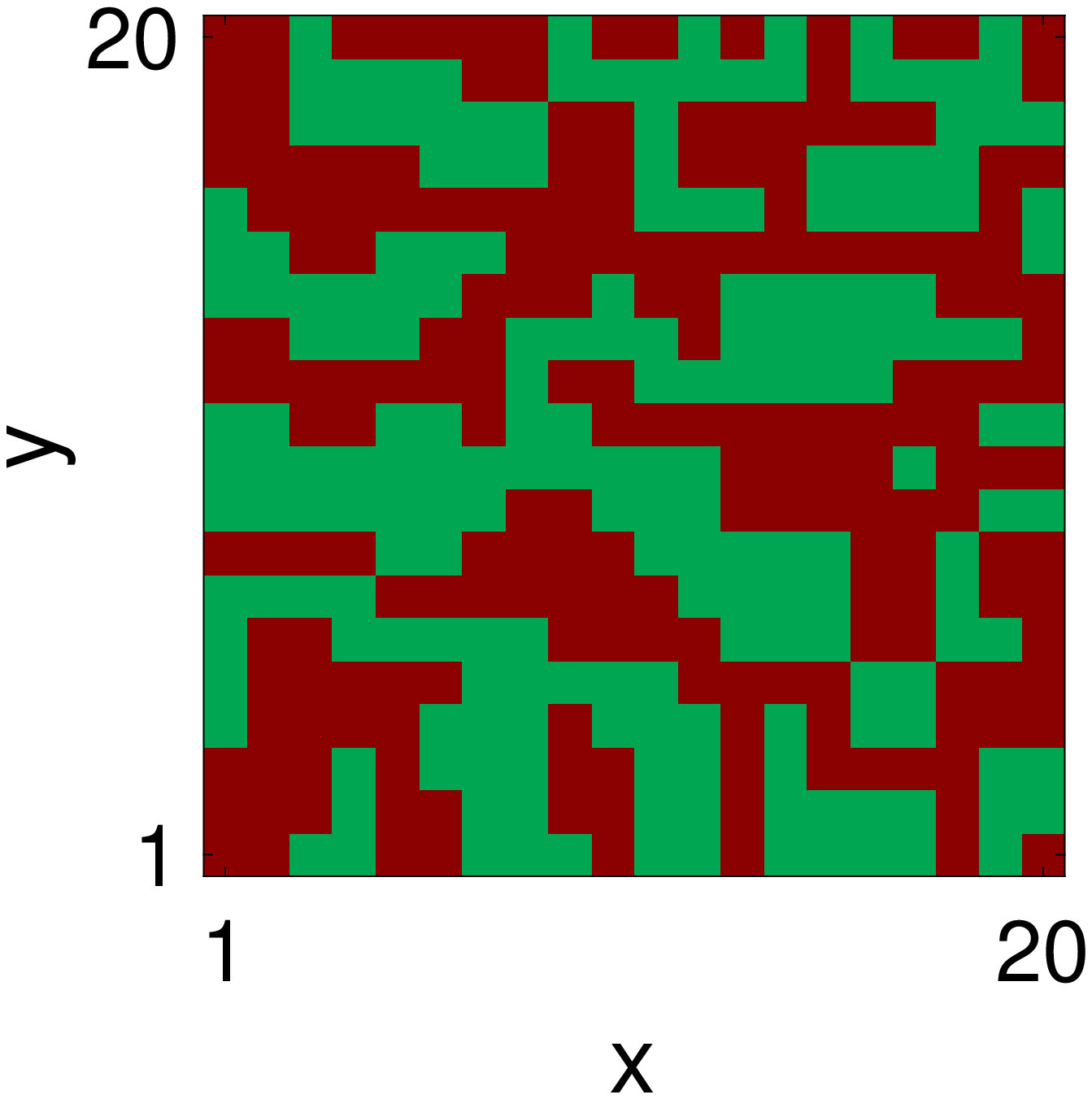}
\label{figure:pattern_c}
}

\end{center}
\caption{Pattern formation in a crowded environment. The domain is initialised as a square lattice with $L=20$ sites in the horizontal and vertical directions, respectively. Initially, the domain is fully populated by type-A and type-B agents at a density of 0.5 each, where the positions of the agents are assigned uniformly at random (as shown in \subref{figure:pattern_t=0}). We present the state of the lattice at $t=15,000$ for two different sets of adhesion strengths: $p=0.90$ and $q=0.70$ \subref{figure:pattern_b} and $p=q=0.98$ \subref{figure:pattern_c}. In both cases, we let $\rho=1$ and the movement rates of the two species $r_A=r_B=1$.}
\label{figure:patterns}
\end{figure}

In Figure \ref{figure:patterns}, we present some results that demonstrate the importance and impact of swapping in a model of adhesion-mediated pattern formation. For this, we consider a square lattice with $L=20$ sites in both the horizontal and vertical direction. As before, one compartment can accommodate no more than a single agent at a time. We impose periodic boundary conditions. We seed the lattice with type-A and type-B agents at a density of 0.5 each, where the initial positions of the agents on the lattice are assigned uniformly at random (Figure \ref{figure:pattern_t=0}). We let positions of the agents evolve according to the kinetics described above using the movement rate $r_A = r_B=1$ and $\rho=1$.

Snapshots of the evolving lattice occupancy at $t=15,000$ for two different sets of $p$ and $q$ values are shown in Figure \ref{figure:pattern_b} and \subref{figure:pattern_c}. We can see self-organisation of agents into clusters of like type agents. The characteristic size of the aggregates is sensitive to the magnitude of the adhesion strengths. The clusters are bigger where self-adhesion within one species is stronger than within the other ($p=0.9$ and $q=0.7$) whereas we see more labyrinthine patterns when both the adhesion strengths are very strong and equal ($p=q=0.98$). It is evident from these figures that in densely crowded domains, swapping plays a vital role in allowing agents to organise themselves to form patterns. Without swapping, no agent movement and hence no pattern formation would be possible.

\section{Discussion and conclusion}\label{sec:discussion}

Cell movement is often modelled as a volume-exclusion process. However, in reality cell movement is not completely inhibited volume exclusion. In this paper, motivated by real-life examples, we developed an ABM that allows a pair of neighbouring agents to swap positions with each other. Our model maintains the important carrying capacity component of volume-exclusion models but allows the flexibility of movement observed even amongst some densely packed configurations. We considered a two-species system and allowed it to evolve according to the dynamics of our model. We found that swapping enhances the movement of agents by allowing agents to mix more compared to the pure volume-exclusion model. Comparing the ABM to the population-level model we found excellent agreement between the two as long as the swapping probability was sufficiently large.

To understand how swapping affects agent movement at an individual level, we analysed simulated agent tracks to determine the time-uncorrelated individual-level diffusion coefficient. We found that swapping enhanced the movement of agents in all cases compared to the volume-exclusion model. Using the probability master equation, we were able to analytically derive an expression for the diffusion coefficient that confirms the relationship obtained via the simulated tracks.

In Section \ref{sec:examples}, we demonstrated the importance of swapping via a couple of examples. In the first application, we considered a cell migration model with proliferation. We found that swapping accelerates the proliferation process by allowing the agents to disperse more and breaking spatial correlations. We found that the time to reach the domain's carrying capacity varies inversely with the swapping probability. Deriving the PLM and comparing it to the average density of the ABM, we showed that there is a good agreement between the two profiles for sufficiently large swapping probability. For the $\rho=0$ case, we note discrepancies that are caused by the build up of spatial correlations between sites.

For the second example, by incorporating swapping into a cell-cell adhesion model, we showed that agents can spontaneously rearrange themselves into clusters to form patterns even on densely populated domains. We stress that without swapping these patterns would not be realised. Biological patterns involving cell-cell adhesion have been observed amply in experimental work. \citet{honda1986tpc} reported chequered patterns in Japanese quail oviduct epithelium which were later modelled mathematically by \citet{glazier1993sda} using a cellular Potts model. \citet{armstrong1971lem} observed sorting of neural and pigmented retinal epithelial cells in chick embryo where neural cells completely engulfed the pigmented cells from an initially disordered arrangement. Engulfment was modelled mathematically using a cellular Potts model \citep{glazier1993sda,graner1992sbc} and by \citet{armstrong2006cam} using a continuous approach.

Models incorporating swapping may prove useful when analysing multi-species systems in which cells need to migrate in a crowded epithelium or tissue parenchyma. These systems are often interrogated experimentally using the generation of chimaeras or mosaics, analysing the resulting patterns generated by cell mixing \citep{chang2011tos,landini2000mmp,petit2009lca,mort2011eap,mort2009mas}. For example the generation of corneal epithelial stripes in X-inactivation mosaic female mice hemizygous for an X-linked copy of the LacZ gene, expressing the enzyme $\beta$-galactosidase \citep{tan1993xio}. Here, a mixture of $\beta$-gal positive and $\beta$-gal negative limbal epithelial cells are specicifed as limbal stem cells whose progeny then migrate centripetally to generate $\beta$-gal positive and $\beta$-gal negative corneal epithelial stripes \citep{collinson2002cap}. Experiments suggest that the degree of mixing of the limbal progenitors will generate stripes of different width composed of multiple clones of the same colour \citep{tan1993xio}. Swapping would be useful in understanding how cell mixing at the limbus impacts the final stripe pattern generated.

In conclusion, motivating our study by real-life examples, we have developed a cell migration model that incorporates swapping as a viable movement process. As well as adding biological realism, our model has the added benefit of better agreement between the corresponding continuum description and the ABM compared to the classical volume-exclusion process. We also saw that the ABM with swapping and cell-cell adhesion, when applied to cells in densely crowded environments, leads to pattern formation. We once again stress that the patterns would be unattainable under the traditional volume-exclusion model. The results in this paper hint that swapping is an important and overlooked mechanism in the context of modelling real biological scenarios and merits further explorations in conjunction with experimental data.

\section*{Acknowledgements}

Shahzeb Raja Noureen is supported by a scholarship from the EPSRC Centre for Doctoral Training in Statistical Applied Mathematics at Bath (SAMBa), under the project EP/S022945/1. This research made use of the Balena High Performance Computing (HPC) Service at the University of Bath. The images in Figure \ref{fig:cell_swapping_1} were taken by Dr. Matthew Ford (Centre for Research in Reproduction and Development, McGill, Canada). Richard Mort was supported by North West Cancer Research (NWCR Grant CR1132) and the NC3Rs (NC3Rs grant NC/T002328/1). We would like to thank the members of Christian A. Yates’ mathematical biology journal club for constructive and helpful comments on a preprint of this paper. Finally, we would like to thank Fraser Waters for his help in managing some of the code related to this project.

\pagebreak

\appendix

\section*{Appendices}

\section{Single-species PDE vs. ABM results}{\label{app:appendixA}}

Below we summarise the single-species swapping model and present results comparing the discrete and continuum models.

The single-speices model is a simplified version of the two-species model in the sense that there is only species on the domain of interest at density $c$ and hence only a single movement rate. We summarise the the model here.

We let $r$ be the rate of movement of an agent such that $r \delta t$ is the probability that the agent attempts to move during a finite time interval of duration $\delta t$. The agent attempts to move into one of its four neighbouring sites with equal probability. If the chosen neighbouring site is empty, the focal agent successfully moves and its position is updated. If another agent already occupies the site, the agent at site $(i,j)$ attempts to swap positions with the neighbouring agent with probability $\rho$. If the swap is successful, the two agents exchange positions with each other. Otherwise, the move is aborted.

\begin{figure}[h!]

\begin{center}

\subfigure[]{
\includegraphics[width=0.475\textwidth]{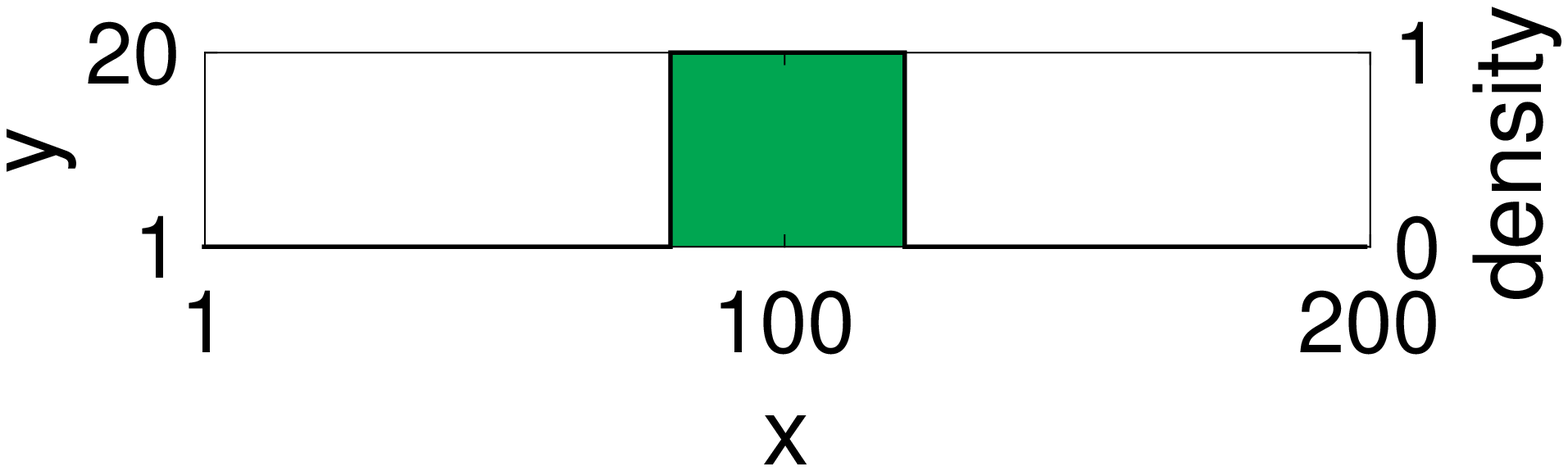}
\label{figure:sss_Ps=0_t=0}
}
\setcounter{subfigure}{3} 
\subfigure[]{
\includegraphics[width=0.475\textwidth]{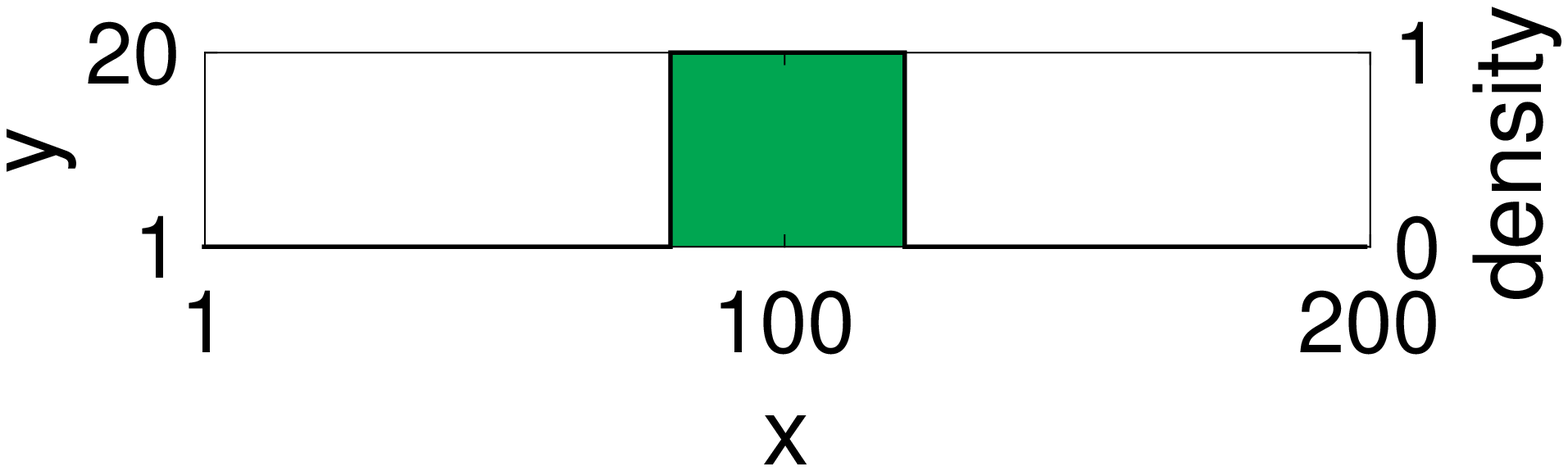}
\label{figure:sss_Ps=0.5_t=0}
}

\setcounter{subfigure}{1}
\subfigure[]{
\includegraphics[width=0.475\textwidth]{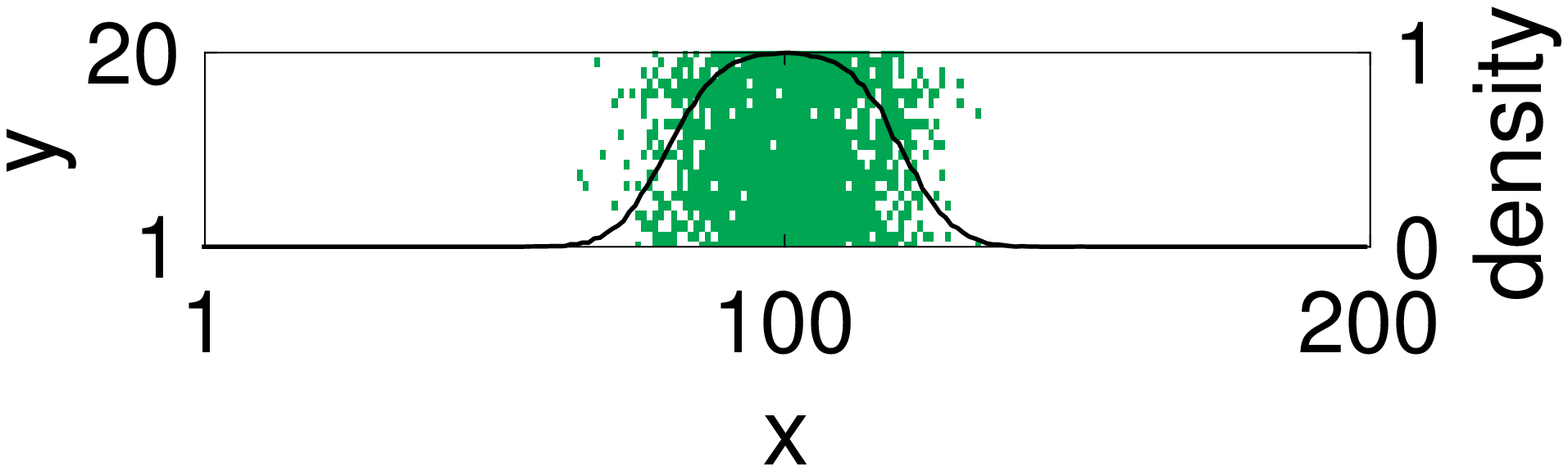}
\label{figure:sss_Ps=0_t=100}
}
\setcounter{subfigure}{4}
\subfigure[]{
\includegraphics[width=0.475\textwidth]{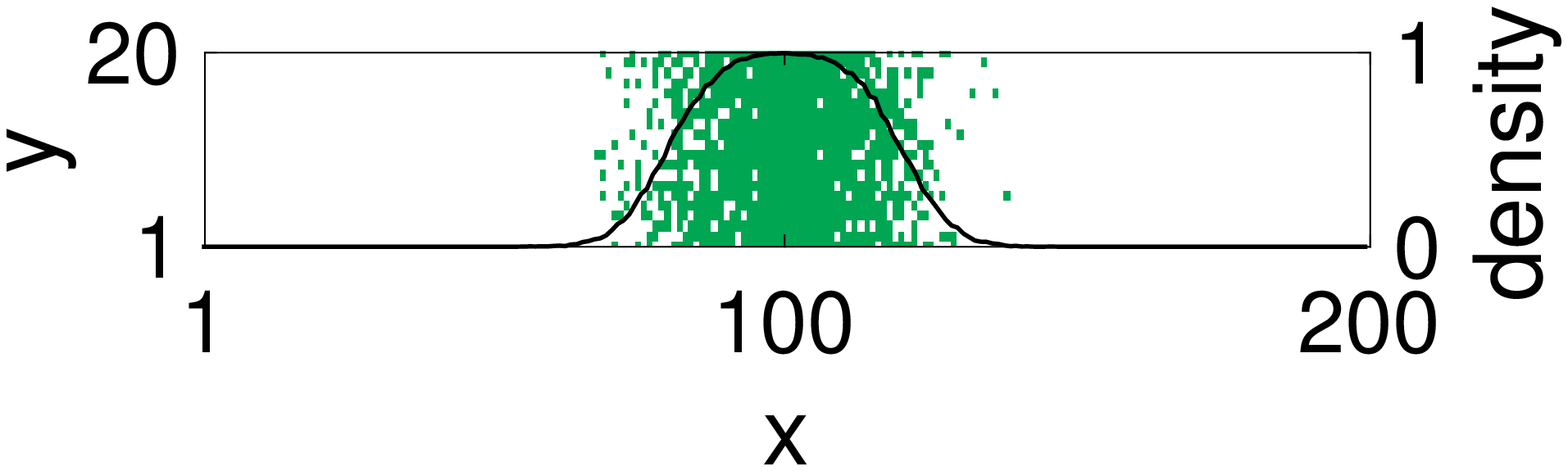}
\label{figure:sss_Ps=0.5_t=100}
}

\setcounter{subfigure}{2}
\subfigure[]{
\includegraphics[width=0.475\textwidth]{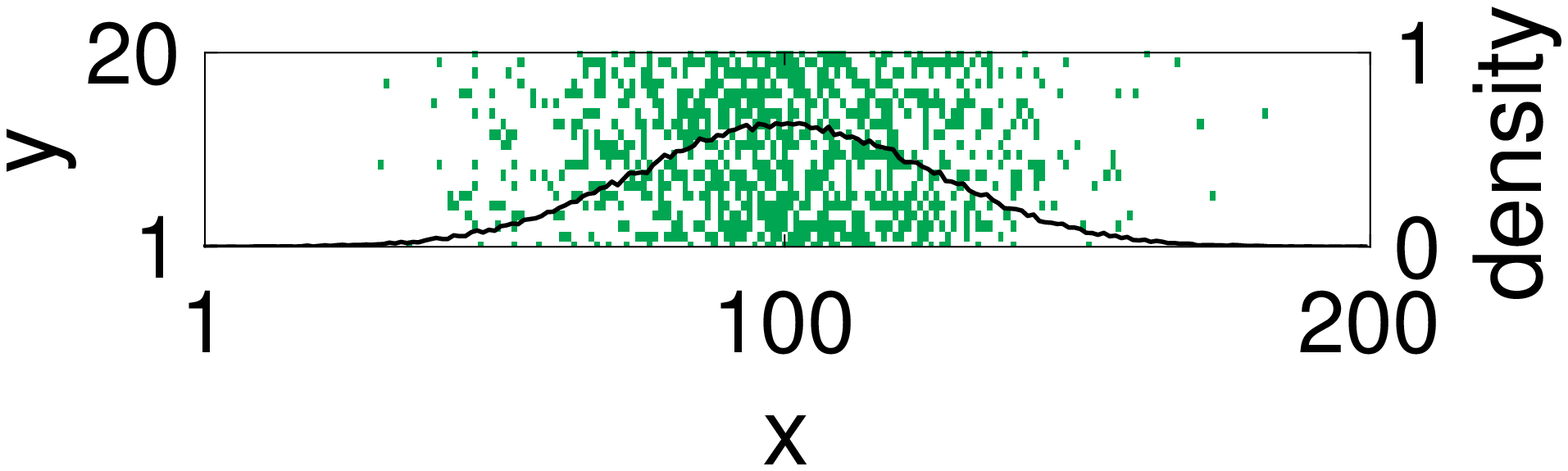}
\label{figure:sss_Ps=0_t=1000}
}
\setcounter{subfigure}{5}
\subfigure[]{
\includegraphics[width=0.475\textwidth]{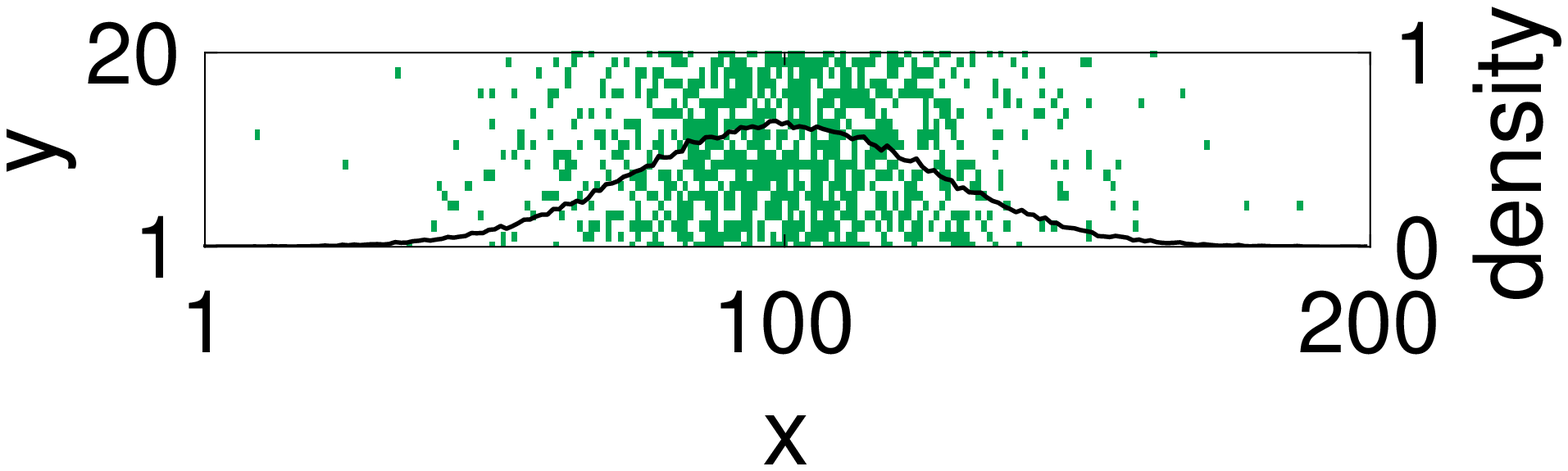}
\label{figure:sss_Ps=0.5_t=1000}
}

\subfigure[]{
\includegraphics[width=0.475\textwidth]{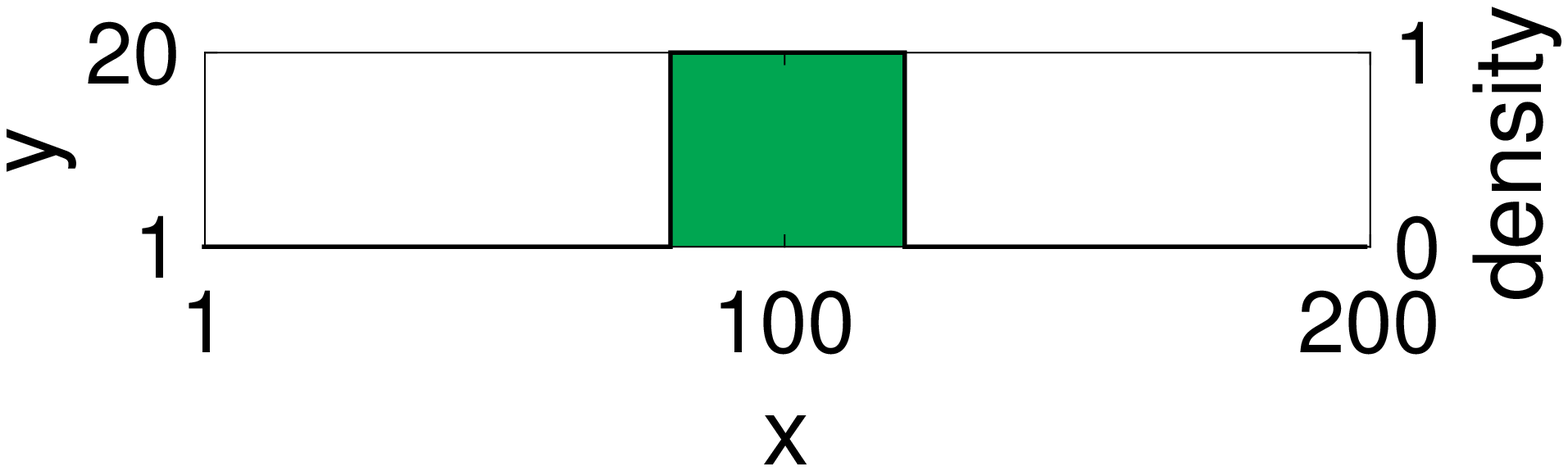}
\label{figure:sss_Ps=1_t=0}
}

\subfigure[]{
\includegraphics[width=0.475\textwidth]{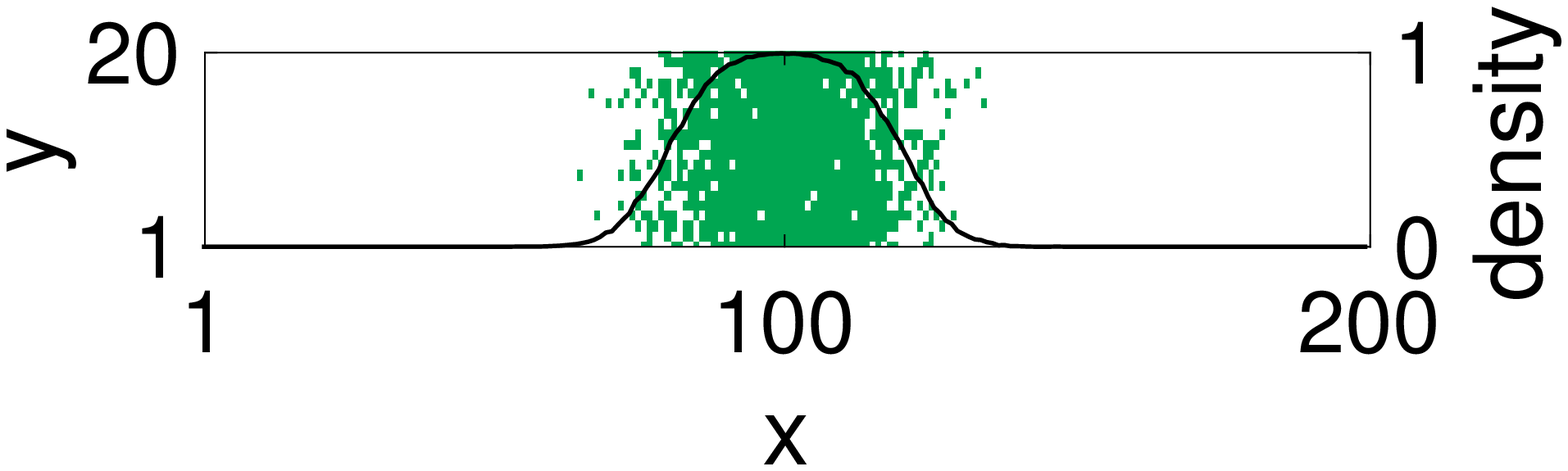}
\label{figure:sss_Ps=1_t=100}
}

\subfigure[]{
\includegraphics[width=0.475\textwidth]{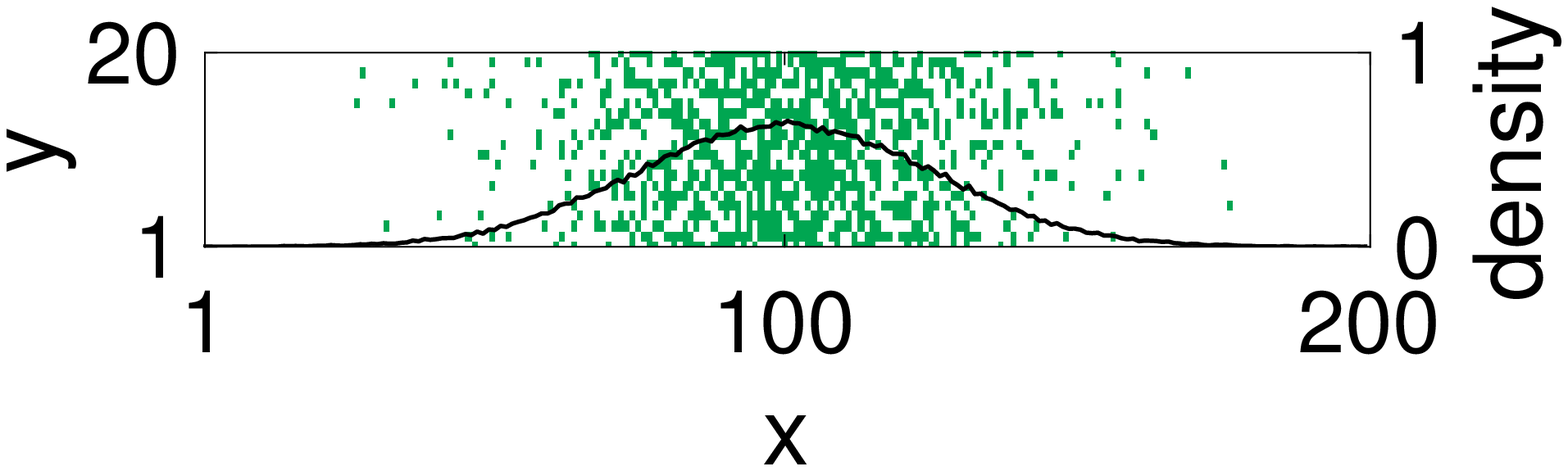}
\label{figure:sss_Ps=1_t=1000}
}

\end{center}
\caption{Snapshots of lattice occupancy for the single-species swapping model at $t=0, 100,1000$ with $r=1$ for swapping probabilities $\rho=0$ [\subref{figure:sss_Ps=0_t=0}-\subref{figure:sss_Ps=0_t=1000}], $\rho=0.5$ [\subref{figure:sss_Ps=0.5_t=0}-\subref{figure:sss_Ps=0.5_t=1000}] and $\rho=1$ [\subref{figure:sss_Ps=1_t=0}-\subref{figure:sss_Ps=1_t=1000}]. Agents are initialised on a domain with dimensions $L_x=200$ and $L_y=20$ such that all the sites in the range $81 \leqslant x \leqslant 120$ are occupied with agents (green) [\subref{figure:sss_Ps=0_t=0},\subref{figure:sss_Ps=0.5_t=0},\subref{figure:sss_Ps=1_t=0}]. Further snapshots of the IBM at $t=100$ and $t=1000$ show the dispersal of agents with time. The column-averaged density of agents over 100 runs of the ABM is also plotted (shown in black). We impose reflective boundary conditions on all four boundaries of the domain.}
\label{figure:sss_profiles}
\end{figure}

In Figure \ref{figure:sss_profiles} we present snapshots of the lattice occupancy for the single-species model at $t=0,100,1000$ and for swapping probabilities $\rho=0,0.5,1$ with $r=1$ with reflective boundary conditions. We see that swapping seems to have no effect on the dispersion of the agents at a macroscopic scale as the density profiles are indistinguishable regardless of the swapping probability. If two agents are identical to each other and unlabelled then exchanging positions by swapping is equivalent to an aborted movement attempt in the volume-exclusion process and produces no change in the state of the system.

In the next section, we derive the macroscopic PDE describing the evolution of the mean lattice occupancy in the single-species model.

\subsection{Single-species continuum model}\label{sec:appendix_single_species_pde}

Let $C_{ij}^k(t)$ be the occupancy of site $(i,j)$ on the $k$th repeat at time $t$, where $C_{ij}^k(t)=1$ if the site $(i,j)$ is occupied and $C_{ij}^k(t)=0$ otherwise. The average occupancy of site $(i,j)$ at time $t$ after $K$ runs is then given by,

\begin{equation}
    \langle C_{ij}(t)\rangle=\frac{1}{K}\sum_{k=1}^K C_{ij}^k(t).
\end{equation}
Let $C_{ij}(t) = C_{ij}$ for conciseness. By considering the possible movement events of the agent at the site $(i,j)$ during the small time step $\delta  t$ \citep{yates2015ipe,chappelle2019pmc}, we can write down the master equation for the occupancy of the site at time $t+\delta t$,

\begin{align}
    \langle C_{ij} (t+\delta t)\rangle-\langle C_{ij} \rangle&=\frac{r}{4}\delta t[ (1-\langle C_{ij} \rangle)(\langle C_{i-1,j} \rangle + \langle C_{i+1,j} \rangle + \langle C_{i,j-1} \rangle + \langle C_{i,j+1} \rangle) \nonumber \\
    & \quad -\langle C_{ij} \rangle(4-\langle C_{i-1,j} \rangle - \langle C_{i+1,j} \rangle - \langle C_{i,j-1} \rangle - \langle C_{i,j+1} \rangle)] \nonumber \\
    & \quad + \frac{r}{4}\rho \delta t \langle C_{ij} \rangle(\langle C_{i-1,j} \rangle + \langle C_{i+1,j} \rangle + \langle C_{i,j-1} \rangle + \langle C_{i,j+1} \rangle) \nonumber\\
    & \quad + \frac{r}{4}\rho \delta t \langle C_{ij} \rangle(\langle C_{i-1,j} \rangle + \langle C_{i+1,j} \rangle + \langle C_{i,j-1} \rangle + \langle C_{i,j+1} \rangle) \nonumber\\
    & \quad - \frac{r}{4}\rho \delta t \langle C_{ij} \rangle(\langle C_{i-1,j} \rangle + \langle C_{i+1,j} \rangle + \langle C_{i,j-1} \rangle + \langle C_{i,j+1} \rangle) \nonumber\\
    & \quad - \frac{r}{4}\rho \delta t \langle C_{ij} \rangle(\langle C_{i-1,j} \rangle + \langle C_{i+1,j} \rangle + \langle C_{i,j-1} \rangle + \langle C_{i,j+1} \rangle). \label{eqn:ss_occupancy_eqn}
\end{align}

\begin{figure}[t!]
\begin{center}

\subfigure[]{
\includegraphics[width=0.31\textwidth]{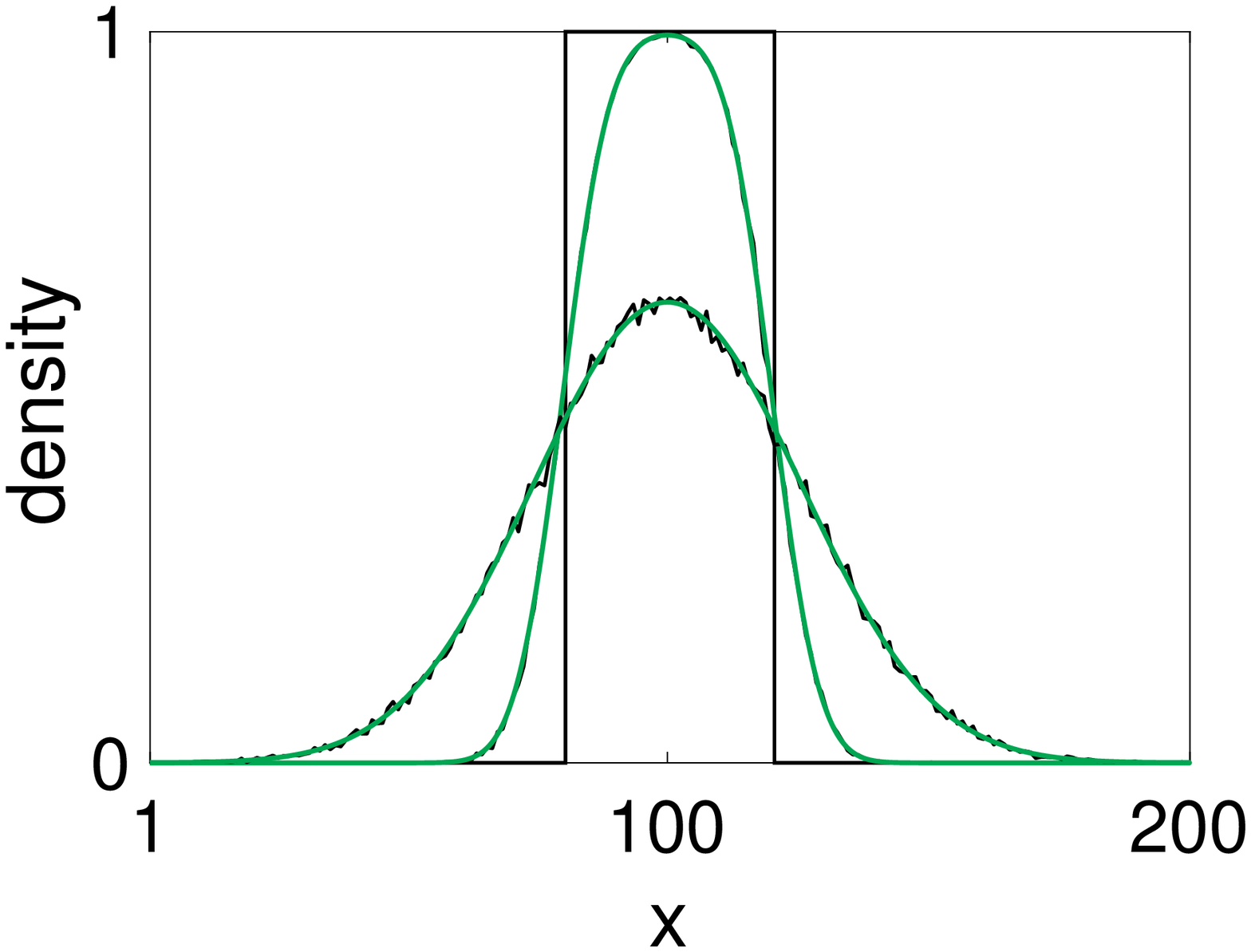}
\label{figure:sss_density_only_rho=0}
}
\subfigure[]{
\includegraphics[width=0.31\textwidth]{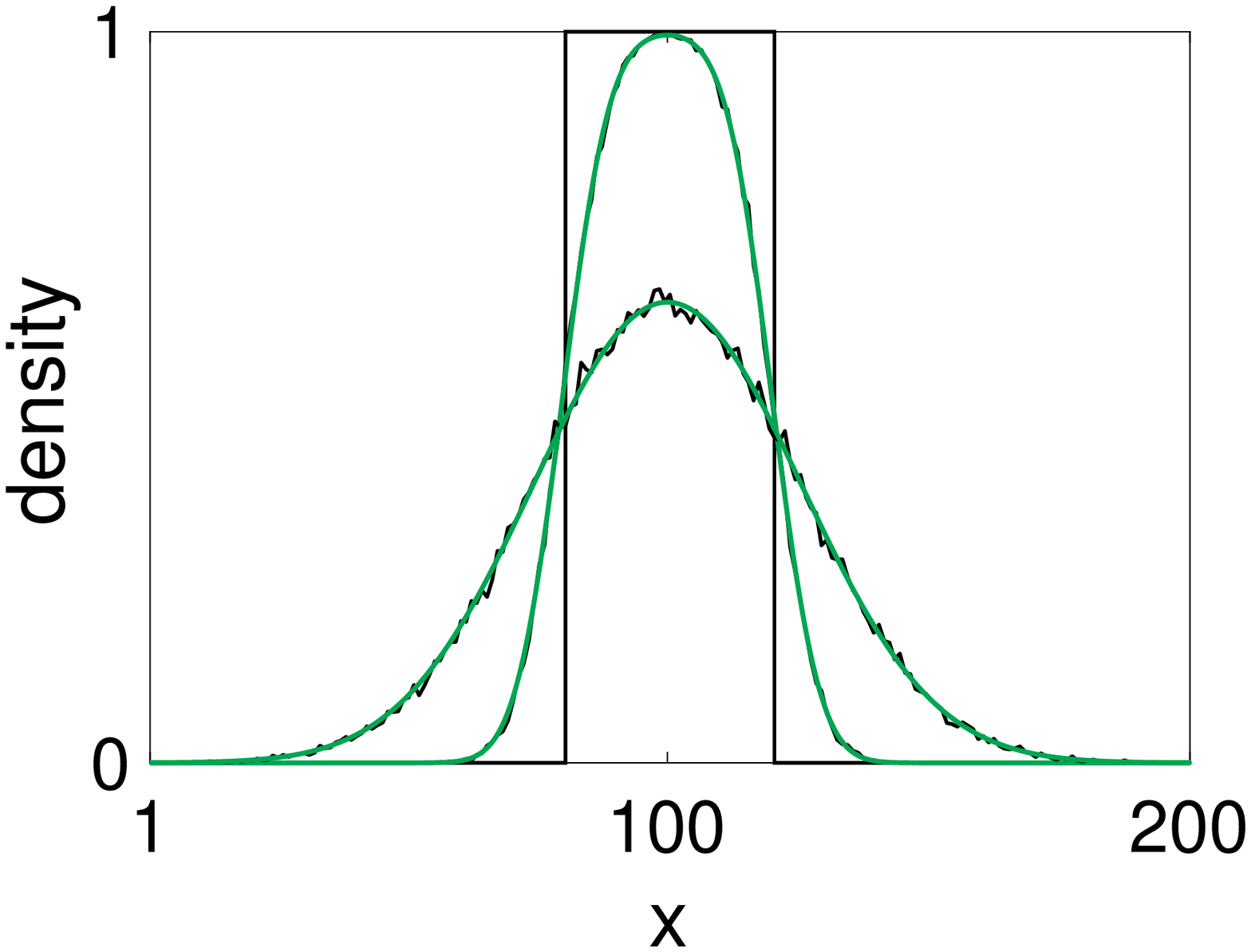}
\label{figure:sss_density_only_rho=0.5}
}
\subfigure[]{
\includegraphics[width=0.31\textwidth]{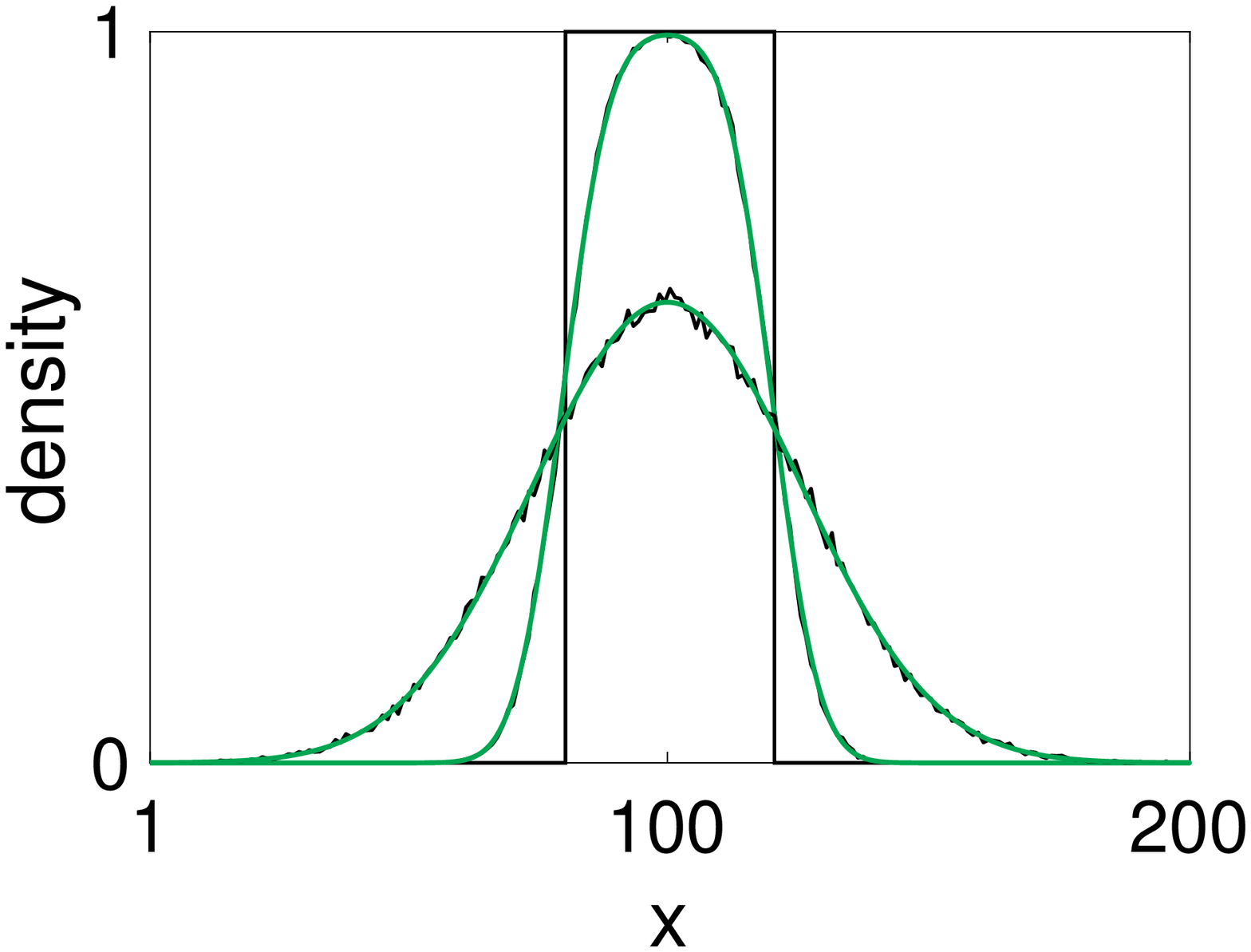}
\label{figure:sss_density_only_rho=1}
}
\end{center}
\caption{A comparison between the numerical solution of the one-dimensional version of the PDE \eqref{eqn:diffusion_eqn} and the averaged behaviour of the ABM with $r=1$ for $\rho=0$ \subref{figure:sss_density_only_rho=0}, $\rho=0.5$ \subref{figure:sss_density_only_rho=0.5} and $\rho=1$ \subref{figure:sss_density_only_rho=1}. The averaged densities are shown in black and the corresponding PDE approximation is shown in green. We present solutions at $t=0$, $t=100$ and $t=1000$. The black arrows show the direction of increasing time.}
\label{figure:density_compar_single_species}
\end{figure}

The site $(i,j)$ can gain occupancy if it is unoccupied at time $t$ and the agent from a neighbouring site moves in (first line on the RHS of Equation \eqref{eqn:ss_occupancy_eqn}). Similarly, the site $(i,j)$ can lose occupancy if the site is occupied at time $t$ and the residing agent jumps out to a neighbouring compartment, leaving the site $(i,j)$ empty (second line in Equation \eqref{eqn:ss_occupancy_eqn}). Movement of agents due to a successful swapping event is captured by the lines 3-6 in Equation \eqref{eqn:ss_occupancy_eqn}. However, since the agents are assumed to be identical and unlabelled lines 3-6 cancel each other out, eliminating the effect of swapping. Consequently, we are left with lines 1 and 2 only. Taylor expanding these remaining terms around the site $(i,j)$ up to second-order and taking the limit as $\mathrm{\Delta} \to 0$ gives the diffusion equation, as expected:

\begin{equation}\label{eqn:diffusion_eqn}
    \D C t = D \nabla^2  C.
\end{equation}
Here,

\begin{equation*}
    D = \lim_{\mathrm{\Delta} \to 0} \frac{r \mathrm{\Delta}^2}{4},
\end{equation*}
is the diffusion coefficient given that $\mathrm{\Delta^2}/\delta t$ is held constant in the diffusive limit. Equation \eqref{eqn:diffusion_eqn} describes the evolution of the lattice occupancy over time. It is well-known that for the simple exclusion process, the occupancy is described by the diffusion equation \citep{simpson2009mse,gavagnin2018sdm}. It makes sense therefore that in Figure \ref{figure:sss_profiles} swapping made no difference to the overall occupancy of the lattice. In Figure \ref{figure:density_compar_single_species} we compare the average column density of the ABM which is given by,

\begin{equation*}
    \overline{C}_i(t)=\frac{1}{L_y}\sum_{j=1}^{L_y} C_{ij}(t),
\end{equation*}
to the solution of the one-dimensional analogue of Equation \eqref{eqn:diffusion_eqn} with reflective boundary conditions by averaging the PDEs over the $y$ direction.

As expected, we see an excellent agreement between the two density profiles. We also see that there is perfect agreement even in the zero-swapping case which we did not see in the two-species example in Section \ref{sec:sm_pde}. This is because in a single-species system there is no way of distinguishing between a successful swapping of the position of a pair of neighbouring agents and an aborted movement event due to the volume-exclusion principle. This leads to identical profiles for different $\rho$ regardless of its magnitude.

In contrast to the two-species case, here there is perfect agreement between the ABM and the PDE profiles since in the single-species case the agents are indistinguishable from each other. In the two-species scenario, because the agents are competing for space, in the no-swapping situation one species affects the occupancy of the other, leading to spatial correlation and discrepancy between the ABM and PDE descriptions. Swapping serves to break up the correlation and improve agreement between the discrete and continuum models.

\section{Two-species individual-level diffusion coefficients}\label{app:two_species}

Let $P^{A}_{ij}(t)=P_{ij}$ be the probability that a type-A agent occupies the site $(i,j)$ and let $P^{B}_{ij}(t)=P^{B}_{ij}$ be the equivalent for species B. We can write down the master equation for species A and species B at time $t+\delta t$ where $\delta t$ is a small change in time,

\begin{align}
P^{A}_{ij}(t+\delta t)&=\frac{r_{A}}{4}\delta t(1-c)\left[P^{A}_{i-1,j}+P^{A}_{i+1,j}+P^{A}_{i,j-1}+P^{A}_{i,j+1} \right] + \frac{r_{A}}{4}c_{A}\rho\delta t \bigg[2P^{A}_{i-1,j} + 2P^{A}_{i+1,j} \nonumber\\
& \quad  + 2P^{A}_{i,j-1} + 2P^{A}_{i,j+1}\bigg] + \frac{r_{A}}{4}c_{B}\rho\delta t \bigg[P^{A}_{i-1,j} + P^{A}_{i+1,j} + P^{A}_{i,j-1} + P^{A}_{i,j+1}\bigg] \nonumber\\
& \quad  + \frac{r_{B}}{4}c_{B}\rho\delta t \bigg[P^{A}_{i-1,j} + P^{A}_{i+1,j} + P^{A}_{i,j-1} + P^{A}_{i,j+1}\bigg] + P^{A}_{ij}\bigg[1-r_{A}(1-c)\delta t \nonumber\\
& \quad - 2r_{A}\delta t c_{A}\rho - r_{A}\delta t c_{B}\rho - r_{B}\delta t c_{B}\rho  \bigg] \label{eqn:twospecies_master_eqn_m}\\
&= \left(\frac{r_{A}}{4}(1-c)+\frac{r_{A}}{2}c_{A}\rho+\frac{(r_Ac_B+r_Bc_B)}{4}\rho \right)\delta t (P^{A}_{i-1,j}+P^{A}_{i+1,j}+P^{A}_{i,j-1}+P^{A}_{i,j+1}) \nonumber\\
&\quad + P^{A}_{ij}+(-r_{A}(1-c) - 2r_{A}c_{A}\rho - (r_Bc_B + r_Ac_B)\rho)\delta t P^{A}_{ij} \nonumber
\end{align}

\begin{align}
P^{B}_{ij}(t+\delta t)&=\frac{r_{B}}{4}\delta t(1-c)\left[P^{B}_{i-1,j}+P^{B}_{i+1,j}+P^{B}_{i,j-1}+P^{B}_{i,j+1} \right] + \frac{r_{B}}{4}c_{B}\rho\delta t \bigg[2P^{B}_{i-1,j} + 2P^{B}_{i+1,j} \nonumber\\
& \quad  + 2P^{B}_{i,j-1} + 2P^{B}_{i,j+1}\bigg] + \frac{r_{B}}{4}c_{A}\rho\delta t \bigg[P^{B}_{i-1,j} + P^{B}_{i+1,j} + P^{B}_{i,j-1} + P^{B}_{i,j+1}\bigg] \nonumber\\
& \quad  + \frac{r_{A}}{4}c_{A}\rho\delta t \bigg[P^{B}_{i-1,j} + P^{B}_{i+1,j} + P^{B}_{i,j-1} + P^{B}_{i,j+1}\bigg] + P^{B}_{ij}\bigg[1-r_{B}(1-c)\delta t \nonumber\\
& \quad - 2r_{B}\delta t c_{B}\rho - r_{B}\delta t c_{A}\rho - r_{A}\delta t c_{A}\rho  \bigg] \label{eqn:twospecies_master_eqn_x}\\
&= \left(\frac{r_{B}}{4}(1-c)+\frac{r_{B}}{2}c_{B}\rho+\frac{(r_Bc_A+r_Ac_A)}{4}\rho \right)\delta t (P^{B}_{i-1,j}+P^{B}_{i+1,j}+P^{B}_{i,j-1}+P^{B}_{i,j+1}) \nonumber\\
&\quad + P^{B}_{ij}+(-r_{B}(1-c) - 2r_{B}c_{B}\rho - (r_Ac_A + r_Bc_A)\rho)\delta t P^{B}_{ij} \nonumber
\end{align}

We explain the meaning of the terms in the discrete-time master Equation \eqref{eqn:twospecies_master_eqn_m}. Equation \eqref{eqn:twospecies_master_eqn_x} can be interpreted similarly. The equations describe the evolution of the probability that a focal agent of type A is occupying site $(i,j)$ by considering possible movement events of the focal agent at site $(i,j)$ and agents at neighbouring sites. Firstly, the terms which correspond movements that increase the probability that the focal agent sits at position $(i,j)$ are as follows:

\begin{enumerate}
	\item The focal type-A agent residing at a neighbouring site moves into the empty site $(i,j)$ (line 1, term 1).
	\item The focal type-A agent at a neighbouring site initiates and successfully completes a swap with a type-A agent at site $(i,j)$ or, alternatively a type-A agent at site $(i,j)$ initiates the swap and exchanges position with the focal type-A agent at the neighbouring site (hence the multiplier 2 for the two equally likely probabilities) (term 2 on line 1 through line 2 up to first closing square bracket).
	\item The focal type-A agent at a neighbouring site initiates and successfully swaps with a type-B agent at site $(i,j)$ (term 2 on line 2). 
	\item A type-B agent at site $(i,j)$ initiates and successfully swaps positions with the focal type-A agent at a neighbouring site (term 1 on line 3).
	\end{enumerate}
	Secondly the terms which correspond to the site already being occupied and no event occurring to change that state correspond to the terms inside the pair of square brackets which spans lines 3 and 4. Recalling that the probability of nothing happening in the time interval $[t,t+\delta t)$ is unity minus the probability that something happens, the terms in square brackets can be described as follows:
	\begin{enumerate}
	\item The unit corresponds to the probability that the focal agent occupied site $(i,j)$ at time $t$.
	\item The second term correspond to the focal type-A agent at site $(i,j)$ moving to an unoccupied neighbouring site.
	\item The third term corresponds to the probability that the focal agent of type A at site $(i,j)$ initiates and successfully swaps with a type-A agent at a neighbouring site and the probability that a type-A agent at a neighbouring site initiates and successfully swaps with the focal agent of type A at site $(i,j)$ (hence the factor of 2 for these equally probably events).
	\item The fourth term corresponds to the probability that the focal type-A agent at position $(i,j)$ initiates and successfully swaps with a neighbouring agent of type B.
	\item Finally, the fifth term in the square brackets corresponds to a type-B agent at a neighbouring site successfully swapping positions with the type-A focal agent occupying site $(i,j)$.
\end{enumerate}

Subtracting $P^{A}_{ij}$ from both side and dividing by $\delta t$ and taking the limit as $\delta t \to 0$ gives,

\begin{align}
\frac{d P^{A}_{ij}}{dt}&= \left(\frac{r_{A}}{4}(1-c)+\frac{r_{A}}{2}c_A\rho+\frac{(r_{A}+r_{B})}{4}c_{B}\rho \right) (P^{A}_{i-1,j}+P^{A}_{i+1,j}+P^{A}_{i,j-1}+P^{A}_{i,j+1}) \nonumber\\
&\quad + (-r_{A}(1-c) - 2r_{A}c_A\rho - (r_{A} + r_{B})c_{B}\rho)P^{A}_{ij},
\end{align}

\begin{align}
\frac{d P^{B}_{ij}}{dt}&= \left(\frac{r_B}{4}(1-c)+\frac{r_B}{2}c_B\rho+\frac{(r_B+r_{A})}{4}c_{A}\rho \right) (P^{B}_{i-1,j}+P^{B}_{i+1,j}+P^{B}_{i,j-1}+P^{B}_{i,j+1}) \nonumber\\
&\quad + (-r_B(1-c) - 2r_Bc_B\rho - (r_{A} + r_{B})c_{A}\rho)P^{B}_{ij}.
\end{align}

After using the definitions in Equation \eqref{eqn:mth_moment_i} and \eqref{eqn:mth_moment_j}, and suitably transforming the indices $i$ and $j$, after simplifying it can be shown that the equations describing the evolution of the second moment of the position $i$ and $j$ of species A and B are given by,

\begin{align}
\frac{d (\langle i^{2}\rangle+\langle j^{2}\rangle)_{A}}{dt}&= r_{A}(1-c)+2r_{A}c_A\rho+(r_{A}+r_B)c_B\rho,
\end{align}

\begin{align}
\frac{d (\langle i^{2}\rangle+\langle j^{2}\rangle)_{B}}{dt}&= r_B(1-c)+2r_Bc_B\rho+(r_{A}+r_{B})c_A\rho. 
\end{align}

Given that $\langle i(0)\rangle=\langle j(0)\rangle=0$,

\begin{equation*}
(\langle i^{2}\rangle+\langle j^{2}\rangle)_{A} = r_{A}(1-c)+2r_{A}c_{A}\rho+(r_{A}+r_{B})c_B\rho t,
\end{equation*}

\begin{equation*}
(\langle i^{2}\rangle+\langle j^{2}\rangle)_{B} = r_B(1-c)+2r_Bc_B\rho+(r_A+r_{B})c_A\rho t.
\end{equation*}

Thus,

\begin{align}
    D_{A}^\star&=\frac{1}{4}(r_A(1-c)+(r_Ac+r_Ac_A+r_Bc_B)\rho),\\
    D_{B}^\star&=\frac{1}{4}(r_B(1-c)+(r_Bc+r_Ac_A+r_Bc_B)\rho).
\end{align}

are the individual-level time-uncorrelated diffusion coefficients of species A and B, respectively.

\bibliographystyle{apsrev4-1} 
\bibliography{my_Library.bib}



\end{document}